\documentstyle[graphicx,wrapfig,epsfig,subfigure,multirow,longtable,cite,,fancyhdr,12pt]{iiscthes}
\begin{document}
\begin{frontmatter}
%
%
\title{Robust and Precision Satellite Formation Flying Guidance Using Adaptive Optimal Control Techniques}

\author{Girish Joshi}

\submitdate{July 2013} \dept{Aerospace Engineering Department}
\enggfaculty \degreein{Aerospace Engineering}

\me
\iisclogotrue 
\tablespagetrue 
\maketitle
\begin{dedication}
\begin{center}

\begin{dedication}
\begin{center}
\Large \textbf{Dedicated} \\[0.7em]
\Large \textbf{to}\\[0.7em]
\Large \Large \textbf{My wife Sushma \\and\\ My Parents}\\
\end{center}
\end{dedication}

\ \ \ \ \ \ \

\thispagestyle{empty}\setcounter{page}{0} 
\end{center}
\end{dedication}
\acknowledgements

I would like to thank my research advisor Dr. Radhakant Padhi for his constant guidance, discussions, support and constructive
criticism. I also want to thank all my course instructors for having enriched my knowledge on various research topics.

I am also grateful to Shivendra, Arnab, Pradeep, Nitin, Chandrashekar and other members of Integrated Control, Guidance and Estimation Laboratory(ICGEL) for creating a constructive research environment in the lab. It is through the discussions with lab members that i have gained the knowledge on various problems in the areas of aerospace and control design.

I would like to take the opportunity to thank our Chairman Prof.D.Ghose and Prof M. S. Bhat for all the research facilities at department and his motivating talks. I am thankful to GD, SMG and ISAC (ISRO) for providing me this opportunity and encouragement to take up my Masters degree at IISc. I will not miss to thank the staff at aerospace main office and STC who have been most humble and have always shown readiness to help.

Last but not the least, I am immensely grateful to my wife Sushma and my parents for understanding me and giving me motivation and moral strength.

\begin{abstract}
A emerging trend in space technology is small satellite formation flying. The small satellite technology has opened a new era of satellite engineering by decreasing space mission cost, without reducing the performance. Advancement in electronic miniaturization, data compression and data handling, imaging technology and autonomous intelligence has rocketed the small satellite technology by leaps and bounds. Mission involving formation flying of small satellite are providing economical alternative for one single large spacecraft missions. The new technologies such as, formation flying algorithms, constellation self-reconfiguration, accurate precision algorithms, developed for small satellites are often later used on major missions, involving large spacecrafts. Formation flying mission support diverse application areas, from mission involving distributed monitoring in space which involves mission like geomagnetic study of earth, solar observatory, deep space observatories to close coordinated flying mission like remote sensing, Geo-positional systems etcetera.

The main focus of the work presented in this thesis is to develop a optimal control based formation flying control strategy for high precision formation flying of small satellites which have restricted computation and storage capacity. Using the recently developed model predictive static programming (MPSP), and Generalized MPSP algorithm a suboptimal guidance logic is presented for formation flying of small satellites. Due to the inherent nature of the problem formulation, MPSP does not require the system
dynamics to be linearized. The proposed guidance scheme is valid both for high eccentricity chief satellite orbits as well as large separation distance between chief and deputy satellites. Moreover, since MPSP poses the desired conditions as a set of `hard constraints, the final accuracy level achieved is very high. Comparative study with standard Linear Quadratic
Regulator (LQR) solution (which serves as a guess solution for MPSP) and another nonlinear controller, Finite time State Dependent Ricatti Equation (SDRE) reveals that MPSP guidance achieves the objective with higher accuracy and with lesser amount of control usage.

Another innovative nonlinear online trajectory optimization technique is presented in this thesis which utilizes the well-known linear quadratic regulator (LQR) theory and augmenting it with online trained neural networks, . Two sets of neural networks are used. One to drive the LQR controller towards the optimal control for the nonlinear system and other is used to capture the unmodeled dynamics. Both sets of neural networks are trained online using `Closed form expressions' and do not require any iterative process. The overall structure leads to robust optimal control synthesis and works well despite the presence of
unmodeled dynamics. This control strategy is experimented with formation flying catering to large initial separation, high eccentricity orbits, uncertain semi-major axis of chief satellite and under influence of external perturbation such as  $J2$ gravitational effects. The online optimized LQR controller successfully drives the deputy satellite to desired final orbit under influence of uncertainties with very minimum tracking errors.
\end{abstract}
\publications
{\LARGE Conferences}
\begin{enumerate}
\item Girish Joshi, Radhakant Padhi, Formation Flying of Small Satellites using Suboptimal MPSP Guidance, {\em American control conference,IEEE}, Washington DC, 2013
\item Girish Joshi, Arnab Maity, Radhakant Padhi, Formation Flying of Small Satellites using Suboptimal G-MPSP Guidance {\em Guidance, Navigation, and Control Conference, AIAA }, Boston, 2013
\item Girish Joshi, Radhakant Padhi, Robust Satellite Formation Flying Using Dynamic Inversion with Modified State Observer, {\em Multi System Conference, Computer-Aided Control System Design and Systems with Uncertainty(CACSD-SU) IEEE}, Hyderabad 2013
\end{enumerate}\vspace{5mm}
{\LARGE Journals}
\begin{enumerate}
\item Girish Joshi, Arnab Maity, Radhakant Padhi, Formation Flying of Small Satellites using Suboptimal G-MPSP Guidance,  {\em Journal of Guidance, Control and Dynamics, AIAA} (to be submitted)
\item Girish Joshi, Arnab Maity, Radhakant Padhi, Closed-Form Solution to Finite-Horizon Suboptimal Guidance for Formation Flying of Small Satellites, {\em Optimal Control Applications and Methods} ( to be submitted )
\item Girish Joshi, Radhakant Padhi, Robust Satellite Formation Flying using Neural Networks Augmented LQR, {\em Acta Astronautica} ( to be submitted )
\end{enumerate}
\makecontents
\notations
\begin{tabular}{p{3cm}p{10cm}}
    $X$  & State vector of the dynamical system \\
    $U$  & Input (guidance) vector of the dynamical system \\
    $Y$  & Output vector of the dynamical system \\
    $\lambda$  & Lagrange multiplier for MPSP algorithm \\
    $t$  & Time, $s$ \\
    $\triangle t$ & Sample time\\
    $\delta X$  & Error of state \\
    $\delta U$  & Error of control \\
    $R$  & Positive definite weighting matrix for G-MPSP algorithm \\
    $U^{p}$  & Previous control history \\
    $x$  & Position state x in Hills Reference frame \\
    $\dot x$ & Velocity in x direction in Hills Reference frame \\
    $y$  & Position state y in Hills Reference frame \\
    $\dot y$ & Velocity in y direction in Hills Reference frame \\
    $z$  & Position state z in Hills Reference frame \\
    $\dot z$ & Velocity in z direction in Hills Reference frame \\
    $\rho$  & Spatial separation of deputy with respect to chief satellite in Hills frame of Reference\\
    $\theta$ & Angle of satellite position vector with respect to
    chief satellite velocity vector\\
    $a,b$  & Center offset of ellipse traced by deputy satellite with
    respect to chief satellite in Hills frame\\
\end{tabular}
\newpage
\thispagestyle{fancyplain} \fancyhead[L]{\bfseries NOTATION AND
ABBREVIATIONS} \fancyhead[R]{\thepage}
\begin{tabular}{p{3cm}p{10cm}}
    $m,n$ & Slopes of the line formed by rotation of ellipse about minor and major axis respectively\\
    $\nu$  & true anomaly \\
    $\dot \nu$ & True anomaly rate \\
    $\ddot \nu$ & True anomaly acceleration \\
    $\mu$ & Gravitational Parameter: $\mu = GM{\rm{ =
    398601}}\;{\rm{k}}{{\rm{m}}^{\rm{3}}}{\rm{/}}{{\rm{s}}^{\rm{2}}}$\\
    $r_{c}$ & Radius of chief satellite orbit measured from center of
    earth\\
    $\omega$ & $\sqrt {\frac{\mu }{{r_c^3}}}$\\
    $E$ & State Error\\
    $e_i$ & Channel wise state Error\\
    $W_i$ & $i^{th}$ Network weight\\
    $\tilde W_i$ & Approximate $i^{th}$ network weights\\
    $\hat W_i$ & Error in $i^{th}$ network weights\\
    $ \phi $ & Basis Function or Activation function\\
    Subscript \\
    $f$  & Terminal value \\
    $N$  & At the time step $N$ \\
    $e$ & Error\\
    $f$ & Final values\\
    $c$ & Chief Satellite\\
    $d$  & Deputy Satellite \\
    $i$  & Channel \\
     Superscript \\
    $0$ & Previous value \\
    $*$  & Desired value \\
\end{tabular}
\end{frontmatter}
\chapter{Introduction} \label{Chapter:Introduction}

An emerging trend across the globe is to have missions involving many small, distributed and largely inexpensive
satellites flying in formation to achieve a common objective. Satellite formation flying enables new application areas such as
spar antenna arrays for remote sensing, distributed sensing for solar and extra-terrestrial observatories, interferometry
synthetic aperture radar and many more. Missions involving conventional large satellites are usually quite expensive to design, fabricate, launch and operate as they require massive investment on infrastructure and support system. In addition, in general they require large control forces and moments for their trajectory and attitude corrections, which has been an important factor for the limited life span of the satellites as well. Consequently, an emerging trend across the globe is to have missions involving many small, distributed and largely inexpensive satellites. Since it is feasible to do so, many space research projects in university laboratories are also focused on the development of small to very small satellites (i.e. micro, nano and even pico satellites). Note that new technologies such as formation flying and reconfiguration algorithms developed for small satellites can be used on major missions involving large spacecrafts as well

Due to their limited size and weight, small satellites can not achieve many missions on their own. Hence, there is a strong need to have missions involving multiple small satellites. In view of this, Satellite Formation Flying (SFF) has become popular because of the potential to perform coordinated missions enhancing their overall capability substantially. Some applications require distributed systems such as employing constellations of small satellites optimally configured to achieve global cover. Yet, other space missions need fairly centralized systems (e.g., remote sensing of wide area, communications systems etc.), where high precision formation flying with close proximity is a strong requirement. Satellite formation flying enables distributed sensing and spar antenna arrays for remote sensing, gravitational mapping, solar observatories, interferometry synthetic aperture radar and many more. In extra-terrestrial applications, SFF enables variable baseline interferometry and large scale distributed sensors that can probe origin and structure of stars and galaxies with high precision.

\section{Small Satellite mission}
Over last few decades there has been immense advancement on the miniaturization of electronics through advances in semiconductor technology. A a result of this there is ever growing interest in development of small to micro payloads and above all miniature satellites themselves. Primarily the interest in the smaller and lighter payloads and satellites in driven by the capital involved in launch and operation of the conventional large satellites. Currently around $25000\$$ per kg of payload at liftoff is the launch cost.

Traditional satellites building and launch are budgeted in few million dollars, hence any on orbit failure or malfunction cost huge capital and are single point failures. Hence for this reason the conventional satellites are build to be highly reliable using conventional and well proven technology hence leaving very less space for experimentation and innovation. The satellite realization time is long and hence there is no flexibility in design of mission, since the objective of the mission is already frozen at the inception of the satellite payload, layout and design.

Where as the small satellites provide the advantage over the conventional satellites is there launch cost is reduced tremendously as they fly as ``piggy back`` with conventional prime mission satellites. Small satellites mission duration are restricted to couple of months to over a year the material used in building these satellites can be ``Conventional of the shelf`` (COTS) materials unlike in big satellites which need test and proven material for space usage, hence the satellite realization cost and time is much lower. Besides time and money involved in the small satellite development, the main motivation in using the small satellites is opportunity to enable mission that a conventional satellite with added advantage of on orbit mission objective flexibility, redundancy and multi-point observation, which a conventional satellite fall short to achieve.

\section{Small satellite classification on their launch mass}
\begin{itemize}
\item Mini-Satellite $\left(100-500kg\right)$:Mini-satellites are also termed as small satellites. The technologies used in building the mini satellite are usually borrowed from the conventional satellites, where as the difference might be in the number of payloads such as transponders or over all capability and longevity of the life of satellite due to reduced payload and power generation capability. These satellites are used in the application of communication, remote sensing, weather monitoring, solar and geomagnetic observation and many more such application. These satellites are equipped with chemical/ION thruster for orbit and station keeping activities.
\item Micro-Satellites $\left(10-100kg\right)$: Micro-satellite are name coined for satellites between $10kg$ to $100kg$ of mass. However some time micro satellite can be marginally over $100kg$ as well. Usually these satellites are used for remote sensing purpose. They use cold gas thruster or spin stabilization technique.
\item Nano-Satellite $\left(1-10kg\right)$ : Nano-satellite are the one classified as $\left(1-10kg\right)$ range.
\item Pico-Satellite $\left(0.1-1kg\right)$ : Pico-Satellite or picosat are $\left(0.1-1kg\right)$ range. CUBESAT and Palmsat are few of the examples for picosatellite.
\item Femto-Satellite $\left(<0.1kg\right)$: Femto satellites are $<0.1kg$ satellites. these satellites are used in mission where array of hundreds of satellite are needed for discrete measurement and sparse sensing for mission like observation of sola activity, multi-point sensing in remote sensing missions. The advents in micro and nanotechonolgy has made it possible to replicate the functionality of entire satellite on a printed circuit board. Considerable effort is being made in this direction of developing femto-satellites, example PCBsat.
\end{itemize}
\section{ Satellite Formation Flying: Classification}
Depending on the configuration, mode of operation etc., SFF can be classified into several categories. Three most formation flying architectures that are most commonly used are as follows:
\begin{itemize}
\item Trailing/ Leader Follower (Figure \ref{TRAILING}): Trailing formations are one where spacecrafts share same orbit and follow each other on same path at specified distance. The follower spacecraft will follow a path defined by the leader's position. A follower spacecraft may have the ability to operate without the intervention of the leader, controlling and maintaining a desired relative position. Generally leader-follower architecture has generally been implemented whereby the leader satellite follows a natural orbit trajectory, and controllers on-board the follower spacecraft to maintain the formation based on relative position measurements. The nature of the communications between spacecraft is dependent on formation control hierarchy. For most applications, the transmitted data would be in the form of guidance functions from the leader to the follower spacecraft, for example, desired relative orbit trajectories and relative position measurements.
\begin{figure}
\begin{center}
\includegraphics[width=4.5in]{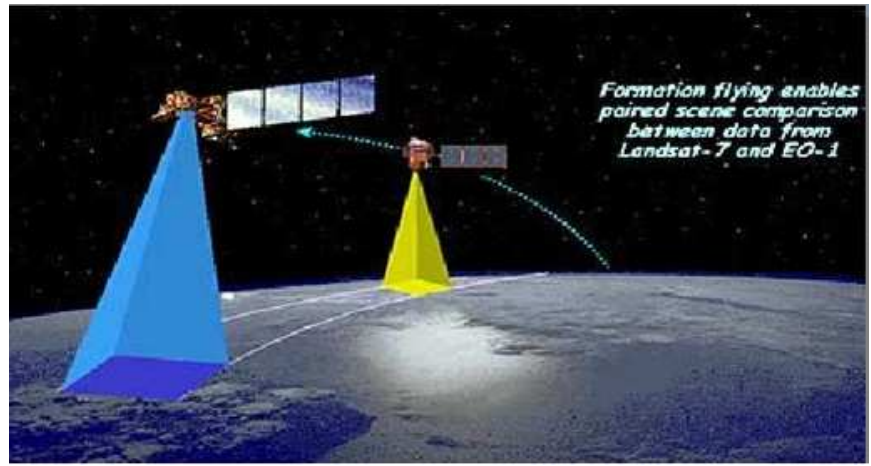}\\
\caption{Pictorial representation of Trailing formation \cite{Wikipedia}}\label{TRAILING}
\end{center}
\end{figure}
\item Constellation: (Figure \ref{const_form}) Regularly spaced satellites with separation on a global scale. Constellation normally consists of set of satellites in organized orbital plane that cover entire earth. Note that the global positioning system (GPS) is the most prominent example of constellation flying.
\begin{figure}
\begin{center}
\includegraphics[width=4in]{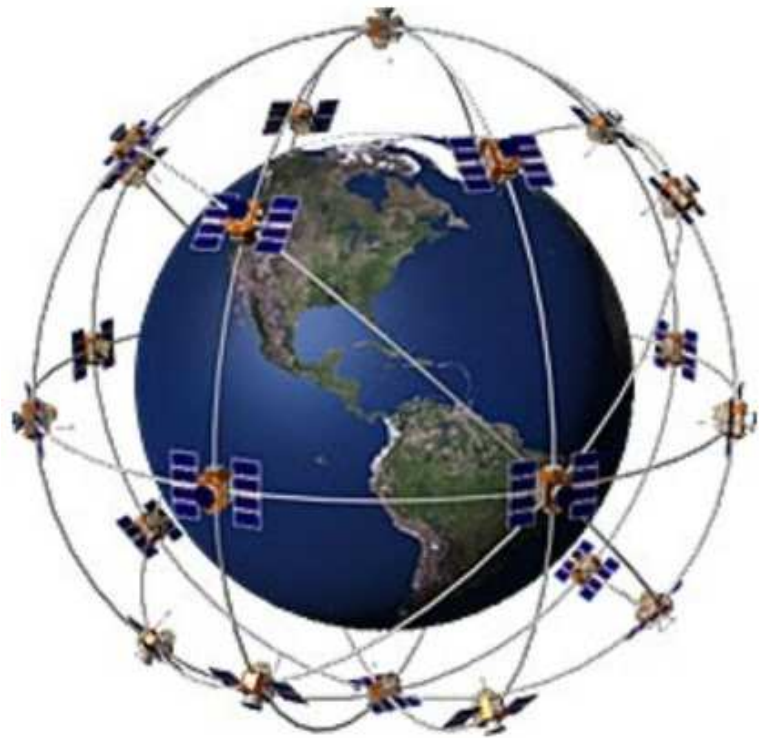}\\
\caption{Pictorial representation of Constellation formation \cite{Wikipedia}}\label{const_form}
\end{center}
\end{figure}

\item Cluster:(Figure \ref{CLUSTER}) Group of satellites are located in formation close to each other and are placed in orbits such that they remain in cluster. A 'Cluster' includes any group of two or more spacecraft whose cooperation and knowledge of relative position is essential for completion of the mission. The term generally implies a level of spacecraft inter-dependency, but does not imply that precision formation keeping is required. While a cluster is not a constellation formation, it is not possible to specify an upper limit to spacecraft separation distance for this definition, although a cluster would usually operate in a closer formation than a constellation. For a cluster formation (of more than two spacecraft), the followers may require little on board processing capability, but sufficient to obey the commands of the master spacecraft.
\begin{figure}
\begin{center}
\includegraphics[width=4.5in]{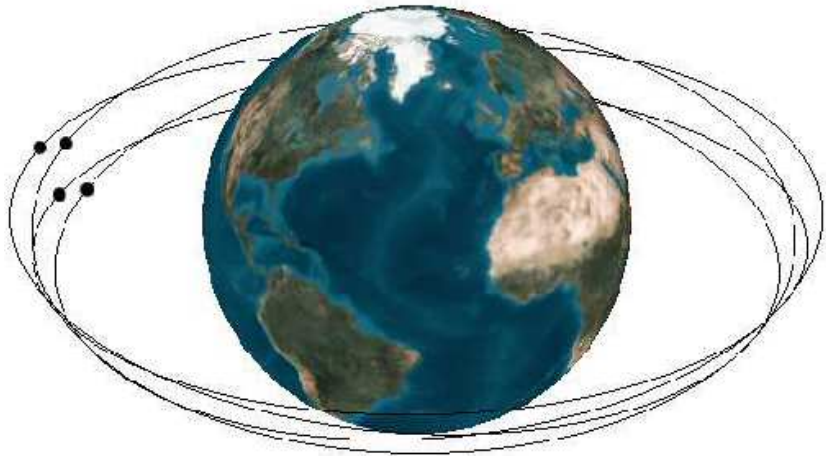}\\
\caption{Pictorial representation of Cluster formation \cite{Lorenz}}\label{CLUSTER}
\end{center}
\end{figure}
\end{itemize}
\section{Satellite Formation Flying Control Approaches}
Formation flying can be achieved by two different approaches:
\begin{itemize}
\item Ground Based Control: In Ground based control, satellite orbital parameters and current conditions are first communicated to ground based stations. These are then utilized in doing necessary computations and finally each satellite in formation is controlled through ground control center by transmitting back the necessary commands to the orbiting satellites to put the satellites into appropriate position in formation. However, this approach is adequate for formation with separation between spacecrafts is relatively large (e.g. of order of few kilometers) and more or less restricted only to those missions that do not require dynamic adjustment of formation orbits
\item Autonomous Control: In Autonomous flying the orbital parameters and current conditions are shared between spacecrafts. Next, the necessary computations are done onboard the satellites to generate the necessary commands of various satellites in formation. Note that autonomous formation flying algorithms can be implemented either in 'centralized' and 'de-centralized' control architectures (see Figure \ref{centralized}).
\begin{itemize}
\item : Centralized Control : In centralized architecture, a central node does all the necessary computations and transmits the necessary control actions to the other nodes.
\item De-Centralized Control : On the other hand, in de-centralized architecture, each satellite processes the available information and determines its own control actions. Looking from a small satellite mission point of view, the decentralized architecture is more appealing in general as the onboard processing power of each satellites is limited.
\item Distributed Control:  A Local Control solution should use model of neighboring satellites in order to compute the control strategy, to form a new formation or to maintain the formation flying, with respect to its neighboring systems. In ideal case satellite should perform computation to solve the complete nonlinear local problem and thereby collectively develop a globally stabilizing distributed controller with good performance from local controllers.

    The constructed distributed controllers are stabilizing and are dependent on local controller tuning parameters of cost function. (Details are included in numerical results discussion section). This design philosophy leads to a controller for finite number of dynamically identical coupled systems, where local tuning parameters can be chosen to achieve a desirable global performance.

    \vspace{5mm}
    This design approach has following advantages
\begin{itemize}
\item Global controller is asymptotic stable.
\item Basic design is simple dealing with one controller (SDRE/MPSP) at a time to compose global distributed controller for given identical coupled dynamical systems.
\item Solution scheme requires solution of low dimensional problem (Characterized by number of neighboring satellites considered in formation. For our study only one deputy satellite is considered in formation with a chief satellite. The idea of single satellite in formation is more near to practical application in many scenarios. In sparsely placed formation like constellation flight, the separation is large and hence only two body problem suffices the required design constraints) compared to full centralized problem(n-body problem) which renders itself quite complex to be solved on small-spacecrafts onboard.
\item As global controller in constructed from collection of many single local controllers, the design approach is modular. Adding or removing satellite from the formation does not require change in controller design, as long as maximum number of neighbors does not increase.
\end{itemize}

\begin{figure}
\begin{center}
\includegraphics[width=5in]{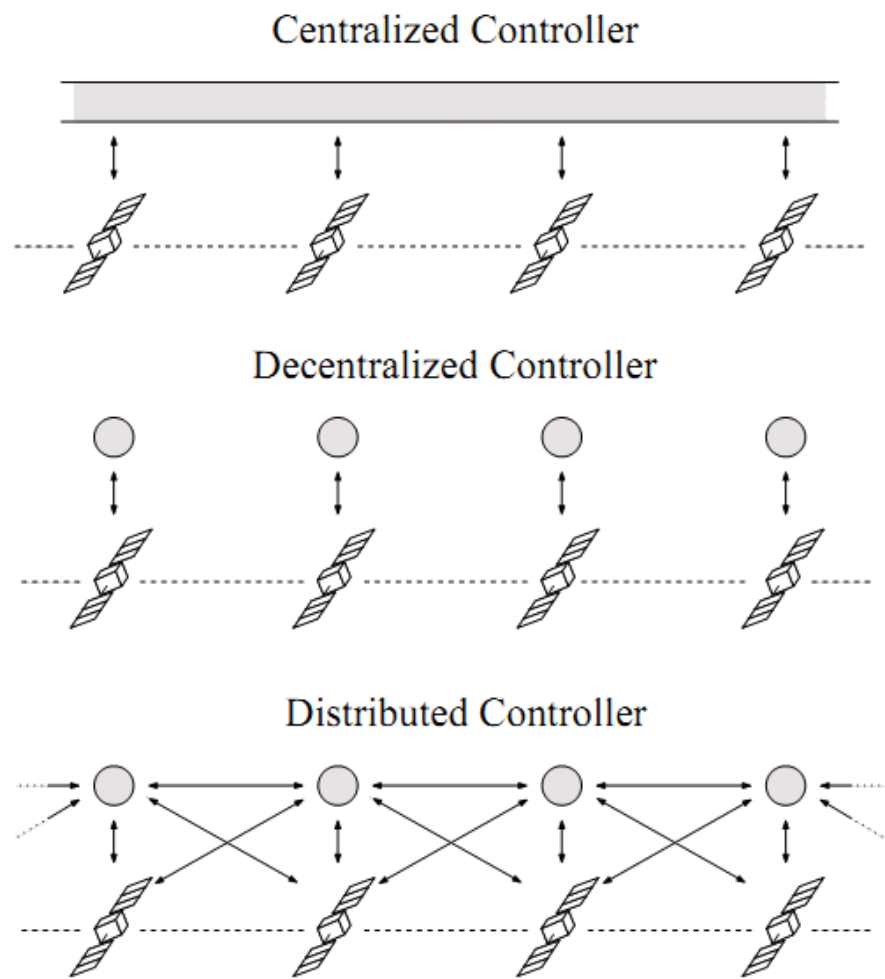}\\
\caption{Pictorial representation of Centralized, De-centralized and Distributed computing in SFF \cite{massioni2008new}}\label{centralized}
\end{center}
\end{figure}

The distributed controller requires minimal communication link between satellites for their relative position update, unlike in centralized control, the communication channel requirement is heavy as the control to be actuated in deputy craft and relative position information is updated from chief satellite. The distributed controller proposed in this document is of modular nature. Addition or Loss of any single satellite (which is common scenario in small satellite formation flying) can be accommodated easily which ensures the mission flexibility. Onboard computation, power requirement and inter satellite communication needs are not high which is more suited for small spacecraft which have limitation on power generation and large payload carrying capabilities
\end{itemize}
\end{itemize}

\section{Advantages of the Satellite formation Flying}
\quad Some of the important advantages of formation flying of satellites can be summarized as follows:
\begin{itemize}
\item Higher redundancy across the formation and improved fault tolerance;
\item On-orbit reconfiguration within the formations offers multi-mission capability by integrating new technology during mission and design flexibility
\item Mission improvements through the ability to view objects from multiple angles or at multiple times;
\item Lower individual launch mass and smaller spacecraft volume translates into a reduced launch cost and an increased launch flexibility;
\item Minimal financial lost in case of failure
\end{itemize}

\section{Organization of the Report}
Here, a brief outline of the report is given, which highlights the contribution of each chapter.

In \textbf{Chapter~$\mathbf{2}$}, A review of the existing literature is presented. This chapter describes the literature addressing both situations of ``\textit{what have been done}" and ``\textit{what have to do}". Particularly, the present scenario of the literature along with the related work (to our problem), is discussed.

In \textbf{Chapter~$\mathbf{3}$}, A brief discussion on orbital dynamics is introduced in this chapter, related orbital dynamics terminologies and definition are presented. Equation on motion of two body problem and relative motion in ECI frame and non inertial Hills frame is introduced. Linearization technique under special assumption of equation of motion of relative satellite dynamics is presented. This chapter concludes by introducing the details of perturbation forces on satellite and giving details on mathematical approach to $J_2$ modeling.

In \textbf{Chapter~$\mathbf{4}$}, Infinite time LQR controller for satellite formation flying is discussed. The linear plant model introduced in previous chapter is used to synthesize the linear controller. Simulation results are discussed to elaborate the effectiveness of LQR controller for circular and small $\rho$ formation problem.

In \textbf{Chapter~${5}$}, A state Dependent Ricatti Equation solution to nonlinear model is discussed. Two State Dependent Coefficient (SDC) formulation of the nonlinear plant model is discussed. A comparative study is done of SDRE solution in infinite and finite time domain solution with two distinct SDC models.

\textbf{Chapter~$\mathbf{6}$} A Suboptimal guidance logic is presented for satellite formation flying problem using a MPSP algorithm. A comparative study is presented of the simulation results of MPSP and SDRE control techniques.

\textbf{Chapter~$\mathbf{7}$} Another suboptimal guidance logic is presented for satellite formation flying problem using a G-MPSP algorithm. A comparative study is presented of the simulation results of G-MPSP and SDRE control techniques.

\textbf{Chapter~$\mathbf{8}$} This chapter discusses a novel robust controller using LQR base line controller is presented. This chapter introduces the online optimized controller with baseline linear LQR controller. The unmodeled dynamics and external perturbation is considered as state dependent disturbance term. Neural networks are implemented to approximate the unknown disturbance term and augment the line LQR control to cater to the nonlinear plant and $J_2$ perturbation.

\textbf{Chapter~$\mathbf{9}$} This chapter concludes the thesis with presenting a brief summary and future scope of the work.

\section{Summary and Conclusions}
This chapter primarily introduces the concept and necessity of Satellite formation flying mission. A brief discussion on the classification of formation and satellites involved in this mission are discussed. The various control strategy like ground based , Autonomous control are discussed. This chapter motivates the control strategy development for formation flying of satellites and lays foundation for further chapters. Next chapter discusses the available literature in formation flying and optimal control strategies.

\cleardoublepage 
\chapter{Literature Survey} \label{Chapter:LIT_SURVEY}
One of the key issues in successful small satellite missions is to come up with efficient and robust guidance logics. In fact,
some interesting guidance strategies for reconfiguration and formation flying have been reported in the recent literature. Few
to mention are, in the framework of optimal control Vadali et al. \cite{Vadali} have proposed an optimal control theory based
solution for the problem of formation flying of satellites. H.Ahn et al. \cite{Ahn} have developed a robust periodic learning
control for trajectory keeping in SFF under time periodic influence of external disturbance such as gravitational
perturbation, solar radiation pressure and magnetic field. Park et al. \cite{Park} have developed a state dependent Ricatti equation
(SDRE) solution for SFF reconfiguration and station keeping. Linear quadratic performance study is done on formation flying in presence of gravity perturbation by Sparks \cite{sparks2000satellite}. Lyapunov based adaptive nonlinear control law for multi-spacecraft formation flying under influence of disturbance force is developed by V.kapila et.al \cite{de2000adaptive}. Schaub and Alfriend have developed near optimal impulsive feedback control, to establish specific relative orbit of the spacecraft formation flying using Gauss variational equation of motion \cite{schaub2001impulsive}. A optimal control based satellite formation guidance under atmospheric drag and $J_2$ perturbation is developed by Mishne \cite{mishne2004formation}, Minimum fuel, Multiple Impulse optimal control strategy is developed by Prussing et.al for circle to circle rendezvous and time vs. fuel optimization for time constraint mission like rescue and collision avoidance \cite{prussing1986optimal}. Irvin \cite{Irvin} has carried out some interesting comparison studies
for various linear and nonlinear control technique applied to SFF such as LQR, SDRE and sliding mode control.The SDC formulation introduced is valid only for circular chief satellite orbits. A SDRE based control technique for non-coplanar formation flying with constant separation distance and in-plane formation with large separation is developed by Won and Ahn \cite{won2003nonlinear}, the problem is also extended to elliptical chief satellite orbits.

Optimal control theory is quite widely used and it is a powerful technique for solving many challenging real-life problems. In
fact, the optimal control theory is the driving force of a large part of the research in aerospace engineering. Optimal control
theory based guidance schemes are available for many real-life problems~\cite{Kirk:1970Book, Roberts:1972Book, Betts:2001Book}.
However, such a formulation often leads to a ``two point boundary value problem'' (like gradient method~\cite{Kirk:1970Book},
shooting method~\cite{Roberts:1972Book}, transcription method~\cite{Betts:2001Book} etc.), which in turn lead to large
computational requirements that are infeasible to implement in real time. Moreover, it results in `open-loop' (off-line)
solutions. Since open-loop solutions are not good to account for unwanted inputs (like wind disturbances, for example), the idea
then is to augment it with a ``neighboring optimal controller''~\cite{bryson_book} which is essentially a Linear
Quadratic Regulator (LQR)~\cite{Naidu} or State Dependent Ricatti Equation (SDRE) controller~\cite{Mracek} based
on the linearized dynamics about the nominal trajectory. Another real-time optimal control design technique is the ``approximate
dynamic programming'', followed by ``adaptive critic'' approach, where the optimal control problem is solved using two neural
networks~\cite{Werbos, Balakrishnan:1996Paper}. Upon mutual consistent training (which is typically done off-line), the
action network eventually leads to a state feedback solution which in turn can be used online.

Hence, an important aspect that perhaps needs special attention is the reduction of computational time, if the computational time can be reduced substantially. That way the effects of unwanted disturbances can directly be accounted for to compute new optimal trajectories onboard. In this thesis, the aim is to implement and demonstrate a computationally efficient optimal steering law for satellite Formation Flying missions and meeting the terminal constraints in presence of external perturbation.

\cleardoublepage 
\chapter{Orbital Dynamics and Relative Satellite Dynamics}
\label{Chapter2}

Orbital dynamics is primarily concerned with the orbital motion of one or more usually smaller bodies around the bigger primary body. Most generic case of orbital dynamics problem is that of the two body problem defined in the Keplerian motion frame work. In this chapter we briefly introduce the orbital mechanics of two body problem under the influence of each others gravitation effects. The governing equation of motion is derived using Newton's laws of motion and in frame work of three keplerian laws of planetary motion

\section{The Two body problem}
In this section we briefly elucidate the two body problem and derive the equation of motion of the two point masses $m_1$ and $m_2$ under influence of the each others gravitational field. The equation of motion in inertial frame is derived using Newton laws of motion and keplerian laws of planetary motion ~\cite{Curtis},\cite{JHOW}

\begin{figure}
\begin{center}
  \includegraphics[width=4in]{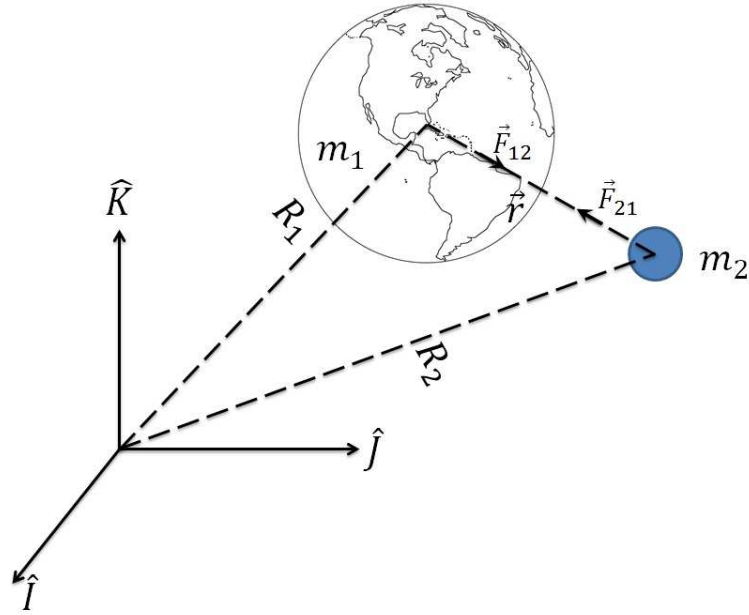}\\
  \caption{Two body problem in Earth Centered Inertial Frame}\label{Twobody_Prob}
  \end{center}
\end{figure}

Figure \ref{Twobody_Prob} shows the two point masses acted upon by the mutual gravitational forces. $\mathbf{R_1}$ and $\mathbf{R_2}$ represent the position vector of the center of masses of point mass $m_1$ and $m_2$  respectively in the inertial frame of reference. Let $X_1$, $Y_1$, $Z_1$ and $X_2$, $Y_2$, $Z_2$ mark the position coordinate of the point mass $m_1$ and $m_2$, hence the radius vector $\mathbf{R_1}$ and $\mathbf{R_2}$ can be written as follows \ref{Inertial_EOM_RADIUS_VECTOR}.
\begin{equation}
\label{Inertial_EOM_RADIUS_VECTOR}
\begin{array}{l}
\mathbf{R_1} = {X_1}\hat i + {Y_1}\hat j + {Z_1}\hat k\\
\mathbf{R_2} = {X_2}\hat i + {Y_2}\hat j + {Z_2}\hat k
\end{array}
\end{equation}
Let the position vector of $m_2$ with respect to $m_1$ be defined as $\mathbf{r} = \mathbf{R_2}-\mathbf{R_1}$. Using \ref{Inertial_EOM_RADIUS_VECTOR} relative position vector $r$ can be written as follows.
\begin{equation}
\mathbf{r} = \left( {{X_2} - {X_1}} \right)\hat i + \left( {{Y_2} - {Y_1}} \right)\hat j + \left( {{Z_2} - {Z_1}} \right)\hat k
\end{equation}
The point mass $m_1$ is acted upon by the gravitational pull from body $m_2$. This gravitational force of attraction $\mathbf{F_{12}}$ acts along the line joining the center of the two masses $\mathbf{\hat r}$. \\
Where $\mathbf{\hat r}$ is the unit vector in the direction of the relative vector $\mathbf{r}$.
\begin{equation}
\label{UNIT_RADIUS_VEC}
\mathbf{\hat r} = \frac{\mathbf{r}}{{\left| \mathbf{r} \right|}}
\end{equation}
Therefore the force acted on $m_1$ by $m_2$ is given in \ref{GRAV_FORCE} \cite{Curtis}
\begin{equation}\label{GRAV_FORCE}
{\mathbf{F_{12}}} = \frac{{G{m_1}{m_2}}}{{{r^2}}}\mathbf{\hat r}
\end{equation}
Where $G$ is universal Gravitational constant, $G = 6.672 \times {10^{ - 11}}\frac{{N.{m^2}}}{{k{g^2}}}$\\
And from Newton's third law of motion, that is action and reaction are equal and opposite we can write the expression for gravitational force exerted on $m_2$ by $m_1$ $\mathbf{F_{21}}$ as follows
\begin{equation}
{\mathbf{F_{21}}} = -\frac{{G{m_1}{m_2}}}{{{r^2}}}\mathbf{\hat r}
\end{equation}
From Newton's second law of motion, we can write the absolute acceleration of the point mass $m_1$ and $m_2$ with respect to the inertial frame of reference as
\begin{eqnarray}
\label{NEWTON_LAW}
  \sum {\cal F}_{12}&=&{m_1}{{\mathbf{\ddot R}}_1} \\
  \sum {\cal F}_{12}  &=& \mathbf{F}_{12} + \mathbf{F}_{c1} + \mathbf{F}_{p1} \nonumber
\end{eqnarray}
Where
\begin{itemize}
\item $\mathbf{F}_{12}$: Gravitational attraction between $m_1$ and $m_2$ \ref{GRAV_FORCE}
\item $\mathbf{F}_{c1}$ : Controlling force.
\item $\mathbf{F}_{p1}$ : Any perturbation forces due to Atmospheric drag, J2 perturbation, Gravitational interaction of third body
\end{itemize}
Therefore from \ref{GRAV_FORCE} and \ref{NEWTON_LAW} the absolute acceleration can be written as
\begin{equation}
\label{NEWTON_EQN_M1}
{m_1}{{\mathbf{\ddot R}}_1} = \frac{{G{m_1}{m_2}}}{{{r^2}}}\mathbf{\hat r} + {\mathbf{F}_{c1}} + {\mathbf{F}_{p1}}
\end{equation}
Similarly we can write the above set of equation for point mass $m_2$ as follows.
\begin{equation}\label{NEWTON_EQN_M2}
{m_2}{{\mathbf{\ddot R}}_2} = -\frac{{G{m_1}{m_2}}}{{{r^2}}}\mathbf{\hat r} + {\mathbf{F}_{c2}} + {\mathbf{F}_{p2}}
\end{equation}
Diving through out by $m_1$ and $m_2$ in \ref{NEWTON_EQN_M1} and \ref{NEWTON_EQN_M2} respectively the above equations can be re-written as follows,
\begin{eqnarray}
\label{NEWTON_EQN_MASS_DIVIDED1}
{{\mathbf{\ddot R}}_1} &=& \frac{{G{m_2}}}{{{r^2}}}\mathbf{\hat r} + \frac{{\mathbf{F}_{c1}}}{m_1} + \frac{{\mathbf{F}_{p1}}}{m_1}\\
\label{NEWTON_EQN_MASS_DIVIDED2}
{{\mathbf{\ddot R}}_2} &=& -\frac{{G{m_1}}}{{{r^2}}}\mathbf{\hat r} + \frac{{\mathbf{F}_{c2}}}{m_2} + \frac{{\mathbf{F}_{p2}}}{m_2}
\end{eqnarray}
The relative distance between the $m_1$ and $m_2$ is given by $\mathbf r$ and the relative acceleration can be written as
\begin{equation}\label{RELATIVE_ACCN}
    \mathbf{\ddot r} = {{\mathbf{\ddot R}}_2} - {{\mathbf{\ddot R}}_1}
\end{equation}
Substituting for $\mathbf{\ddot R}_2$ and $\mathbf{\ddot R}_1$ from \ref{NEWTON_EQN_MASS_DIVIDED1} and \ref{NEWTON_EQN_MASS_DIVIDED2} we get the following expression for relative acceleration.
\begin{equation}
   {\mathbf{\ddot r}} = -\frac{{G{m_1}}}{{{r^2}}}{\mathbf \hat r} + \frac{{{\mathbf F}_{c2}}}{m_2} + \frac{{{\mathbf F}_{p2}}}{m_2}-\frac{{G{m_2}}}{{{r^2}}}\mathbf{\hat r} - \frac{{\mathbf{F}_{c1}}}{m_1} - \frac{{\mathbf{F}_{p1}}}{m_1}
\end{equation}
Rewriting the above equation by combining the terms, following equation is obtained.
\begin{equation}
\label{RELATIVE_DYNAMICS_GENERIC}
   {\mathbf{\ddot r}} = -\frac{{G({m_1+m_2})}}{{{r^2}}}\mathbf{\hat r} + \frac{{\mathbf{F}_{c2}}}{m_2} + \frac{{\mathbf{F}_{p2}}}{m_2}-\frac{{\mathbf{F}_{c1}}}{m_1} - \frac{{\mathbf{F}_{p1}}}{m_1}
\end{equation}
Considering the point mass  $m_1$ as the primary body that is Earth and $m_2$ as secondary body, satellite orbiting the primary body.
\begin{itemize}
\item $m_1 = M_{Earth} = 5.972 \times10^{24} kg$
\item $m_2 = M_{Sat} \ll m_{Earth}$
\end{itemize}
Equation \ref{RELATIVE_DYNAMICS_GENERIC} can be written for Earth, Satellite pair as follows.
\begin{equation}
\label{RELATIVE_DYNAMICS_EARTH}
   {\mathbf{\ddot r}} = -\frac{{G({M_{Earth}+M_{Sat}})}}{{{r^2}}}\mathbf{\hat r} + \frac{{\mathbf{F}_{c2}}}{M_{Sat}} + \frac{{\mathbf{F}_{p2}}}{M_{Sat}}-\frac{{\mathbf{F}_{c1}}}{M_{Earth}} - \frac{{\mathbf{F}_{p1}}}{M_{Earth}}
\end{equation}
Since $M_{Earth} \gg M_{Sat}$ we can write $M_{Earth}+M_{Sat} \approx M_{Earth} $ and $\frac{{\mathbf{F}_{c1}}}{M_{Earth}}$ that is control force on Earth is zero and perturbation acceleration $\frac{{\mathbf{F}_{p1}}}{M_{Earth}}$ is negligible. Hence the earth can be considered as the inertial frame of reference for satellite.
Using \ref{UNIT_RADIUS_VEC} Equation \ref{RELATIVE_DYNAMICS_EARTH} can be rewritten as follows,
\begin{equation}
\label{INERTIAL_EOM}
   {\mathbf{\ddot r}} = -\frac{\mu}{{{r^3}}}\mathbf{r} + U + a_p
\end{equation}
Where
\begin{itemize}
\item $\mu = GM_{Earth} = 398601 \frac{km^3}{s^2}$
\item $U = \frac{{\mathbf{F}_{c2}}}{M_{Sat}}$ (Control Acceleration)
\item $a_p = \frac{{\mathbf{F}_{p2}}}{M_{Sat}}$ (Disturbance acceleration on the satellite)
\end{itemize}
It is to be noted here that the ``Earth centered inertial`` (ECI) reference frame considered in the above derivation, strictly speaking is not a inertial frame. We have made a assumption for all practical purpose mass of the satellite is always negligible compared to that of earth's mass. Thus the orbital motion of a satellite around the earth is a restricted two-body problem and earth is assumed to be inertially fixed in space.\\
\quad The system model developed in this section includes the presence of disturbing force comprising of the gravitational perturbation due to oblateness of earth( $J_2$ perturbation), aerodynamic drag, solar radiation pressure and third body gravitational pull on the satellites. $J_2$ geo-potential perturbation is the dominant source of disturbance compared to aerodynamic drag solar radiation and third body gravitational effects hence the effect of other three are neglected in the problem formulation. The mathematical model of the  $J_2$ effects on the satellite are explained in the section \ref{J2_modelling}.

\section{Relative Satellite Dynamics}

\begin{figure}
\begin{center}
  \includegraphics[width=4in]{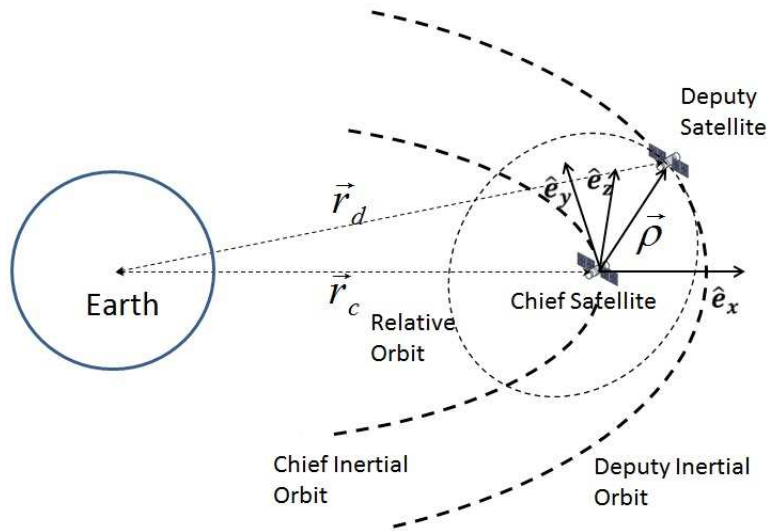}\\
  \caption{Hill's reference frame for satellite relative motion}\label{Hills_ref_frame}
  \end{center}
\end{figure}
Relative motion in orbit, that is mission involving formation flying, rendezvous mission usually involve two satellites orbiting the primary body, one of the orbiting satellite is know as target vehicle or chief satellite and other deputy or chase vehicle. The chief satellite is considered to be passive or non-maneuvering and deputy satellite is active controlled vehicle which can perform maneuvers to bring itself into the desired formation with respect to the chief satellite. The satellite relative dynamics or satellite formation flying problem is defined in the scope of two coordinate frames namely Earth Centered Inertial frame and Hill's Reference frame \cite{Hill}. The definition of the two frames is done the following subsection \ref{ECI_DETAILS} and \ref{HILLS_FRAME_DETAILS}
\subsection{Earth centered inertial reference frame}
\label{ECI_DETAILS}
\begin{figure}
\begin{center}
  \includegraphics[width=3.5in]{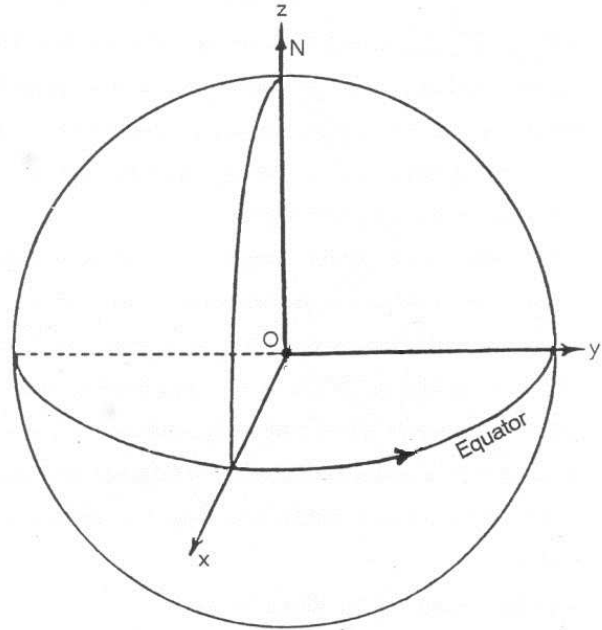}\\
  \caption{Earth Centered Inertial Reference Frame}\label{ECI}
  \end{center}
\end{figure}

Earth Centered Inertial (ECI) \cite{JHOW,Curtis} reference frame has its origin at the center of mass of earth. $\cal X$ axis is in the direction on the vernal equinox, $\cal Z$ towards the north pole and $\cal Y$ completes the triad. The ECI frame is shown in the Figure \ref{ECI}.
\subsection{Hill's Reference Frame}
\label{HILLS_FRAME_DETAILS}
\begin{figure}
\begin{center}
  \includegraphics[width=4in]{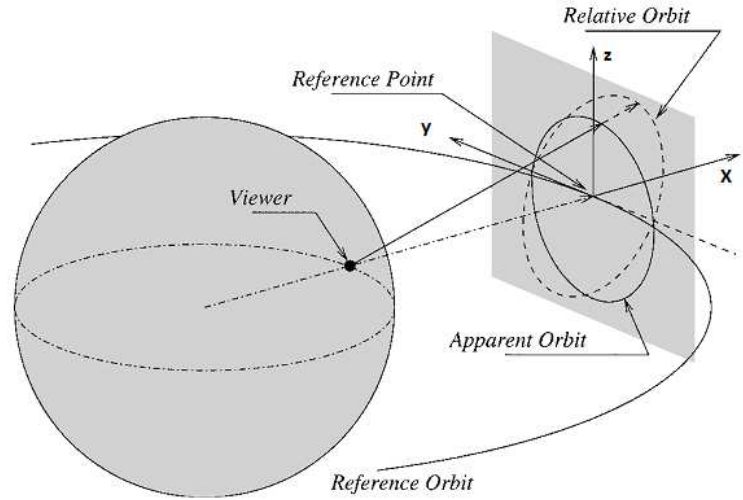}\\
  \caption{Geometric description for Apparent and Relative orbit of deputy satellite in Hill's Frame \cite{chichka2001satellite}}\label{Relative_Orbit}
  \end{center}
\end{figure}

\quad Satellite relative dynamics problem formulation is done in a non-inertial reference frame centered and moving along with chief satellite which is commonly known as the Hills reference frame. This reference frame was first described by G.W Hill in his work on motion of moon about earth \cite{Hill,Clohessy}. (see Figure \ref{Hills_ref_frame} for a description of this reference frame). The origin of the reference frame is chosen as center of chief satellite. Hill's coordinate frame $X$ axis $(\mathbf{\hat e_x})$ is oriented along radius vector $\mathbf{r_c}$ of chief satellite measured from center of the earth, $Z$ $(\mathbf{\hat e_z})$ axis points in the direction along orbital angular momentum vector $(\mathbf{h})$ perpendicular to plane of chief satellite orbit and $Y$ axis $(\mathbf{\hat e_y})$ is cross product of above two and points in the direction of tangent to the reference orbit and in the velocity vector direction of the chief satellite. \cite{JHOW,Park}
\begin{equation}
 \mathbf{\hat e_x} = \frac{{{\mathbf{r_c}}}}{{\left| {{\mathbf{r_c}}} \right|}};\hspace{3 mm}  \mathbf{\hat e_z} = \frac{ \mathbf{ h}}{{\left| \mathbf{ h} \right|}} \hspace{3 mm} \hspace {3 mm}  \mathbf{\hat e_y} =  -\mathbf{\hat e_x} \times  \mathbf{\hat e_z}
\end{equation}
Hills frame facilitates the motion of the deputy satellite to be described with respect to a reference point on the moving chief satellite. The motion of the satellite in this frame will create an relative orbit. However we are primarily interested in the relative motion as it appears to an observer on the planetary surface, whose position is always on the straight line connecting the center of the planet to the reference point on the chief satellite which is origin of the Hill's reference frame. Apparently the observer point on the earth is the point of observation on the surface of earth for both the satellites. Lets consider a point we assumed to be situated on the planetary surface that moves with time such that it is always placed on the line connecting the center of the planet and the origin of Hills frame. The apparent orbit observed from this point is the motion of the satellite relative to the reference point. This relative motion is effect of purely a matter of the line of sight from the viewer to the satellite, however there is no physical meaning to the apparent orbit \cite{chichka2001satellite}. We can visualize it as the trace left by the intersection of the line of sight as it passes through the yz plane in the Hills coordinate frame, as shown in Fig. \ref{Relative_Orbit}.

\subsection{Clohessy-Wiltshire Equation of relative motion of satellite}
\quad Using \ref{INERTIAL_EOM} and assumption that the chief satellite is passive and non-maneuvering the equation of motion for chief satellite in inertial frame of reference is given as
\begin{equation}
\label{chief_inertial_eqn}
\mathbf{\ddot r_c} =  - \frac{\mu }{{r_c^3}}\mathbf{r_c}+ \mathbf{a_{pc}}
\end{equation}
where $\mathbf{r_c}$ is the radius vector of the chief satellite measured from the center of the earth. For circular reference orbit $
\mathbf{r_c}$ is constant value and for elliptical orbits the instantaneous radius vector is calculated as follows \cite{Curtis}.
\begin{equation}
\label{eqn:11}
 \left| \mathbf{{{r_c}}} \right| = \frac{{{a_c}\left( {1 -
e_{_c}^2} \right)}}{{\left( {1 + {e_c}\cos {\nu}} \right)}};
\end{equation}
Similarly equation of motion in the inertial frame can be written for deputy satellite.
\begin{equation}
\mathbf{{{\ddot r}_d}} =  - \frac{\mu }{{r_d^3}}\mathbf{r_d} + \mathbf{U} + \mathbf{a_{pd}}
\end{equation}
Spatial separation between chief and deputy satellite $\mathbf{\rho}$ can be written as, $\mathbf{\rho}=\mathbf{r_d}-\mathbf{r_c}$. Taking double derivative and substituting for $\mathbf{\ddot r}_c$ and $\mathbf{\ddot r}_d$ from Newton's law yields the following expression.
\begin{equation}
\label{eqn:12} \mathbf{\ddot \rho}  =  - \frac{\mu }{{{{\left( {{r_c} + \rho } \right)}^3}}}\left( \mathbf{{r_c} + \rho } \right) + \frac{\mu
}{{r_c^3}}\mathbf{r_c} + \mathbf{U} + \mathbf{a_{p}}
\end{equation}
\begin{itemize}
\item $\mathbf{r_c}$:    Radius vector for chief satellite
\item $\mathbf{r_d} = \mathbf{r_c + \rho} $:    Radius vector for chief satellite
\item $ \mathbf{a_{pc}}$ : Disturbance acceleration on chief satellite
\item $ \mathbf{a_{pd}}$ : Disturbance acceleration on deputy satellite
\item $\mathbf{a_{p}} = \mathbf{a_{pd}} - \mathbf{a_{pc}}$
\item $a_c$:    Semi-major axis of chief satellite
\item $e_c$:    Eccentricity of Chief satellite orbit
\item $\nu$: True anomaly
\end{itemize}
The relative acceleration vector $\mathbf{\ddot {\rho}}$ can be written in the non inertial Hill's reference frame as follows.
\[{\left( {\frac{{{d^2}\rho}}{{d{t^2}}}} \right)^{\cal H}} + 2\omega _{\cal I}^{\cal H} \times {\left( {\frac{{d\rho }}{{dt}}} \right)^{^{\cal H}}} + \left( {\frac{{d\omega _{\cal I}^{\cal H}}}{{dt}}} \right) \times \rho + \omega _{\cal I}^{\cal H} \times \left( {\omega _{\cal I}^{\cal H} \times \rho } \right) - \frac{\mu }{{r_c^3}} {\bf{r_c}}\]
\begin{equation}
\label{nonlinear_equation_SFF}
+\frac{\mu }{{{{\left( {{r_d} }
\right)}^3}}}{\bf{r_d}}+\left(\mathbf U+ \mathbf{a_{p}}\right) = 0
\end{equation}
where, $\omega _{\cal I}^{\cal H} = {\left[
{\begin{array}{*{20}{c}} 0&0&{\dot \nu }
\end{array}} \right]^T}$ denotes angular velocity of
Hill's reference frame relative to inertial reference frame, $\mathbf \rho
= {\left[ {\begin{array}{*{20}{c}} x&y&z
\end{array}} \right]^T}$ and $\mathbf U
= {\left[ {\begin{array}{*{20}{c}} a_x&a_y&a_z
\end{array}} \right]^T}$.
where, $x$, $y$ and $z$ are three component of relative position vector $\mathbf{\rho}$. The terms $\mathbf{a_x}$, $\mathbf{a_y}$  and $\mathbf{a_z}$ are applied control accelerations in the three axes $x$, $y$ and $z$ respectively. The terms $\mathbf{a_p}$ include the external perturbation forces such as gravitational perturbation $\left(\mathbf{a_{J2}}\right)$, solar radiation pressure and atmospheric drag (for low altitude remote sensing satellites). Simplifying \ref{nonlinear_equation_SFF} the following expression is arrived at \cite{Clohessy}
\begin{equation}
\label{EQN_SFF}
\begin{small}
\left[ {\begin{array}{*{20}{c}}
{\ddot x}\\
{\ddot y}\\
{\ddot z}
\end{array}} \right] = \left[ {\begin{array}{*{20}{c}}
{2\dot \nu \dot y + \ddot \nu y + {{\dot \nu }^2}x - \frac{{\mu \left( {{r_c} + x} \right)}}{{{{({r_c} + \rho )}^3}}} + \frac{\mu }{{r_c^2}}}\\
{ - 2\dot \nu x - \ddot \nu x + {{\dot \nu }^2}y - \frac{{\mu y}}{{{{({r_c} + \rho )}^3}}}}\\
{ - \frac{{\mu z}}{{{{({r_c} + \rho )}^3}}}}
\end{array}} \right]
+\left[ {\begin{array}{*{20}{c}}
1&0&0\\
0&1&0\\
0&0&1
\end{array}} \right]\left( \mathbf{U + {a_{p}}} \right)
\end{small}
\end{equation}

The above non-linear equation of motion can be written $\dot X = f(X) + BU$ form as follows.
\begin{equation}
\label{EQN_SFF_FUNC_FORM}
\left[ {\begin{array}{*{20}{c}}
{{{\dot x}_1}}\\
{{{\dot x}_2}}\\
{{{\dot x}_3}}\\
{{{\dot x}_4}}\\
{{{\dot x}_5}}\\
{{{\dot x}_6}}
\end{array}} \right] = \left[ {\begin{array}{*{20}{c}}
{{x_2}}\\
{2\dot \nu {x_4} + \ddot \nu {x_3} + {{\dot \nu }^2}{x_1} - \frac{\mu }{\gamma }\left( {{x_1} + {r_c}} \right) + \frac{\mu }{{r_c^2}}}\\
{{x_4}}\\
{ - 2\dot \nu {x_2} - \ddot \nu {x_1} + {{\dot \nu }^2}{x_3} - \frac{\mu }{\gamma }{x_3}}\\
{{x_6}}\\
{ - \frac{\mu }{\gamma }{x_5}}
\end{array}} \right] + \left[ {\begin{array}{*{20}{c}}
0&0&0\\
1&0&0\\
0&0&0\\
0&1&0\\
0&0&0\\
0&0&1
\end{array}} \right]\left( \mathbf{U + {a_{p}}} \right)
\end{equation}
where,
\begin{equation}
\begin{array}{l} \mathbf{X} = [\begin{array}{*{20}{c}} x&{\dot
x}&y&{\dot y}&z&{\dot z}
\end{array}]^T{\rm{=}}\left[ {\begin{array}{*{20}{c}}
{{x_{1}}}&{{x_{2}}}&{{x_{3}}}&{{x_{4}}}&{{x_{5}}}&{{x_{6}}}
\end{array}} \right]^T
\end{array}\nonumber
\end{equation}
$\mu = GM{\rm{ = 398601}}\;{\rm{k}}{{\rm{m}}^{\rm{3}}}{\rm{/}}{{\rm{s}}^{\rm{2}}}$
is gravitational parameter, where $G$ is universal Gravitational constant and $M$ is mass of earth, $\nu$ is the true anomaly and
\[\gamma = {\left| {\mathbf{{\vec r}_c} + \vec \rho } \right|^3} = {\left( {{{\left( {{\mathbf r_c} + x} \right)}^2} + {y^2} + {z^2}} \right)^{\frac{3}{2}}}\]
Angular velocity of co-moving reference frame $(\dot \nu)$ \cite{Curtis}
\begin{equation}
\dot \nu  = \frac{{\sqrt {\mu a\left( {1 - {e^2}} \right)}
}}{{r_c^2}}\nonumber
\end{equation}
Angular acceleration of co-moving reference frame $(\ddot \nu)$ \cite{Curtis}
\begin{equation}
\ddot \nu  = \frac{{ - 2\mu e{{\left( {1 + e\cos \nu }
\right)}^3}\sin \nu }}{{{a^3}{{\left( {1 - {e^2}} \right)}^3}}}
\nonumber \\
\end{equation}
\subsection{Hill's Equation : Linearized Clohessy Wiltshire Equation}
\label{Linear_Model}
Hills Equation are linearized form of Clohessy-Wiltshire equation of relative motion of satellite in the Hill's frame of reference \cite{Hill}. The Linearization of the equations \ref{EQN_SFF} is done under following assumptions.
\begin{itemize}
\item Circular reference orbit (Chief satellite orbit around earth).\\
$\ddot \nu = 0$ and mean anomaly for circular orbit $ \omega  = \dot \nu = \sqrt {\frac{\mu }{{r_c^3}}}$
\item Radial separation between chief and deputy satellite $\left(\mathbf \rho \right) $ is very small compared to radius vector $\mathbf{r_c}$ of the chief satellite $\left(\mathbf \rho \ll \mathbf{r_c} \right) $
\end{itemize}
Using the definition of $\gamma$ and first assumption the Clohessy-Wiltshire equation \ref{EQN_SFF} can be rewritten as follows \cite{Irvin}.
\begin{eqnarray}
\label{L_EOM_STEP1}
\ddot x - 2\omega \dot y - {\omega ^2}({\mathbf r_c} + x)\left[ {1 - \frac{{\mathbf r_c^3}}{{{{\left( {{{\left( {{\mathbf r_c} + x} \right)}^2} + {y^2} + {z^2}} \right)}^{\frac{3}{2}}}}}} \right] - {a_x} &=& 0 \nonumber \\
\ddot y + 2\omega \dot x - {\omega ^2}y\left[ {1 - \frac{{\mathbf r_c^3}}{{{{\left( {{{\left( {{\mathbf r_c} + x} \right)}^2} + {y^2} + {z^2}} \right)}^{\frac{3}{2}}}}}} \right] - {a_y} &=& 0\\
\ddot z + {\omega ^2}z\left[ {\frac{{\mathbf r_c^3}}{{{{\left( {{{\left( {{\mathbf r_c} + x} \right)}^2} + {y^2} + {z^2}} \right)}^{\frac{3}{2}}}}}} \right] - {a_z} &=& 0 \nonumber
\end{eqnarray}
The nonlinear term in the above equations can be written as
\begin{eqnarray}
\label{SIGMA_DEFINITION}
{\sigma _z} &=& \frac{{r_c^3}}{{{{\left( {{{\left( {{r_c} + x} \right)}^2} + {y^2} + {z^2}} \right)}^{\frac{3}{2}}}}}\nonumber\\
{\sigma _y} &=& 1 - {\sigma _z}\\
{\sigma _x} &=& \left( {\frac{{{r_c}}}{x} + 1} \right){\sigma _y}\nonumber
\end{eqnarray}
\begin{equation}
\label{L_EOM_STEP2}
\gamma  = {\left( {{{\left( {{r_c} + x} \right)}^2} + {y^2} + {z^2}} \right)^{\frac{3}{2}}} = {\left( {{r_c^2} + 2{r_c}x + {x^2} + {y^2} + {z^2}} \right)^{\frac{3}{2}}}
\end{equation}
Rewriting the above equation by factoring out the common $\mathbf r_c$ term, the following equation is obtained.
\begin{equation}
\label{L_EOM_STEP3}
\gamma  = r_c^3{\left( {1 + \frac{{2x}}{{{r_c}}} + \frac{{{x^2}}}{{r_c^2}} + \frac{{{y^2}}}{{r_c^2}} + \frac{{{z^2}}}{{r_c^2}}} \right)^{\frac{3}{2}}}
\end{equation}
Using binomial expansion the above term can be written in the power series form as follows,
\begin{equation}
\label{L_EOM_STEP4}
\gamma  = r_c^3\left( {1 + \frac{3}{2}\left( {\frac{{2x}}{{{r_c}}} + \frac{{{x^2}}}{{r_c^2}} + \frac{{{y^2}}}{{r_c^2}} + \frac{{{z^2}}}{{r_c^2}}} \right) +  \ldots  + HOT} \right)
\end{equation}
Neglecting higher order terms in the binomial expansion and using second assumption which states $\left(\mathbf \rho \ll \mathbf{r_c} \right) $ and hence it can be inferred $x$, $y$ and $z$ are very small compared to radius of the reference orbit, hence the following ratios are approximated to zero.
\[\frac{{{x^2}}}{{r_c^2}} \approx \frac{{{y^2}}}{{r_c^2}} \approx \frac{{{z^2}}}{{r_c^2}} \approx 0\]
With above simplification the nonlinear term $\gamma$ can be written as follows.
\begin{eqnarray}
\label{L_EOM_STEP5}
\gamma  &=& r_c^3\left( {\frac{{{r_c} + 3x}}{{{r_c}}}} \right)
\end{eqnarray}
Substituting \ref{L_EOM_STEP5} in \ref{L_EOM_STEP1} and carrying out the further algebraic simplification the following linearized equation of motion of relative dynamics is obtained.
\begin{eqnarray}
\label{L_EOM_STEP6}
{\ddot x} - 2\omega \dot y - \left[ {\frac{{{\omega ^2}({r_c} + x)3x}}{{\left( {{r_c} + 3x} \right)}}} \right] - \mathbf{a_x} &=& 0 \nonumber \\
{\ddot y} + 2\omega \dot x - \left[ {\frac{{3{\omega ^2}yx}}{{\left( {{r_c} + 3x} \right)}}} \right] - \mathbf{a_y} &=& 0\\
{\ddot z} + \left[ {\frac{{{\omega ^2}z r_c}}{{\left( {{r_c} + 3x} \right)}}} \right] - \mathbf{a_z} &=& 0 \nonumber
\end{eqnarray}
Further with following approximations
\begin{eqnarray}
\label{LINEARIZATION_ASSUMPTION1}
{r_c} + 3x &\approx& {r_c} \nonumber \\
{r_c} + x &\approx& {r_c}\\
\frac{{yx}}{{\left( {{r_c} + 3x} \right)}} &\approx& 0 \nonumber
\end{eqnarray}
The final linearized form of Clohessy-Wiltshire equation of motion can be written as follows.
\begin{eqnarray}
\label{LINEAR_EOM}
\mathbf {\ddot x} &=& 2\omega \dot y + 3{\omega ^2}x + {a_x} \nonumber \\
\mathbf {\ddot y} &=&  - 2\omega \dot x + {a_y}\\
\mathbf {\ddot z} &=&  - {\omega ^2}z + {a_z} \nonumber
\end{eqnarray}
Writing in the state space form, $\mathbf{\dot X} = A \mathbf X + B \mathbf U$ and defining the state vector as follows
\begin{equation}
\begin{array}{l} \mathbf{X} = [\begin{array}{*{20}{c}} x&{\dot
x}&y&{\dot y}&z&{\dot z}
\end{array}]^T{\rm{=}}\left[ {\begin{array}{*{20}{c}}
{{x_{1}}}&{{x_{2}}}&{{x_{3}}}&{{x_{4}}}&{{x_{5}}}&{{x_{6}}}
\end{array}} \right]^T
\end{array}\nonumber
\end{equation}
we get
\begin{equation}
\label{LINEAR_EOM_STATE_SPACE_FORM}
\left[ {\begin{array}{*{20}{c}}
{{{\dot x}_1}}\\
{{{\dot x}_2}}\\
{{{\dot x}_3}}\\
{{{\dot x}_4}}\\
{{{\dot x}_5}}\\
{{{\dot x}_6}}
\end{array}} \right] = \left[ {\begin{array}{*{20}{c}}
0&1&0&0&0&0\\
{3{\omega ^2}}&0&0&{2\omega }&0&0\\
0&0&0&1&0&0\\
0&{ - 2\omega }&0&0&0&0\\
0&0&0&0&0&1\\
0&0&0&0&{ - {\omega ^2}}&0
\end{array}} \right]\left[ {\begin{array}{*{20}{c}}
{{x_1}}\\
{{x_2}}\\
{{x_3}}\\
{{x_4}}\\
{{x_5}}\\
{{x_6}}
\end{array}} \right] + \left[ {\begin{array}{*{20}{c}}
0&0&0\\
1&0&0\\
0&0&0\\
0&1&0\\
0&0&0\\
0&0&1
\end{array}} \right]U
\end{equation}
This linear form of the equation of motion is used in Linear Quadratic tracking controller which is explained in the following chapters.
\section{$J_2$ Perturbation model}
\label{J2_modelling}
Earth's equatorial radius is 21 km larger than polar radius this flattening of the poles is known as oblateness of the earth \cite{Curtis}. This lack of symmetry causes force of gravity on the orbiting satellites not to pass through the center of the earth. Oblateness causes the variation in gravitational pull with angular distance (latitude) of the orbiting body. This effect is known as zonal variation, the dimensionless quantity which quantifies the effects of oblateness on orbit is called J2(Second zonal Harmonics).

\begin{figure}
\begin{center}
  \includegraphics[width=3.5in]{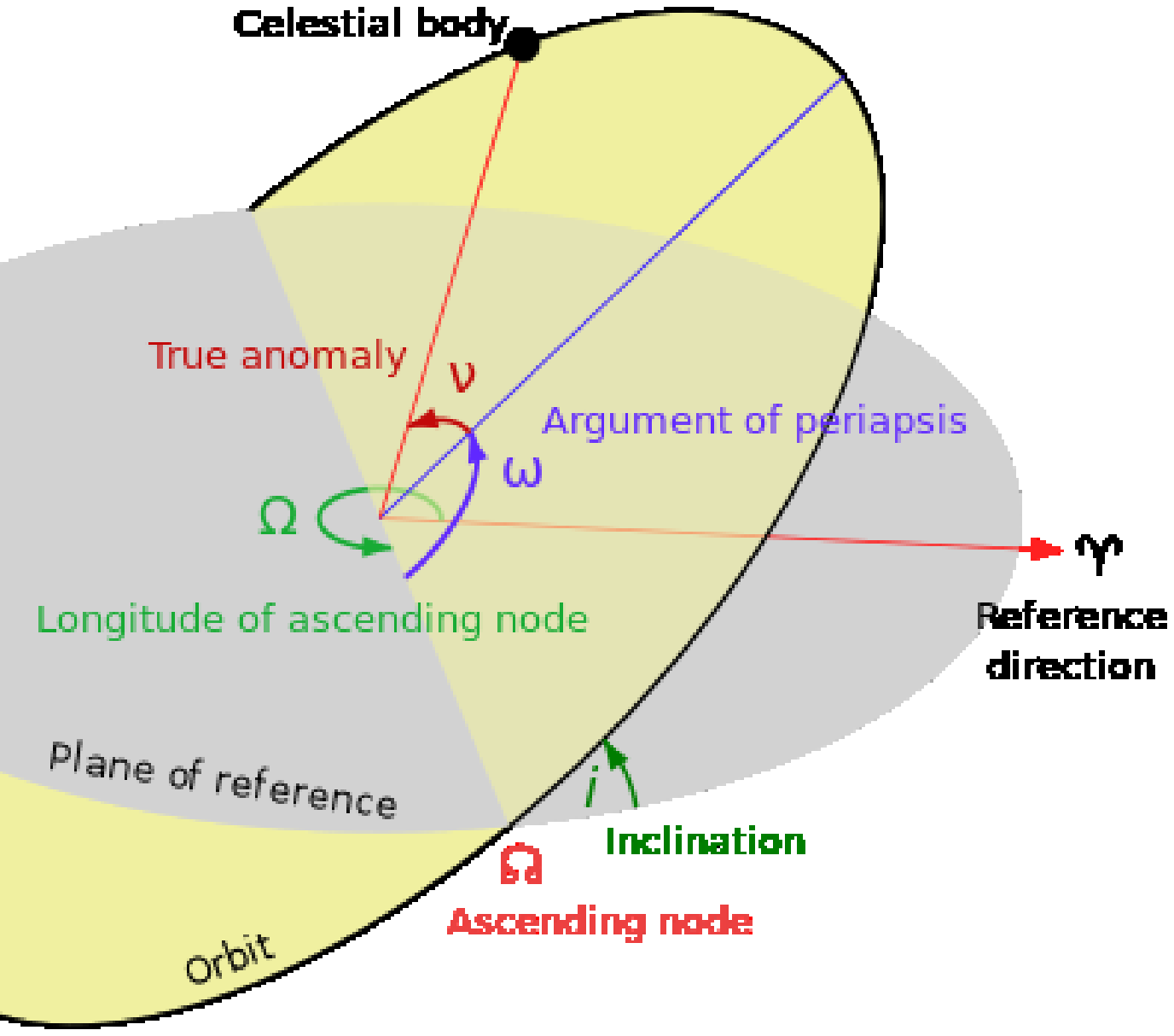}\\
  \caption{Orbital Parameters \cite{Wikipedia}}\label{ORB_PARAM}
  \end{center}
\end{figure}

Since gravitational force is a conservative force, it can be derived from the gradient of the scalar potential function. Using equation \ref{INERTIAL_EOM} for case of zero control on the satellite we can write the equation of motion as follows.
\begin{equation}
   {\mathbf{\ddot r}} = -\frac{\mu}{{{r^3}}}\mathbf{r} + a_p
\end{equation}
$\mathbf r$ is measured in the ECI frame, that is $\mathbf r = X\hat i+Y\hat j + Z\hat k$ and $ r = \sqrt {X^2+Y^2 + Z^2}$
Using the definition of $\mathbf r$ and $r$ and rewriting the above equation we get the gravitational force in the inertial frame component wise as follows.
\begin{equation}
\mathbf{\ddot r} = \left( { - \frac{{\mu X}}{{\sqrt {{X^2} + {Y^2} + {Z^2}} }} + {a_{J2X}}} \right)\hat i + \left( { - \frac{{\mu Y}}{{\sqrt {{X^2} + {Y^2} + {Z^2}} }} + {a_{J2Y}}} \right)\hat j \nonumber
\end{equation}
\begin{equation}
\label{INERTIAL_EOM_COMP}
\hspace{8mm} +\left( { - \frac{{\mu Z}}{{\sqrt {{X^2} + {Y^2} + {Z^2}} }} + {a_{J2Z}}} \right)\hat k
\end{equation}
It is to be note here that selection of the function as follows,
\begin{equation}
\left(\frac{\mu}{\mathbf r} + G_p\right)
\end{equation}
qualifies as the potential function whose gradient in all three direction are equivalent to the \ref{INERTIAL_EOM_COMP} component wise.
\begin{eqnarray}
\frac{\partial }{{\partial X}}\left( {\frac{\mu }{r} + G_p} \right) &=& \left( { - \frac{{\mu X}}{{\sqrt {{X^2} + {Y^2} + {Z^2}} }} + {a_{J2X}}} \right)\\
\frac{\partial }{{\partial Y}}\left( {\frac{\mu }{r} + G_p} \right) &=& \left( { - \frac{{\mu Y}}{{\sqrt {{X^2} + {Y^2} + {Z^2}} }} + {a_{J2Y}}} \right)\\
\frac{\partial }{{\partial Z}}\left( {\frac{\mu }{r} + G_p} \right) &=& \left( { - \frac{{\mu Z}}{{\sqrt {{X^2} + {Y^2} + {Z^2}} }} + {a_{J2Z}}} \right)
\end{eqnarray}
The term $G_p$ is gravitational potential function expressed as infinite sum series derived from oblate earth model \cite{Irvin}
\begin{equation}
{G_p} =  - \frac{\mu }{r}\left\{ {\sum\limits_{n = 2}^\infty  {\left[ {{{\left( {\frac{{{R_e}}}{r}} \right)}^n}{J_n}{P_n}\sin (\phi ) + \sum\limits_{m = 1}^n {{{\left( {\frac{{{R_e}}}{r}} \right)}^n}\left( {{C_{nm}}\cos \varphi  + {S_{nm}}\sin \varphi } \right){P_{nm}}\sin \phi } } \right]} } \right\}
\end{equation}
Where
\begin{itemize}
\item $\varphi  = m\lambda  + {\omega _e}{t_e}$
\item $\lambda$ : Geographical longitude measure from prime meridian
\item $\phi$ : Geocentric latitude of satellite measured from equator.
\item $R_e$ : Mean Equatorial radius of earth
\item $\omega_e$ : Rotation rate of earth.
\item $t_e$ : Time since Greenwich meridian lined up with X axis of ECI  .
\item $J_n$ : Zonal harmonics of order zero.
\item $P_n, P_{nm}$ : Legendre polynomial of degree $n$ and order $0,m$ respectively
\item $C_{nm}$ : tesseral harmonic coefficient for $n \ne m $
\item $S_{nm}$ : Sectorical harmonic coefficient for $n = m $
\end{itemize}
Measurement of zonal, tesseral, sectorical coefficients, it is found that effects of $J_2$ is at least $400$ times larger than the next most significant term. Hence for satellite formation, reconfiguration problem where the control application and reconfiguration happens over shorter period of the time hence all higher terms can be ignored. There for the gradient function $G_p$ can be written as \cite{Irvin}
\[{G_p} =  - \frac{\mu }{r}{\left( {\frac{{{R_e}}}{r}} \right)^2}{J_2}{P_2}\sin (\phi )\]
where
\begin{itemize}
\item $J_2$: 0.0010826
\item $P_2$ : $2^{nd}$ Legendre polynomial of the form $P_2(X) = \frac{1}{2}\left(3X^2+1\right)$
\end{itemize}
Taking the gradient of $G_p$ the perturbation acceleration in all three axis ECI frame is obtained further the components of the $a_{J2}$ can be modeled in Hills frame as follows.
\begin{equation}
{a_{J2}} =  - \frac{{3\mu R_e^2{J_2}}}{{r_c^4}}\left[ {\begin{array}{*{20}{c}}
{\left( {\frac{1}{2} - \frac{{3{{\sin }^2}i{{\sin }^2}\theta }}{2}} \right){{\hat e}_x}}\\
{({{\sin }^2}i\sin \theta \cos \theta ){{\hat e}_y}}\\
{(\sin i\sin \theta \cos i){{\hat e}_z}}
\end{array}} \right]
\end{equation}
Where
\begin{itemize}
\item $i$ : Chief satellite orbit inclination
\item $\theta$: $\nu  + \omega $ (True Anomaly + Argument perigee)
\item ${R_e}$: Equatorial Earth Radius
\item ${J_2}$: J2 Zonal Harmonic Coefficient (0.00108629)
\end{itemize}
(Refer Figure \ref{ORB_PARAM})
$J_2$ perturbation term is function of satellite orbit inclination and $i$ and $\theta$ it is beneficial from point of view of controller synthesis that the disturbance is modeled in the terms of state variables of deputy satellite $\mathbf X$. The total disturbance term $\mathbf{a_p} = \mathbf{a_{J2}}$ is defined as difference in the disturbance acceleration of deputy and chief satellite \cite{Park}
\begin{equation}
\mathbf{a_{J2}} = \mathbf{a_{J2d}} - \mathbf{a_{J2c}}\\
\end{equation}
\begin{equation}
\label{J2_MODEL1}
\begin{small}
{a_{J2}} =  - \frac{{3\mu R_e^2{J_2}}}{2}\left\{ {\frac{1}{{r_c^4}}\left( {\begin{array}{*{20}{c}}
{{J_{{{\hat e}_x}}}({i_c},{\theta _c})}\\
{{J_{{{\hat e}_y}}}({i_c},{\theta _c})}\\
{{J_{{{\hat e}_z}}}({i_c},{\theta _c})}
\end{array}} \right) - \frac{1}{{r_d^4}}\left( {\begin{array}{*{20}{c}}
{{J_{{{\hat e}_x}}}({i_d},{\theta _d})}\\
{{J_{{{\hat e}_y}}}({i_d},{\theta _d})}\\
{{J_{{{\hat e}_z}}}({i_d},{\theta _d})}
\end{array}} \right)} \right\}
\end{small}
\end{equation}
\hspace{2mm}where,
\begin{itemize}
\item $\left(i_c,\theta_c\right)$ : Orbital elements for chief satellite
\item $\left(i_d,\theta_d\right)$ : Orbital elements for deputy satellite
\item ${{J_{{{\hat e}_x}}} = (1 - 3{{\sin }^2}i{{\sin }^2}\theta ){{\hat e}_x}}$
\item ${{J_{{{\hat e}_y}}} = ({{\sin }^2}i\sin \theta \cos \theta ){{\hat e}_y}}$
\item ${{J_{{{\hat e}_z}}} = (\sin i\sin \theta \cos i){{\hat e}_z}}$
\end{itemize}
Equation \ref{J2_MODEL1} is to be transformed to state dependent form, to carry out this transformation the orbital elements $\left(i,\theta \right)$ needs to be expressed in the state variables $\mathbf X$. A transformation matrix $\Sigma (t)$ introduced by \cite{Alf} is used to convert the orbital elements into state variables of relative motion under J2 perturbation. Using transformation matrix $\Sigma (t)$ state vector can be written as $ \vec X = \Sigma (t)\delta \xi$. Where $\vec X$ is the state vector and $\delta \xi
= \xi_d - \xi_c$  that is difference between the orbital elements of deputy and chief satellite.
\begin{equation}
\xi  = {[\begin{array}{*{20}{c}} a&\theta &i&{e\cos \omega }&{e\sin \omega }&\Omega \end{array}]^T}
\end{equation}
Is vector of orbital elements, for brevity the details of the transformation matrix are omitted and details can be found in \cite{Park,Alf,JHOW}. The relation between state vector and difference in orbital elements of chief and deputy can be written as follows,
\begin{eqnarray}
X &=& \Sigma {(t)\delta \xi } \\
\delta \xi  &=& {\Sigma {(t)} ^{ - 1}}X
\end{eqnarray}
Since the orbital elements for chief satellite are known and state vector and orbital elements for deputy satellite is known, hence the $J_2$ model \ref{J2_MODEL1} is known completely and is function of state alone.
We can write the orbital elements of deputy in terms of orbital elements of chief and state vectors as follows.
\begin{eqnarray}
\delta \xi  &=& {\Sigma{(t)}} ^{ - 1}X \nonumber\\
{\xi _d} - {\xi _c} &=& {\Sigma {(t)}} ^{ - 1}X\\
{\xi _d} &=& {\xi _c} + {\Sigma{(t)}} ^{ - 1}X
\end{eqnarray}
Hence the rewriting the equation, \ref{J2_MODEL1} as function of chief satellite orbital elements $\left(i_c,\theta_c \right)$  and $\delta \xi$ the $J_2$ perturbation model is expressed in the state dependent form as follows.
\begin{equation}
\label{J2_MODEL2}
{a_{J2}} = \frac{{3\mu R_e^2{J_2}}}{2}\left\{ {\frac{1}{{{{\left( {{r_c} + \rho } \right)}^4}}}\left( {\begin{array}{*{20}{c}}
{{J_{{{\hat e}_x}}}({i_c} + \delta i,{\theta _c} + \delta \theta )}\\
{{J_{{{\hat e}_y}}}({i_c} + \delta i,{\theta _c} + \delta \theta )}\\
{{J_{{{\hat e}_z}}}({i_c} + \delta i,{\theta _c} + \delta \theta )}
\end{array}} \right) - \frac{1}{{r_c^4}}\left( {\begin{array}{*{20}{c}}
{{J_{{{\hat e}_x}}}({i_c},{\theta _c})}\\
{{J_{{{\hat e}_y}}}({i_c},{\theta _c})}\\
{{J_{{{\hat e}_z}}}({i_c},{\theta _c})}
\end{array}} \right)} \right\}
\end{equation}
where
\begin{eqnarray}
\delta \theta  &=& \Sigma _{21}^{ - 1}{x_1} + \Sigma _{22}^{ - 1}{x_2} + \Sigma _{23}^{ - 1}{x_3} + \Sigma _{24}^{ - 1}{x_4} + \Sigma _{25}^{ - 1}{x_5} + \Sigma _{26}^{ - 1}{x_6}\\
\delta i &=& \Sigma _{31}^{ - 1}{x_1} + \Sigma _{32}^{ - 1}{x_2} + \Sigma _{33}^{ - 1}{x_3} + \Sigma _{34}^{ - 1}{x_4} + \Sigma _{35}^{ - 1}{x_5} + \Sigma _{36}^{ - 1}{x_6}
\end{eqnarray}

The terms of transformation matrix $\Sigma^{ - 1}$ are given in Appendix of Reference \cite{JHOW}

\section{Summary and Conclusions}
Primarily this chapter introduces the concept of two body problem under influence of each other gravitational forces. Further section are concentrated on establishing the concept of relative satellite dynamics, in this effort of deriving the relative dynamic model of two satellite the reference frames earth centered frame (ECI) and Hill's reference frame are introduced. The relative motion of the satellites are derived in ECI frame and further transformation details to Hill's are introduced. The nonlinear model of equation of motion in Hill's frame (Clohessy-Wiltshire equation) is linearized to obtain Hill' equation of satellite formation flying. $J_2$ perturbation is considered as the only perturbing force external to system. The J2 model is derived using potential function concepts and further the model is transformed to Hill's frame of reference. This chapter forms the basis for further chapter for control synthesis techniques.

\cleardoublepage
\chapter{Infinite time LQR controller for Satellite Formation Flying}
\label{LQR_chap}
Linear Quadratic Regulator(LQR), is a optimal control approach based on linear approximation of plant model of the form \cite{JHOW},
\begin{equation}
\mathbf{\dot X} = A\mathbf{X} + B\mathbf{U} + N(\mathbf{X})
\end{equation}
where $A \in \Re^{n \times n}, \ B \in \Re^{n \times m}$, and  $N(\mathbf{X})$ denotes the effect due to nonlinearity in the plant model or unmodeled dynamics. The control synthesis using LQR algorithm involves computation of optimal feedback gain matrix $K$ such that the optimal control can be written in state feedback form $\mathbf U=-K\mathbf X$ \cite{bryson_book,Elbert}. The term $N(\mathbf{X})$ is ignored in computation of control law $\mathbf U$ from this approach.

The following quadratic performance index is chosen to be minimized
\begin{equation}
\label{COST}
J = \frac{1}{2}\int\limits_0^{{t_f}} {\left( {{\mathbf X^T}Q\mathbf X + {\mathbf U^T}R\mathbf U} \right)dt}
\end{equation}
where $t_f$ is final time and $Q\geq0$, $R>0$ are respectively state and control weight matrices. For autonomous system, constant weight matrices and $t_f \to \infty$, minimization of the above cost function \ref{COST} is achieved at optimal control value
\begin{eqnarray}
\label{OPTIMAL_CNTRL}
\mathbf U=-K\mathbf X\\
\mathbf U =-R^{-1}B^TP\mathbf X
\end{eqnarray}
Where $P$ satisfies the Algebraic Ricatti Equation(ARE) \ref{ARE} \cite{bryson_book}
\begin{equation}
\label{ARE}
PA+A^TP-PBR^{-1}B^TP+Q = 0
\end{equation}
Given condition on system that pair $\left(A,B\right)$ is controllable and pair $\left(A,C\right)$ is observable where $C$ is given as $C=Q^TQ$ the solution to ARE \ref{ARE} is positive definite \cite{bryson_book,Elbert}. positive definiteness of Ricatti coefficient matrix $P$ guarantees the close loop stability of the plant that is asymptotic stability of the system.

\section{Satellite formation Flying control using LQR}
The Linear plant model for satellite formation flying in Hill's frame of the form
\begin{eqnarray}
\label{LINEAR_DYN}
\mathbf{\dot X} &=& A\mathbf{X} + B\mathbf{U}\nonumber\\
\mathbf Y &=& C \mathbf X
\end{eqnarray}
where the system matrices $A,B,C$ are defined in the section \ref{Linear_Model} and are repeated here for easy reference,
\begin{equation}
A=\left[ {\begin{array}{*{20}{c}}
0&1&0&0&0&0\\
{3{\omega ^2}}&0&0&{2\omega }&0&0\\
0&0&0&1&0&0\\
0&{ - 2\omega }&0&0&0&0\\
0&0&0&0&0&1\\
0&0&0&0&{ - {\omega ^2}}&0
\end{array}} \right], B = \left[ {\begin{array}{*{20}{c}}
0&0&0\\
1&0&0\\
0&0&0\\
0&1&0\\
0&0&0\\
0&0&1
\end{array}} \right], C = I_{6 \times 6}
\end{equation}
Satellite formation flying is basically a tracking problem, where the satellite states has to track a desired states. To solve the SFF tracking problem in frame work of LQR theory, the plant model needs to be remodeled in terms of state errors there by converting the tracking problem to regulator problem and the solution can be obtained using optimal LQR control technique.
Lets consider the vectors
\begin{eqnarray}
\mathbf X &=& \left[ {\begin{array}{*{20}{c}}
x&{\dot x}&y&{\dot y}&z&{\dot z}
\end{array}} \right]\\
\mathbf{X_d} &=& \left[ {\begin{array}{*{20}{c}}
{{x_d}}&{{{\dot x}_d}}&{{y_d}}&{{{\dot y}_d}}&{{z_d}}&{{{\dot z}_d}}
\end{array}} \right]
\end{eqnarray}
and state error vector are defined as,
\begin{eqnarray}
\label{ERR_DYN_EQN1}
\mathbf{\tilde X} &=& \mathbf X - \mathbf{X_d}\\
\label{ERR_DYN_EQN2}
\mathbf{\dot {\tilde X}} &=& \mathbf{\dot X} - \mathbf{{\dot X}_d}
\end{eqnarray}
Substituting \ref{ERR_DYN_EQN1} and \ref{ERR_DYN_EQN2} in \ref{LINEAR_DYN} the system dynamics can be rewritten as follows,
\begin{eqnarray}
\mathbf{\dot {\tilde X}} + \mathbf{{\dot X}_d} &=& A\left( \mathbf{\tilde X + {X_d}} \right) + B\mathbf U\\
\label{LQR_EQN_FINAL}
\mathbf{\dot {\tilde X}} &=& A\mathbf{\tilde X} + B\mathbf U + \left( \mathbf{A{X_d} - {{\dot X}_d}} \right)
\end{eqnarray}
From above equation the truncated system dynamics $\mathbf{\dot {\tilde X}} = A\mathbf{\tilde X} + B\mathbf U$ is used for computation of optimal control using LQR technique and term $N\left(\mathbf X \right) = \left( \mathbf{A{X_d} - {{\dot X}_d}} \right)$ is considered as the known control such that the total control $\mathbf{U_{tot}} = -K\mathbf {\tilde X} + \left( \mathbf{A{X_d} - {{\dot X}_d}} \right)$ ensure $\mathbf X \to \mathbf X_d$ \cite{jin2011formation}
\section{Results and Discussion}
In satellite formation flying the satellite can change the formation geometry through reconfiguration of the length of the base line of formation or a new satellite can be introduced into the formation.

\begin{figure}
\begin{center}
  \includegraphics[width=3.6in]{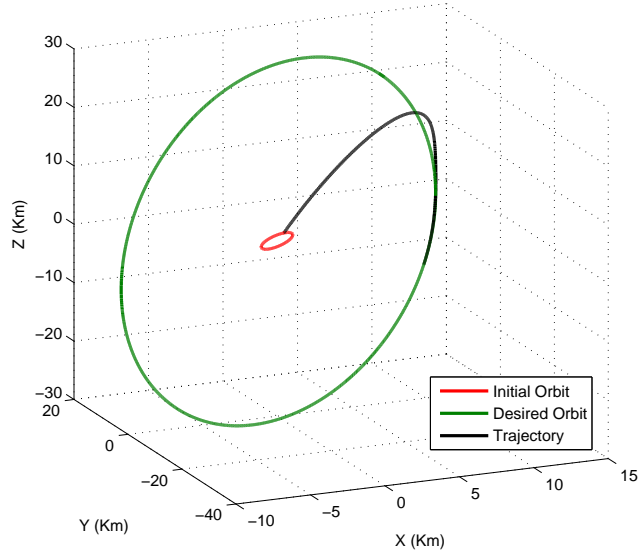}
  \caption{Deputy satellite formation trajectory in Hill's Frame}\label{LQR_PLOT1}
  \end{center}
\end{figure}

\begin{figure}
\begin{center}
  \includegraphics[width=3.6in]{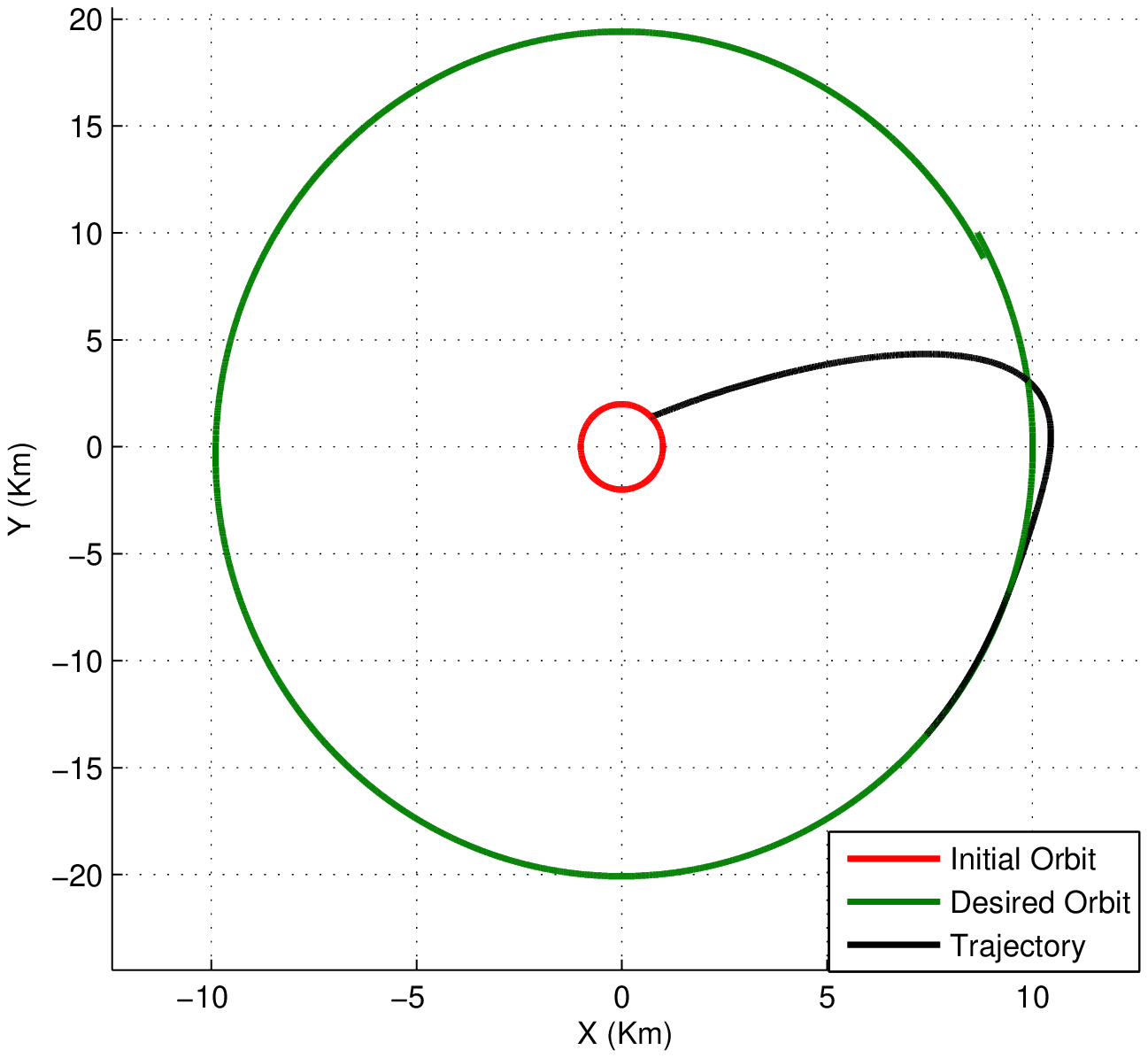}
  \caption{Deputy satellite formation trajectory in XY plane of Hill's Frame}\label{LQR_PLOT2}
  \end{center}
\end{figure}

\begin{figure}
\begin{center}
  \includegraphics[width=3.6in]{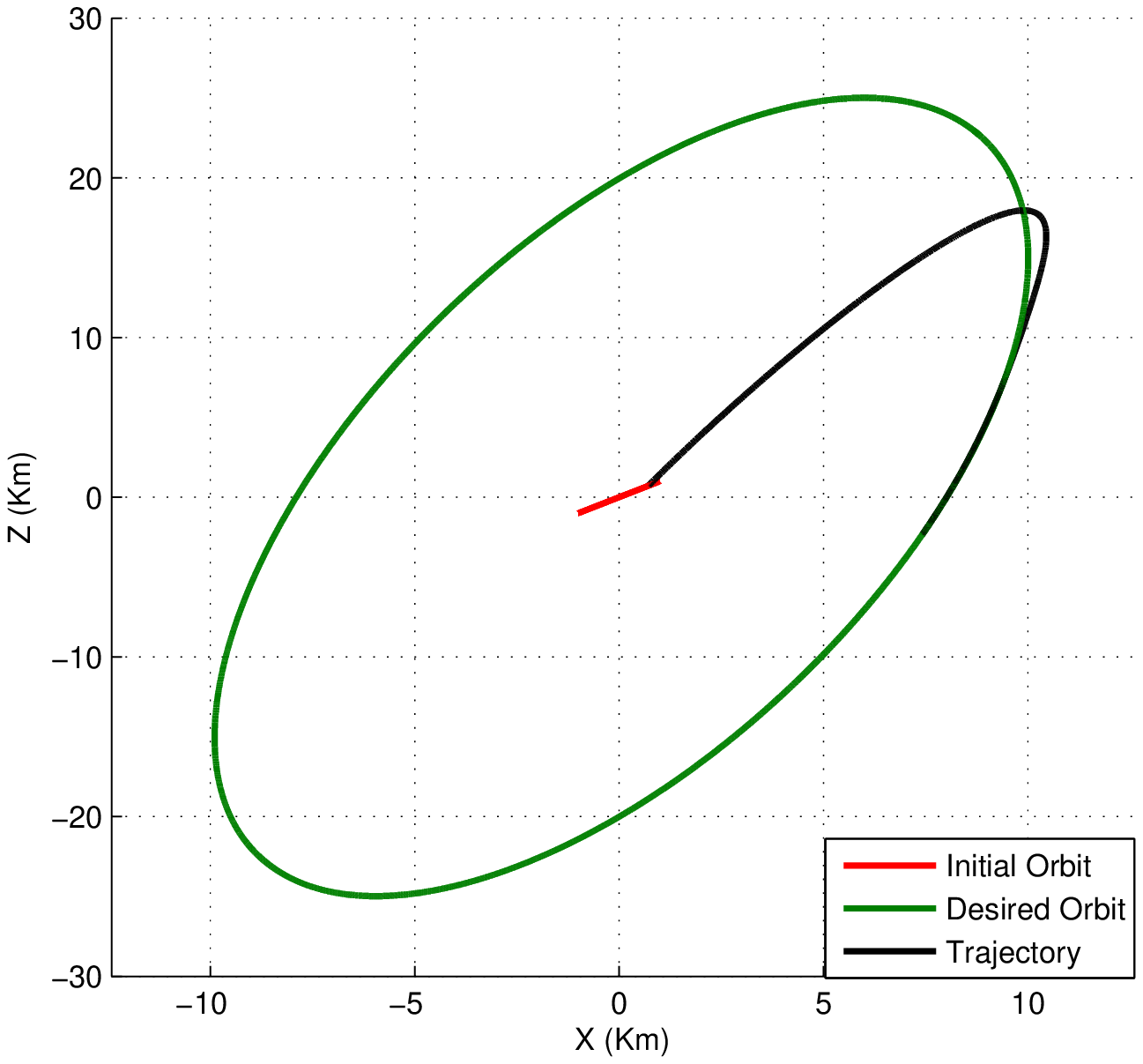}
  \caption{Deputy satellite formation trajectory in XZ plane of Hill's Frame}\label{LQR_PLOT3}
  \end{center}
\end{figure}

\begin{figure}
\begin{center}
  \includegraphics[width=3.6in]{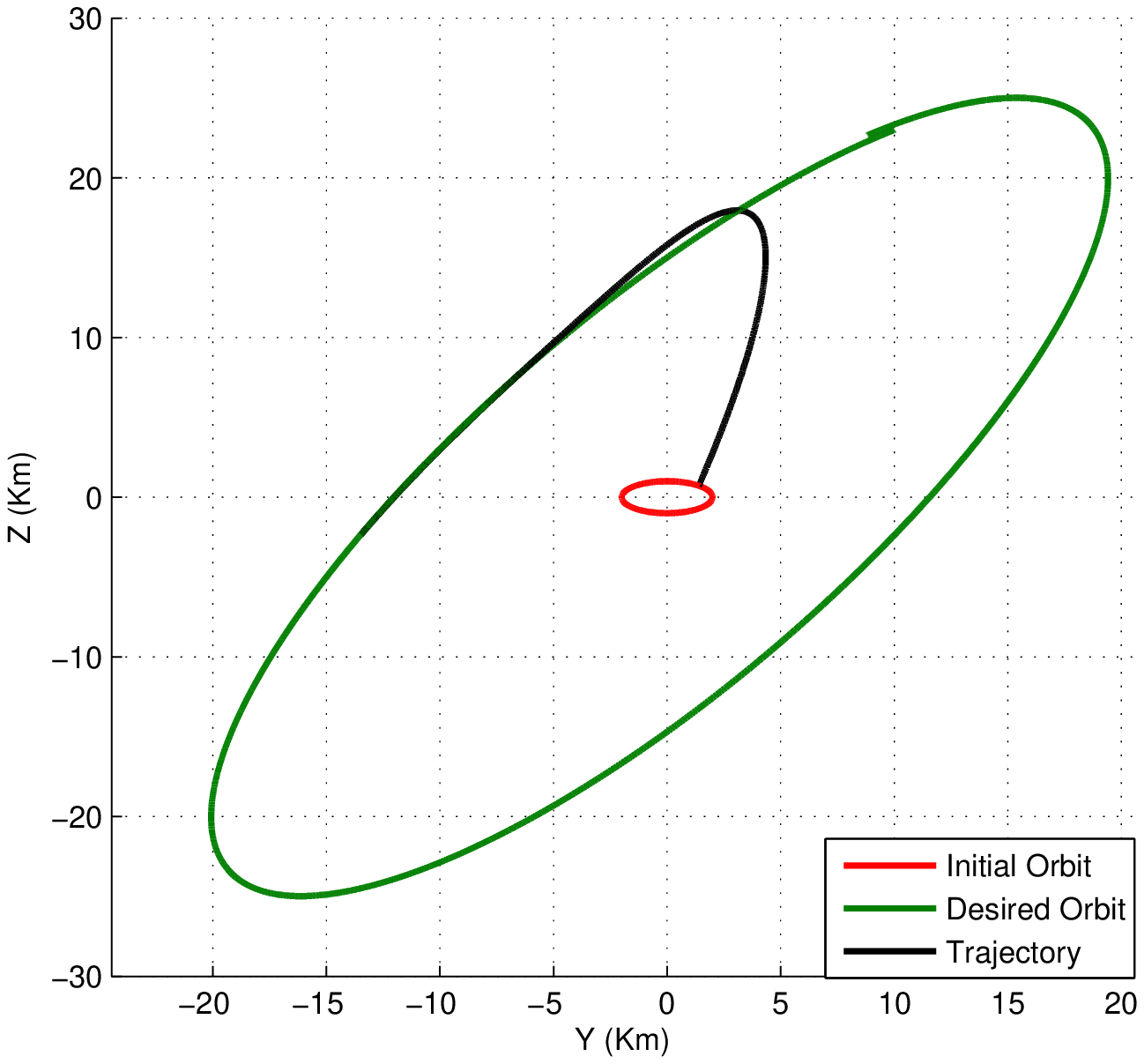}
  \caption{Deputy satellite formation trajectory in YZ plane of Hill's Frame}\label{LQR_PLOT4}
  \end{center}
\end{figure}

\begin{figure}
\begin{center}
  \includegraphics[width=4.5in]{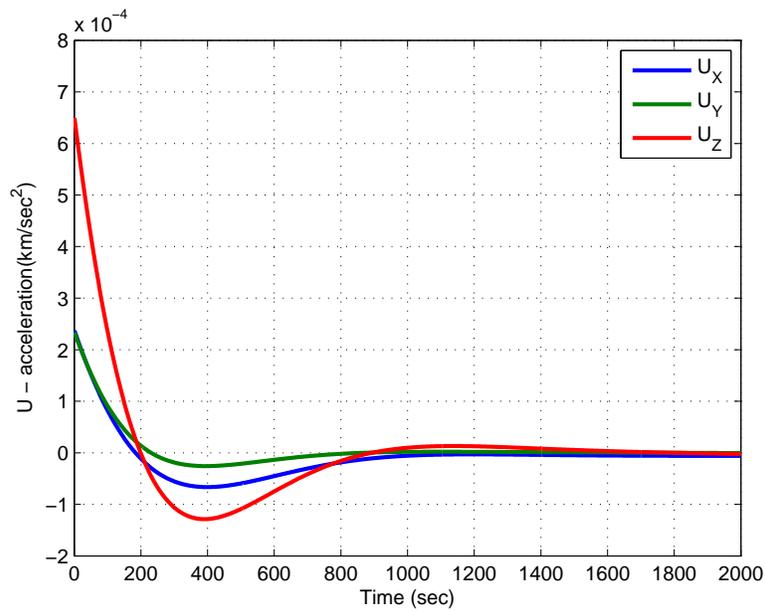}
  \caption{Control History for formation reconfiguration for circular chief satellite orbit}\label{LQR_PLOT5}
  \end{center}
\end{figure}

\begin{figure}
\begin{center}
  \includegraphics[width=4.5in]{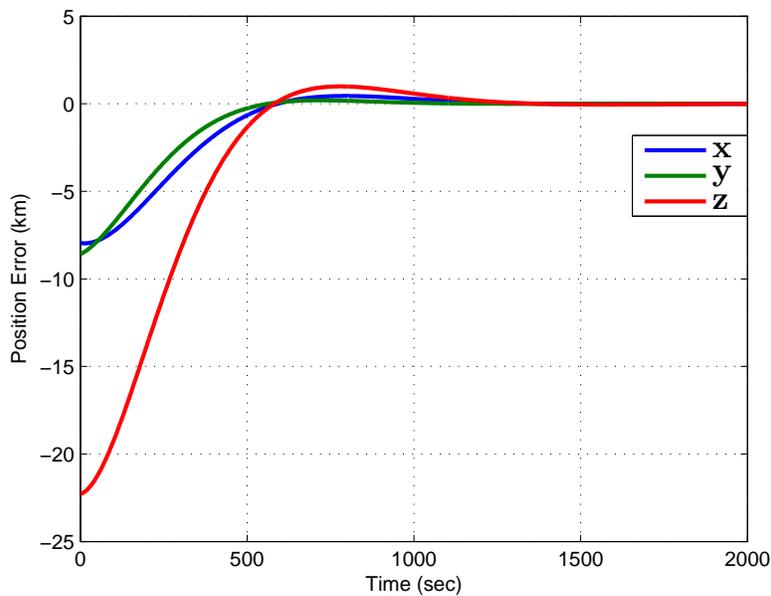}
  \caption{Position Error for formation reconfiguration for circular chief satellite orbit}\label{POS_ERR}
  \end{center}
\end{figure}

\begin{figure}
\begin{center}
  \includegraphics[width=3.6in]{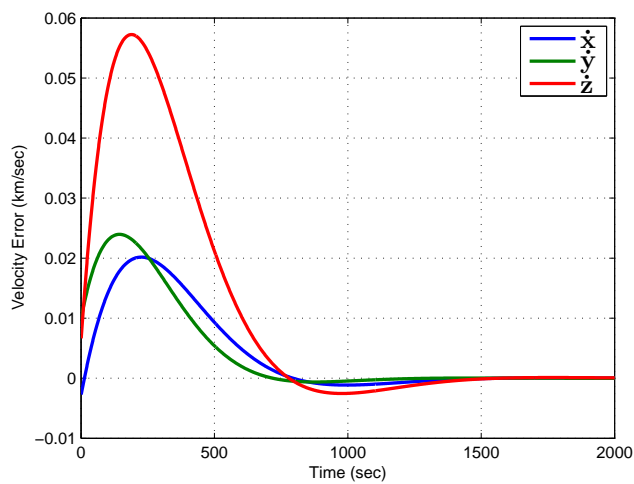}
  \caption{Velocity Error for formation reconfiguration for circular chief satellite orbit}\label{VEL_ERR}
  \end{center}
\end{figure}

For LQR numerical simulation, a formation reconfiguration of deputy satellite with respect to the chief satellite is considered. The deputy satellite is considered to be in a lower baseline length formation and it is desired to place the deputy satellite in the higher baseline length formation with respect to chief satellite. The choice of initial and final $\rho$ are made small enough such that linear equation of motion that is Hill's equation for SFF is close enough to nonlinear Clohessy Wiltshire equation of SFF. Orbital parameters for chief satellite are given in the following Table \ref{ORB_PAR_LQR}. Weight on states and control are selected  as $Q =I_{6 \times 6}$ and $R = 10^9I_{3 \times 3}$. The terminal position and velocity error are given in Table \ref{STATE_ERROR_LQR} and state error history is plotted in Figure \ref{POS_ERR} and \ref{VEL_ERR}
\begin{table}
\caption{\small{Chief Satellite Orbital Parameters, (LQR)}}\label{ORB_PAR_LQR}
\begin{center}\begin{tabular}{|c|c|}
    \hline
    \textbf{Orbital Parameters} & \textbf{Value}\\
    \hline
    \textbf{Semi-major axis} & $10000 km$ \\
    \hline
    \textbf{Eccentricity} & $0$ \\
    \hline
    \textbf{Orbit Inclination} & $0$ \\
    \hline
    \textbf{Argument of Perigee} & $0$ \\
    \hline
    \textbf{Longitude of ascending node} & $0$ \\
    \hline
    \textbf{Initial True Anomaly} & $10$ \\
    \hline
\end{tabular}\end{center}
\end{table}
The initial and final desired relative parameters in terms of orbital elements for deputy satellite are given in Table \ref{Initial_condition_LQR}. The simulation uses $\triangle t = 1sec$ time step.
\begin{table}
\caption{\small{Deputy Satellite Initial condition for LQR solution}}\label{Initial_condition_LQR}
\begin{center}\begin{tabular}{|c|c|c|}
    \hline
    \textbf{Orbital} & \textbf{Initial Value}&\textbf{Final Value}\\
    \textbf{Parameters} &                     &\\
    \hline
    $\rho (km)$ & $1 km$& $10 (km)$\\
    \hline
    $\theta (deg) $& $45^0$ & $60^0$\\
    \hline
    $a (km)$& $0$ & $0$\\
    \hline
    $b (km)$& $0$ & $0$\\
    \hline
    $m$ (slope)& $1$ & $1.5$\\
    \hline
    $n$(slope)& $0$ & $0$\\
    \hline
\end{tabular}\end{center}
\end{table}
The corresponding initial state vector $\mathbf X_0$ for given initial relative orbital elements of the deputy satellite is obtained from using transformation relations \ref{eqnsf}-\ref{eqnsl}, which relates the state parameters in Hill's reference frame to the orbital parameter of deputy satellite.
\begin{eqnarray}
\label{eqnsf}
{x_1} &=& \rho \sin (\omega t + \theta ) + a\\
{x_3} &=& 2\rho \cos (\omega t + \theta ) - \frac{{3\omega }}{2}at + b\\
{x_5} &=& m\rho \sin (\omega t + \theta ) + 2n\rho \cos (\omega t + \theta )\\
{{\dot x}_1} &=& \rho \cos (\omega t + \theta )\\
{{\dot x}_3} &=&  - 2\rho \omega \sin (\omega t + \theta ) - \frac{{3\omega }}{2}a\\
\label{eqnsl}
{{\dot x}_5} &=& m\rho \omega \cos (\omega t + \theta) - 2n\rho \omega \sin (\omega t + \theta )
\end{eqnarray}
The formation trajectory of deputy satellite in Hill's frame is shown in Figure \ref{LQR_PLOT1}. The deputy satellite starts from the inner initial relative formation trajectory and is commanded to outer relative orbit. For better clarity of formation geometry the formation trajectory are also shown in $XY$ (Radial-Cross track), $XZ$ (Radial-Out-of-plane) and $YZ$ (Cross track-Out-of-plane) planes in Figures \ref{LQR_PLOT2}, \ref{LQR_PLOT3}, and \ref{LQR_PLOT4} respectively. Figure \ref{LQR_PLOT5} illustrates the optimal control required in achieving the desired state values $\mathbf X_d$

To have comparative study of LQR controller terminal state accuracy achieved for circular and eccentric chief satellite orbits, a eccentric chief satellite orbit problem is considered. Since the problem with eccentricity in chief satellite orbit is defined in complete nonlinear domain, the term $\left(A\mathbf{X_d} - \mathbf{\dot X_d}\right)$ in \ref{LQR_EQN_FINAL} is no more zero and acts as known controller in addition to the optimal control term \ref{OPTIMAL_CNTRL}. The initial and final condition on the deputy satellite is same as that for circular case given in Table \ref{Initial_condition_LQR}. Orbital parameters of chief satellite are taken to be same as that for circular orbit case but for eccentricity value is considered as $e = 0.15$.

\begin{figure}
\begin{center}
  \includegraphics[width=5in]{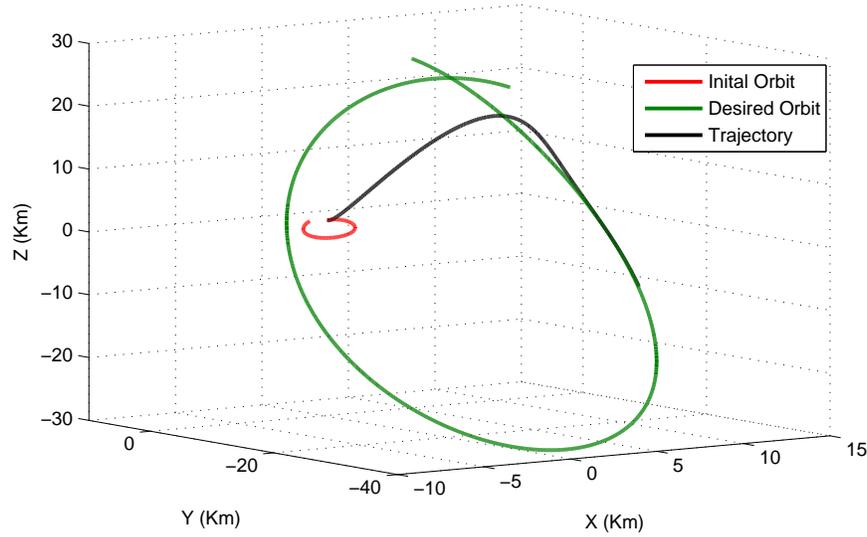}
  \caption{Deputy satellite formation trajectory in Hill's Frame (Eccentric Chief satellite Orbit)}\label{ECC_TRAJ}
  \end{center}
\end{figure}

\begin{figure}
\begin{center}
  \includegraphics[width=4in]{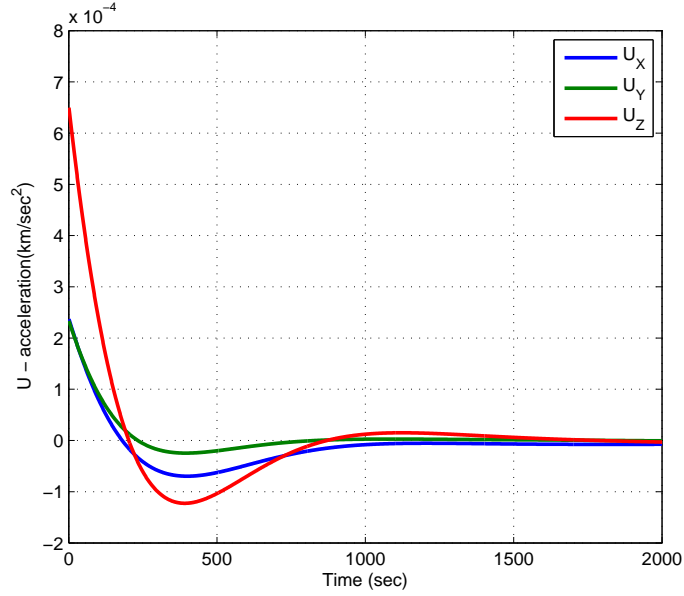}
  \caption{Control Profile,(Eccentric Chief satellite Orbit)}\label{CNTRL_ECC}
  \end{center}
\end{figure}

The formation trajectory for eccentric case and control profile is given in Figures \ref{ECC_TRAJ} and \ref{CNTRL_ECC} respectively. The final state errors for eccentric chief satellite case are given in Table \ref{STATE_ERROR_LQR}. The position error for eccentric case is $35m$ , $230m$ and $333m$ in $x$,$y$ and $z$ respectively as compared to $1.62m$,$-3m$ and $-6.6m$ respectively for circular case for $10km$ base length formation.

\begin{table}
\caption{\small{LQR trajectory State Errors for Circular and Eccentric Chief satellite orbits}}\label{STATE_ERROR_LQR}
\begin{center}\begin{tabular}{|c|c|c|}
    \hline
    \textbf{State} & \textbf{Circular}& \textbf{Eccentric}\\
    \textbf{Error} & \textbf{Orbit Case} & \textbf{Orbit case}  \\
    \hline
    $x(km)$    &     $0.001619$ &0.035      \\
    \hline
    $\dot x (km/sec) $& $6.224 \times 10^{-5}$&$-8.923 \times 10^{-4}$ \\
    \hline
    $y (km)$& $-0.003066$&$0.23$ \\
    \hline
    $\dot y (km/sec)$& $2.692 \times 10^{-5}$&$-1.395 \times 10^{-2}$ \\
    \hline
    $z (km)$& $-0.006593$&$0.3337$ \\
    \hline
    $\dot z (km/sec)$& $-1.039 \times 10^{-5}$&$-1.696 \times 10^{-2}$ \\
    \hline
\end{tabular}\end{center}
\end{table}
\section{Summary and Conclusions}
This chapter primarily introduced the basic concept of generic LQR controller philosophy, works with linear, control affine systems. Linear plant model that is Hill's equation of motion for satellite formation flying introduced in chapter were used. This chapter dealt with details of optimal control computation to achieve the objective of putting a deputy satellite in commanded formation with respect to the chief satellite. A brief comparative study is made for state accuracies achieved using LQR controller for circular and eccentric chief satellite orbit. It was inferred from the above said comparison since LQR controller works with linear state model, final achieved state value accuracy degrades for eccentric chief satellite orbits. Next chapter \ref{SDRE} introduce the concept of a suboptimal nonlinear control namely State Dependent Ricatti Equation (SDRE) control technique. It is shown in Results and Discussion section of next chapter \ref{SDRE} that SDRE technique caters to both circular and eccentric chief satellite orbit formations with improved accuracy in the final states compared to LQR results.

\cleardoublepage
\chapter{State Dependent Ricatti controller for Satellite Formation Flying}
\label{SDRE}

The SDRE technique has been primarily motivated from the standard linear quadratic regulator (LQR) design philosophy. The key idea here is to first write the system dynamics in linear-looking state dependent coefficient (SDC) form, and then by repeatedly solving the corresponding Ricatti equation online at every grid point of time \cite{Cimen,Cloutier}

	Even though the SDRE technique has been primarily developed for infinite-time problems, recently some key ideas have emerged in the literature to extend the concept to finite-time problems as well. A key motivation for that is perhaps the fact that many guidance problems naturally result in finite-time formulations. Some of the prominent techniques that have been reported in the literature are discussed here.
\section{Infinite-time SDRE Formulation ARE Approach}
\label{Infinite_Time_SDRE}
The SDRE technique is primarily valid for control affine systems, the system dynamics for which is given by
\begin{equation}
\label{eqn:21}
\mathbf {\dot X} = f\left( \mathbf X \right) + B\left( \mathbf X \right)\mathbf U
\end{equation}
The key philosophy in this technique is to first convert it to the state dependent coefficient (SDC) form, where the system dynamics is algebraically re-written as
\begin{equation}
\label{eqn:22}
\mathbf {\dot X} = A\left( \mathbf X \right)X + B\left( \mathbf X \right)\mathbf U
\end{equation}
Note that the above expression does not involve any linearization process. Next, the idea is to minimize the following cost function
\begin{equation}
\label{eqn:23}
J = \frac{1}{2}\int\limits_{{t_0}}^\infty  {\left( {\mathbf{X^T}Q\left( \mathbf X \right)\mathbf X + \mathbf{U^T}R\left( \mathbf X \right)\mathbf U} \right)dt}
\end{equation}
Quite obviously, \ref{eqn:22}, along with \ref{eqn:23} appear to be in the LQR form as soon as numerical values of the state vector  $\mathbf X$ is inserted in various matrices. Hence, following the solution procedure of LQR theory, the control solution can be written as
\begin{equation}
\label{eqn:24}
\mathbf U =  - \left[ {{R^{ - 1}}\left(\mathbf X \right){B^T}\left(\mathbf X \right)P\left(\mathbf X \right)} \right]\mathbf X =  - K\left(\mathbf X \right)\,\mathbf X
\end{equation}
where, the Ricatti matrix $P(\mathbf X)$  is repeatedly computed from the following Algebraic Ricatti Equation $(ARE)$
\begin{equation}
\label{eqn:25}
P\left(\mathbf X \right)A\left(\mathbf X \right) + {A^T}\left(\mathbf X \right)P\left(\mathbf X \right) + Q\left(\mathbf X \right) - P\left(\mathbf X \right)B\left(\mathbf X \right){R^{ - 1}}\left(\mathbf X \right){B^T}\left(\mathbf X \right)P\left(\mathbf X \right) = 0
\end{equation}
\\It can be mentioned here that if the objective not $X\rightarrow 0$ , but $X \rightarrow X^*$  (some desired value), then the following expression for the control variable can be used \cite{Cloutier}.
\begin{equation}
\label{eqn:26}
\mathbf U =  - K\left(\mathbf X \right)\left( {\,\mathbf X - \mathbf{X^*}} \right)
\end{equation}
In fact, for better tracking properties, it is also suggested in the literature \cite{Mracek} to incorporate an integral feedback term and use the following expression instead.
\begin{equation}
\label{eqn:27}
\mathbf U =  - {K_P}\left(\mathbf X \right)\left( {\,\mathbf X - \mathbf{X^*}} \right) - {K_I}\left(\mathbf X \right)\int\limits_{{t_0}}^t {\left( {\,\mathbf X - \mathbf{X^*}} \right)\;dt}
\end{equation}
Note that the Ricatti equation \ref{eqn:25} needs to be solved at grid point of time as the matrix values keep on changing. If possible, it can be solved in closed form by long hand algebra, but most of the time it is solved using numerical algorithms. Even though the SDRE technique is obviously a sub-optimal control design and can be carried out under certain conditions, there are certain nice properties of this technique, which can be summarized as follows:
\begin {itemize}
\item Under certain mild assumptions, the SDRE approach produces a closed loop system that is locally asymptotically stable
\item For scalar problems, the resulting SDRE nonlinear controller satisfies all the necessary conditions of optimality and hence results in an optimal controller.
\item Even though initially the solution is sub-optimal, it approaches to the optimal solution with the evolution of time.
\end{itemize}
It can however be noted that Non-uniqueness of the parameterization of the system dynamics poses a major challenge in successful implementation of the SDRE technique. Nevertheless, it has found wide application in a number of problems across the globe. One can find more details about the SDRE technique in \cite{Cloutier}.
\section{Finite-time SDRE Approach}
\label{Finite_Time_SDRE}
As pointed out before, even though the SDRE technique has been primarily developed for infinite-time problems, recently some key ideas have emerged in the literature to extend the concept to finite-time problems as well. Where the cost function to be minimized is as follows
\begin{equation}\label{QUADRATIC_COST}
J = \int\limits_{{t_0}}^{{t_f}} {\left( {\mathbf{X^T}Q\mathbf X + {\mathbf U^T}R\mathbf U} \right)dt}
\end{equation}
subject to the state equation
\[\mathbf{\dot X} = A(\mathbf X)\mathbf X + B\mathbf U\]
and imposing a hard constraint on the final states $\mathbf X\left( {{t_f}} \right) = \mathbf {X_f}$.
Following the classical optimal control theory, in addition to state equation \ref{eqn:22}, the other necessary conditions of optimality are given by \cite{Naidu}
\begin{equation}
\dot \lambda  =  - Q\mathbf X - {A^T}\lambda
\end{equation}
\begin{equation}
\label{Optimal_control}
U =  - {R^{ - 1}}{B^T}\lambda
\end{equation}
Substituting \ref{Optimal_control} in \ref{eqn:22}, the combined state and costate equation can be written in matrix form as Hamiltonian system of state and co-state as
\begin{equation}
\label{Hamiltonian}
\left[ {\begin{array}{*{20}{c}}
\mathbf{\dot X}\\
{\dot \lambda }
\end{array}} \right] = \left[ {\begin{array}{*{20}{c}}
{A(\mathbf X)}&{ - B{R^{ - 1}}{B^T}}\\
Q&{ - A{{(\mathbf X)}^T}}
\end{array}} \right]\left[ {\begin{array}{*{20}{c}}
X\\
\lambda
\end{array}} \right] = H\left[ {\begin{array}{*{20}{c}}
\mathbf X\\
\lambda
\end{array}} \right]
\end{equation}
where
\begin{equation}
H = \left[ {\begin{array}{*{20}{c}}
{A(\mathbf X)}&{ - B{R^{ - 1}}{B^T}}\\
Q&{ - A{{(\mathbf X)}^T}}
\end{array}} \right]
\end{equation}
is known as the 'Hamiltonian matrix'. The solution for the linear equation in \ref{Hamiltonian} is given as
\begin{equation}
\label{STM_SOLN}
\left[ {\begin{array}{*{20}{c}}
{\mathbf  X(t)}\\
{\lambda (t)}
\end{array}} \right] = \left[ {\varphi (t,{t_0})} \right]\left[ {\begin{array}{*{20}{c}}
{\mathbf X({t_0})}\\
{\lambda ({t_0})}
\end{array}} \right] = \left[ {\begin{array}{*{20}{c}}
{{\varphi _{11}}(t,{t_0})}&{{\varphi _{12}}(t,{t_0})}\\
{{\varphi _{21}}(t,{t_0})}&{{\varphi _{22}}(t,{t_0})}
\end{array}} \right]\left[ {\begin{array}{*{20}{c}}
{\mathbf X({t_0})}\\
{\lambda ({t_0})}
\end{array}} \right]
\end{equation}
where $\varphi (t,{t_0})$ is known as the state transition and can be expressed as
\begin{equation}
\label{STM}
\varphi (t,{t_0})\, = {e^{H\left( {t - {t_0}} \right)}}
\end{equation}
One can notice here that whereas $X({t_0})$ is known from the initial condition, $\lambda({t_0})$ is not known. However, from \ref{STM_SOLN}, it is also a fact that the following relationship holds good,
\begin{equation}
\label{STM_SOLN_FINAL_TIME}
\left[ {\begin{array}{*{20}{c}}
{\mathbf X({t_f})}\\
{\lambda ({t_f})}
\end{array}} \right] = \left[ {\begin{array}{*{20}{c}}
{{\varphi _{11}}({t_f},{t_0})}&{{\varphi _{12}}({t_f},{t_0})}\\
{{\varphi _{21}}({t_f},{t_0})}&{{\varphi _{22}}({t_f},{t_0})}
\end{array}} \right]\left[ {\begin{array}{*{20}{c}}
{\mathbf X({t_0})}\\
{\lambda ({t_0})}
\end{array}} \right]
\end{equation}
Hence, from \ref{STM_SOLN_FINAL_TIME} $\lambda({t_0})$ can be calculated as
\begin{equation}
\label{LAMBDA_TO_CALC}
\lambda ({t_0}) = \varphi _{12}^{ - 1}({t_f},{t_0})\left[ {\mathbf {X_f} - {\varphi _{11}}({t_f},{t_0})X({t_0})} \right]
\end{equation}
Note that the hard constraint information $\mathbf X\left( {{t_f}} \right) = \mathbf {X_f}$ is utilized in the expression in \ref{LAMBDA_TO_CALC} to compute  $\lambda({t_0})$. After knowing $\lambda({t_0})$ , $\lambda({t})$  can be calculated from \ref{STM_SOLN} and finally the optimal control is calculated from \ref{Optimal_control}.
Note that the matrices $A,B,Q,R$ are time varying matrices, the expression for $\varphi (t,{t_0})$ where matrix $H(\mathbf X)$  becomes time-varying and the closed form expression for $\varphi (t,{t_0})$ in \ref{STM} is not valid in 'strict sense'. However, following the philosophy of the SDRE framework, the idea is to repeatedly evaluate $\varphi (t,{t_0})$ in \ref{STM} at every grid point of time and then evaluate the optimal control expression.
\section{Satellite Formation Flying SDC formulation}
The SDRE control technique requires the nonlinear equation of motion to be re-written in state dependent coefficient(SDC) form which has following structure.
\[\mathbf{\dot X} = A(\mathbf X)\mathbf X + B(\mathbf X)\mathbf U\]
relative position and relative velocities of the deputy satellite with respect to chief satellite in Hill's frame of reference are chosen as states of the system. It is assumed that the all the states are available through measurement, hence a full state feedback is implemented. The SDC formulation is not unique and dependents on the designer how they reform the nonlinear system equation into the SDC form, many such suboptimal control synthesis is possible depending on the SDC form selected. The best SDC form is selected which preserves the as much as possible the nonlinearity of the problem, avoid singularity and yet rewrite the equation in linear looking form.
Two methods of SDC formulation using the nonlinear SFF equation of motion \ref{EQN_SFF} is discussed in the subsequent sections
\subsection{SDC Formulation $Method : I$}
$Method : I$ uses equation \ref{EQN_SFF} SFF nonlinear equation of motion to be rewritten into SDC form. We rewrite the following nonlinear terms,
${\frac{\mu }{\gamma }{r_c} - \frac{\mu }{{r_c^2}}}$
 in \ref{EQN_SFF} and express them in the linear looking form and at the same time preserving the nonlinear behavior to the extent possible and avoid any singularity. The term is rewritten and simplified as follows \cite{Park}
\begin{eqnarray}
\frac{\mu }{\gamma }{r_c} - \frac{\mu }{{r_c^2}} &=& \mu \left[ {\frac{{{r_c}}}{{{{\left( {{{\left( {{r_c} + x} \right)}^2} + {y^2} + {z^2}} \right)}^3}}} - \frac{1}{{r_c^2}}} \right]\\
 &=& \mu \left[ {\frac{{{r_c}}}{{{{\left( {r_c^2 + 2{r_c}x + {x^2} + {y^2} + {z^2}} \right)}^3}}} - \frac{1}{{r_c^2}}} \right]
\end{eqnarray}

Factorizing the term $r_c^2$ from the denominator term.
\begin{eqnarray}
\frac{\mu }{\gamma }{r_c} - \frac{\mu }{{r_c^2}} &=& \frac{\mu }{{r_c^2}}\left[ {\frac{1}{{{{\left( {1 + 2\frac{x}{{{r_c}}} + \frac{{{x^2}}}{{r_c^2}} + \frac{{{y^2}}}{{r_c^2}} + \frac{{{z^2}}}{{r_c^2}}} \right)}^{\frac{3}{2}}}}} - 1} \right]\\
&=& \frac{\mu }{{r_c^2}}\left[ {{{\left( {1 - \left( { - 2\frac{x}{{{r_c}}} - \frac{{\left( {{x^2} + {y^2} + {z^2}} \right)}}{{r_c^2}}} \right)} \right)}^{ - \frac{3}{2}}} - 1} \right]
\end{eqnarray}
Defining
\begin{eqnarray}
\xi  &=&  - 2\frac{x}{{{r_c}}} - \frac{{\left( {{x^2} + {y^2} + {z^2}} \right)}}{{r_c^2}}\\
&=& \left( { - \frac{2}{{{r_c}}} - \frac{x}{{r_c^2}}} \right)x + \left( { - \frac{y}{{r_c^2}}} \right)y + \left( { - \frac{z}{{r_c^2}}} \right)z
\end{eqnarray}
Further using negative binomial expansion,
\begin{equation}
\label{Binomial_generic}
{\left( {1 + x} \right)^{ - n}} = 1 - nx + \frac{{n\left( {n + 1} \right)}}{{2!}}{x^2} - \frac{{n\left( {n + 1} \right)\left( {n + 2} \right)}}{{3!}}{x^3} +  \ldots
\end{equation}
the term ${\left( {1 - \xi } \right)^{ - \frac{3}{2}}}$ can be written as infinite series sum as follows,
\begin{equation}
{\left( {1 - \xi } \right)^{ - \frac{3}{2}}} = 1 + \frac{3}{2}\xi  + \frac{3}{2}\frac{{\left( {\frac{3}{2} + 1} \right)}}{{2!}}{\xi ^2} + \frac{3}{2}\frac{{\left( {\frac{3}{2} + 1} \right)\left( {\frac{3}{2} + 2} \right)}}{{3!}}{\xi ^3} +  \ldots
\end{equation}
Defining the term
\[\psi  = 1 + {\psi _1} + {\psi _2} + {\psi _3} +  \ldots \]
where
\[{\psi _1} = \frac{{\left( {\frac{3}{2} + 1} \right)}}{2}\xi ,\hspace{5mm}{\psi _2} = \frac{{\left( {\frac{3}{2} + 2} \right)}}{3}{\psi _1}\xi ,\hspace{5mm}{\psi _3} = \frac{{\left( {\frac{3}{2} + 3} \right)}}{4}{\psi _2}\xi \ldots \]
with the series expression the nonlinear term can be expressed as follows.
\begin{equation}
\frac{\mu }{\gamma }{r_c} - \frac{\mu }{{r_c^2}} = \frac{\mu }{{r_c^2}}\left[ {1 + \frac{3}{2}\psi \xi  - 1} \right]
 = \frac{{3\mu }}{{2r_c^2}}\psi \xi
\end{equation}
Using the above state dependent coefficient form of the nonlinear term, the nonlinear equation of relative motion of the satellite in Hill's frame can be rewritten in the SDC form as follows \cite{Park}
\begin{equation}
\left[ {\begin{array}{*{20}{c}}
{{{\dot x}_1}}\\
{{{\dot x}_2}}\\
{{{\dot x}_3}}\\
{{{\dot x}_4}}\\
{{{\dot x}_5}}\\
{{{\dot x}_6}}
\end{array}} \right] + \left[ {\begin{array}{*{20}{c}}
0&1&0&0&0&0\\
{{{\dot \nu }^2} - \frac{\mu }{\gamma } + \frac{{3\mu }}{{2r_c^3}}\left( {2 + \frac{x_1}{{{r_c}}}} \right)\psi }&0&{\ddot \nu  + \frac{{3\mu }}{{2r_c^2}}\psi x_3}&{2\dot \nu }&{\frac{{3\mu }}{{2r_c^2}}\psi x_5}&0\\
0&0&0&1&0&0\\
{ - \ddot \nu }&{ - 2\dot \nu }&{{{\dot \nu }^2} - \frac{\mu }{\gamma }}&0&0&0\\
0&0&0&0&0&1\\
0&0&0&0&{ - \frac{\mu }{\gamma }}&1
\end{array}}\right]\left[ {\begin{array}{*{20}{c}}
{{x_1}}\\
{{x_2}}\\
{{x_3}}\\
{{x_4}}\\
{{x_5}}\\
{{x_6}}
\end{array}} \right]\nonumber
\end{equation}
\begin{equation}
\label{SDC_SFF}
+ \left[ {\begin{array}{*{20}{c}}
0&0&0\\
1&0&0\\
0&0&0\\
0&1&0\\
0&0&0\\
0&0&1
\end{array}} \right]U
\end{equation}
\subsection{SDC Formulation $Method : II$}
$Method : II$ uses the equation \ref{L_EOM_STEP1}, SFF nonlinear equation of motion to be rewritten into SDC form \cite{Irvin}.
\begin{eqnarray}
\ddot x - 2\omega \dot y - {\omega ^2}({r_c} + x)\left[ {1 - \frac{{r_c^3}}{{{{\left( {{{\left( {{r_c} + x} \right)}^2} + {y^2} + {z^2}} \right)}^{\frac{3}{2}}}}}} \right] - {a_x} &=& 0\\
\ddot y + 2\omega \dot x - {\omega ^2}y\left[ {1 - \frac{{r_c^3}}{{{{\left( {{{\left( {{r_c} + x} \right)}^2} + {y^2} + {z^2}} \right)}^{\frac{3}{2}}}}}} \right] - {a_y} &=& 0\\
\ddot z + {\omega ^2}z\left[ {\frac{{r_c^3}}{{{{\left( {{{\left( {{r_c} + x} \right)}^2} + {y^2} + {z^2}} \right)}^{\frac{3}{2}}}}}} \right] - {a_z} &=& 0
\end{eqnarray}
Lets use definition of ${\sigma _x}$, ${\sigma _y}$ and ${\sigma _z}$ as defined in equations \ref{SIGMA_DEFINITION}
\begin{eqnarray}
\label{Method2_Substitution}
{\sigma _z} &=& \frac{{r_c^3}}{{{{\left( {{{\left( {{r_c} + x} \right)}^2} + {y^2} + {z^2}} \right)}^{\frac{3}{2}}}}} \nonumber \\
{\sigma _y} &=& 1 - {\sigma _z} \nonumber \\
{\sigma _x} &=& \left( {\frac{{{r_c}}}{x} + 1} \right){\sigma _y}\nonumber
\end{eqnarray}
substituting the terms in \ref{SIGMA_DEFINITION} into equation \ref{L_EOM_STEP1} we can write the equation \ref{L_EOM_STEP1} as follows,
\begin{eqnarray}
\label{METHOD2_SDC1}
\ddot x &=& 2\omega \dot y + {\omega ^2}{\sigma _x}x + {a_x} \nonumber  \\
\ddot y &=&  - 2\omega \dot x + {\omega ^2}{\sigma _y}y + {a_y}\\
\ddot z &=& -{\omega ^2}{\sigma _z}z + {a_z} \nonumber
\end{eqnarray}
Writing the above equation \ref{METHOD2_SDC1} in state space form $\mathbf{\dot X} = A(\mathbf X)\mathbf X + B(\mathbf X)\mathbf U$ we get the following equation.
\begin{equation}
\label{METHOD2_SDC2}
\left[ {\begin{array}{*{20}{c}}
{{{\dot x}_1}}\\
{{{\dot x}_2}}\\
{{{\dot x}_3}}\\
{{{\dot x}_4}}\\
{{{\dot x}_5}}\\
{{{\dot x}_6}}
\end{array}} \right] = \left[ {\begin{array}{*{20}{c}}
0&1&0&0&0&0\\
{{{\dot \nu }^2}{\sigma _x}}&0&0&{2\dot \nu }&0&0\\
0&0&0&1&0&0\\
0&{ - 2\dot \nu }&{{{\dot \nu }^2}{\sigma _y}}&0&0&0\\
0&0&0&0&0&1\\
0&0&0&0&{ - {{\dot \nu }^2}{\sigma _z}}&0
\end{array}} \right]\left[ {\begin{array}{*{20}{c}}
{{x_1}}\\
{{x_2}}\\
{{x_3}}\\
{{x_4}}\\
{{x_5}}\\
{{x_6}}
\end{array}} \right] + \left[ {\begin{array}{*{20}{c}}
0&0&0\\
1&0&0\\
0&0&0\\
0&1&0\\
0&0&0\\
0&0&1
\end{array}} \right]U
\end{equation}
For circular orbits,\\
The mean motion, $\omega $ $=$ rate of true anomaly (satellite orbital angular velocity $\dot \nu$)
\begin{equation}
\omega  = \sqrt {\frac{\mu }{{r_c^3}}}  = \dot \nu
\end{equation}
$Method:II$ SDC formulation do not approximate the nonlinear equation motion of SFF as closely as approximated by $Method:I$. Since the nonlinear equations \ref{L_EOM_STEP1} of SFF used by $Method:II$ to arrive at the SDC formulation \ref{METHOD2_SDC2} is derived from \ref{EQN_SFF} under assumption that orbit is circular and term $\ddot \nu = 0$. Hence SDC formulation \ref{METHOD2_SDC2} only caters to the circular reference orbit solution. Therefore the SDRE solution accuracy depends on the SDC formulation of the system. $Method:I$ formulation is used for the result generation and as nonlinear controller for comparison with MPSP and G-MPSP solution in chapter.
\section{SDC formulation for $J_2$ perturbation model}
Nonlinear $J_2$ perturbation model details are introduced in the section \ref{J2_modelling}. The nonlinear $J_2$ model is given in the equation \ref{J2_MODEL2}. Redefining the terms $i_d = i_c+ \delta i $ and $\theta_d = \theta_c+ \delta \theta $ in equation \ref{J2_MODEL2} and rewriting the equation \ref{J2_MODEL2},
\begin{equation}
\label{J2_MODEL3}
{a_{J2}} = \frac{{3\mu R_e^2{J_2}}}{2}\left\{ {\frac{1}{{{{\left( {\mathbf{r_c} + \mathbf \rho } \right)}^4}}}\left( {\begin{array}{*{20}{c}}
{{J_{{{\hat e}_x}}}({i_d},{\theta _d})}\\
{{J_{{{\hat e}_y}}}({i_d},{\theta _d})}\\
{{J_{{{\hat e}_z}}}({i_d},{\theta _d})}
\end{array}} \right) - \frac{1}{{r_c^4}}\left( {\begin{array}{*{20}{c}}
{{J_{{{\hat e}_x}}}({i_c},{\theta _c})}\\
{{J_{{{\hat e}_y}}}({i_c},{\theta _c})}\\
{{J_{{{\hat e}_z}}}({i_c},{\theta _c})}
\end{array}} \right)} \right\}
\end{equation}
Expressing the term, $\frac{1}{{{{\left( {\mathbf{r_c} + \mathbf \rho } \right)}^4}}}$ using negative binomial expansion as a infinite sum series,
\begin{eqnarray}
\frac{1}{{{{\left( \mathbf{{r_c} + \rho } \right)}^4}}} &=& \frac{1}{{{{\left( {{{\left( {{r_c} + x} \right)}^2} + {y^2} + {z^2}} \right)}^2}}}\\
&=& \frac{1}{{{{\left( {r_c^2 + 2{r_c}x + {x^2} + {y^2} + {z^2}} \right)}^2}}}\\
&=& \frac{1}{{r_c^4}}\left\{ {{{\left( {1 - \left( {2\frac{x}{{{r_c}}} - \frac{{{x^2} + {y^2} + {z^2}}}{{r_c^2}}} \right)} \right)}^{ - 2}}} \right\}
\end{eqnarray}
Defining
\[\xi  =  - 2\frac{x}{{{r_c}}} - \frac{{\left( {{x^2} + {y^2} + {z^2}} \right)}}{{r_c^2}}
 = \left( { - \frac{2}{{{r_c}}} - \frac{x}{{r_c^2}}} \right)x + \left( { - \frac{y}{{r_c^2}}} \right)y + \left( { - \frac{z}{{r_c^2}}} \right)z\]
Binomial series expansion can be written as,
\begin{equation}
{\left( {1 - \xi } \right)^{ - 2}} = 1 + 2\xi  + \frac{{2\left( {2 + 1} \right)}}{{2!}}{\xi ^2} + \frac{{2\left( {2 + 1} \right)\left( {2 + 2} \right)}}{{3!}}{\xi ^3} +  \ldots
\end{equation}
Defining the term
\begin{equation}
\eta  = 1 + {\eta _1} + {\eta _2} + {\eta _3} +  \ldots
\end{equation}
Where $\eta's $ are defined as follows,
\[{\eta _1} = \frac{{\left( {2 + 1} \right)}}{2}\xi ,\hspace{3mm} {\eta _2} = \frac{{\left( {2 + 2} \right)}}{3}{\psi _1}\xi ,\hspace{3mm} {\eta _3} = \frac{{\left( {2 + 3} \right)}}{4}{\psi _2}\xi \] and so on.

Therefore using the above definition the nonlinear term can be written in linear looking form as follows,
\begin{equation}
\label{J2_simplification_of_NL_term}
\frac{1}{{{{\left( {{r_c} + \rho } \right)}^4}}} = \frac{1}{{r_c^4}} + 2\frac{1}{{r_c^4}}\eta \xi
\end{equation}
substituting \ref{J2_simplification_of_NL_term} in \ref{J2_MODEL3} and rearranging the terms,
\begin{equation}
{a_{J2}} = \frac{3}{2}\frac{{\mu R_e^2{J_2}}}{{r_c^4}}\left\{ {\left( {\begin{array}{*{20}{c}}
{{J_{{{\hat e}_x}}}({i_d},{\theta _d})}\\
{{J_{{{\hat e}_y}}}({i_d},{\theta _d})}\\
{{J_{{{\hat e}_z}}}({i_d},{\theta _d})}
\end{array}} \right) - \left( {\begin{array}{*{20}{c}}
{{J_{{{\hat e}_x}}}({i_c},{\theta _c})}\\
{{J_{{{\hat e}_y}}}({i_c},{\theta _c})}\\
{{J_{{{\hat e}_z}}}({i_c},{\theta _c})}
\end{array}} \right)} \right\}
\end{equation}
\begin{equation}\label{J2_MODEL4}
\hspace{10mm}+ \frac{{3\mu R_e^2{J_2}}}{{r_c^4}}\eta \left\{ {\left( {\begin{array}{*{20}{c}}
{{J_{{{\hat e}_x}}}({i_d},{\theta _d})}\\
{{J_{{{\hat e}_y}}}({i_d},{\theta _d})}\\
{{J_{{{\hat e}_z}}}({i_d},{\theta _d})}
\end{array}} \right)} \right\}\xi
\end{equation}
The Second term in \ref{J2_MODEL4} is explicit function of $\xi$ and hence the explicit function of states values. But where as the first term in \ref{J2_MODEL4} is to be modeled into SDC form as follows.

Consider only the first term of equation \ref{J2_MODEL4},
\begin{equation}
{a_{J2}} = \frac{3}{2}\frac{{\mu R_e^2{J_2}}}{{r_c^4}}\left\{ {\left( {\begin{array}{*{20}{c}}
{{J_{{{\hat e}_x}}}({i_d},{\theta _d})}\\
{{J_{{{\hat e}_y}}}({i_d},{\theta _d})}\\
{{J_{{{\hat e}_z}}}({i_d},{\theta _d})}
\end{array}} \right) - \left( {\begin{array}{*{20}{c}}
{{J_{{{\hat e}_x}}}({i_c},{\theta _c})}\\
{{J_{{{\hat e}_y}}}({i_c},{\theta _c})}\\
{{J_{{{\hat e}_z}}}({i_c},{\theta _c})}
\end{array}} \right)} \right\}
\end{equation}
Using the definition of $J_{{{\hat e}_x}}$, $J_{{{\hat e}_y}}$ and $J_{{{\hat e}_z}}$ and substituting in the above equation we can write the first term in \ref{J2_MODEL4} as follows,
\begin{equation}
\label{J2_MODEL5}
{a_{J2}} = \frac{3}{2}\frac{{\mu R_e^2{J_2}}}{{r_c^4}}\left\{ {\begin{array}{*{20}{c}}
{ - 3{{\sin }^2}(i + \delta i){{\sin }^2}(\theta  + \delta \theta ) + 3{{\sin }^2}(i){{\sin }^2}(\theta )}\\
{{{\sin }^2}(i + \delta i)\sin (2\theta  + 2\delta \theta ) - {{\sin }^2}(i)\sin (2\theta )}\\
{\sin (2i + 2\delta i)\sin (\theta  + \delta \theta ) - \sin (2i)\sin (\theta )}
\end{array}} \right\}
\end{equation}
in the above equation subscript $c$ on the $i$ and $\theta$ is dropped for simplicity. And the orbital elements without subscript are considered to be orbital elements for chief satellite, and orbital elements added with delta variation of quantity signifies the orbital elements for deputy satellite.
Now using Taylor series expansion the trigonometric function can be written in form of infinite series as follows,
\begin{eqnarray}
\sin (i) &=& i - \frac{{{i^3}}}{{3!}} + \frac{{{i^5}}}{{5!}} +  \ldots \\
\sin (i + \delta i) &=& (i + \delta i) - \frac{{{{(i + \delta i)}^3}}}{{3!}} + \frac{{{{(i + \delta i)}^5}}}{{5!}} +  \ldots
\end{eqnarray}
 and similarly we can write Taylor series expansion for
\begin{itemize}
\item $sin(\theta)$
\item $sin(\theta + \delta \theta)$
\item $sin(2i)$
\item $sin(2i + 2\delta i)$
\item $sin(2\theta)$
\item $sin(2\theta + 2\delta \theta)$
\end{itemize}
We can do some algebraic simplification to the to the Taylor series of $sin \left (i + \delta i\right )$ and rewrite the series. Add and subtract $i^2$, $i^4$ and so on ti the term of the series and rewrite the series as follows
\begin{equation}
\sin (i + \delta i) = (i + \delta i)\left( {1 - \frac{{{i^2}}}{{3!}} - \frac{{\left( {{{(i + \delta i)}^2} - {i^2}} \right)}}{{3!}} + \frac{{{i^4}}}{{5!}} + \frac{{\left( {{{(i + \delta i)}^4} - {i^4}} \right)}}{{5!}} +  \ldots } \right)
\end{equation}
\begin{equation}
 = (i + \delta i)\left( {\left( {1 - \frac{{{i^2}}}{{3!}} + \frac{{{i^4}}}{{5!}} +  \ldots } \right) + \left( { - \frac{{\left( {{{(i + \delta i)}^2} - {i^2}} \right)}}{{3!}} + \frac{{\left( {{{(i + \delta i)}^4} - {i^4}} \right)}}{{5!}} +  \ldots } \right)} \right)
\end{equation}
Defining the following quantities
\begin{eqnarray}
{\alpha _i} &=& 1 - \frac{{{i^2}}}{{3!}} + \frac{{{i^4}}}{{5!}} +  \ldots \\
{\alpha _\theta } &=& 1 - \frac{{{\theta ^2}}}{{3!}} + \frac{{{\theta ^4}}}{{5!}} +  \ldots \\
{\beta _i} &=&  - \frac{{\left( {{{(i + \delta i)}^2} - {i^2}} \right)}}{{3!}} + \frac{{\left( {{{(i + \delta i)}^4} - {i^4}} \right)}}{{5!}} +  \ldots \\
{\beta _\theta } &=&  - \frac{{\left( {{{(\theta  + \delta \theta )}^2} - {\theta ^2}} \right)}}{{3!}} + \frac{{\left( {{{(\theta  + \delta \theta )}^4} - {\theta ^4}} \right)}}{{5!}} +  \ldots
\end{eqnarray}
for ${\beta _i}$ and ${\beta _\theta }$ expanding the terms within the brackets and can be rewritten as follows,
\[{\beta _i} = {\eta _i}\delta i,\hspace{5mm}{\beta _\theta } = {\eta _\theta }\delta \theta ,\]
where ${\eta _i}$ and ${\eta _\theta }$ are defined as follows,
\begin{eqnarray}
{\eta _i} &=& \frac{{{}_2{C_1}i + {}_2{C_2}\delta i}}{{3!}} + \frac{{{}_4{C_1}{i^3} + {}_4{C_2}\left( {\delta i} \right){i^2} + {}_4{C_3}{{\left( {\delta i} \right)}^2}i + {}_4{C_4}{{\left( {\delta i} \right)}^3}}}{{5!}}\nonumber\\
{\eta _\theta } &=& \frac{{{}_2{C_1}\theta  + {}_2{C_2}\delta \theta }}{{3!}} + \frac{{{}_4{C_1}{\theta ^3} + {}_4{C_2}\left( {\delta \theta } \right){\theta ^2} + {}_4{C_3}{{\left( {\delta \theta } \right)}^2}\theta  + {}_4{C_4}{{\left( {\delta \theta } \right)}^3}}}{{5!}}\nonumber
\end{eqnarray}
where $C$ is binomial coefficient and ${}_n{C_k} = \frac{{n!}}{{k!\left( {n - k} \right)!}}$.\\
Similar expression can be written for ${\beta _{2i}}$ and ${\beta _{2\theta}}$ \\
Using the above Taylor series approximation of the trigonometric function of $J_2$ model we can write the SDC formulation of \ref{J2_MODEL5} $J_2$ model as follows,
\begin{equation}
{a_{J2}} = \frac{3}{2}\frac{{\mu R_e^2{J_2}}}{{r_c^4}}\left\{ {\begin{array}{*{20}{c}}
{ - 3\left( {{\zeta _{x1}} + {\zeta _{x3}}} \right)\delta i - 3\left( {{\zeta _{x2}} + {\zeta _{x4}}} \right)\delta \theta }\\
{\left( {{\zeta _{y1}} + {\zeta _{y3}}} \right)\delta i + \left( {{\zeta _{y2}} + {\zeta _{y4}}} \right)\delta \theta }\\
{\left( {{\zeta _{z1}} + {\zeta _{z3}}} \right)\delta i + \left( {{\zeta _{z2}} + {\zeta _{z4}}} \right)\delta \theta }
\end{array}} \right\} \nonumber
\end{equation}
\begin{equation}
\label{J2_MODEL_SDC}
\hspace{13mm}+ \frac{{3\mu R_e^2{J_2}}}{{r_c^4}}\eta \left\{ {\left( {\begin{array}{*{20}{c}}
{{J_{{{\hat e}_x}}}({i_d},{\theta _d})}\\
{{J_{{{\hat e}_y}}}({i_d},{\theta _d})}\\
{{J_{{{\hat e}_z}}}({i_d},{\theta _d})}
\end{array}} \right)} \right\}\xi
\end{equation}
Where
\begin{eqnarray}
{\zeta _{x1}} &=& {\left( {{\alpha _i} + {\beta _i}} \right)^2}{\left( {\theta  + \delta \theta } \right)^2}{\left( {{\alpha _\theta } + {\beta _\theta }} \right)^2}\left( {2i + \delta i} \right) \nonumber \\
{\zeta _{x2}} &=& {i^2}{\left( {{\alpha _i} + {\beta _i}} \right)^2}{\left( {{\alpha _\theta } + {\beta _\theta }} \right)^2}\left( {2\theta  + \delta \theta } \right) \nonumber \\
{\zeta _{x3}} &=& {i^2}{\theta ^2}{\left( {{\alpha _\theta } + {\beta _\theta }} \right)^2}\left( {2{\alpha _i} + {\beta _i}} \right){\eta _i} \nonumber \\
{\zeta _{x4}} &=& {i^2}{\theta ^2}{\left( {{\alpha _i}} \right)^2}\left( {2{\alpha _\theta } + {\beta _\theta }} \right){\eta _\theta } \nonumber \\
{\zeta _{y1}} &=& 2{\left( {{\alpha _i} + {\beta _i}} \right)^2}\left( {\theta  + \delta \theta } \right)\left( {{\alpha _{2\theta }} + {\beta _{2\theta }}} \right)\left( {2i + \delta i} \right) \nonumber \\
{\zeta _{y2}} &=& 2{i^2}\left( {\theta  + \delta \theta } \right)\left( {{\alpha _{2\theta }} + {\beta _{2\theta }}} \right){\left( {2{\alpha _i} + {\beta _i}} \right)^2}{\eta _i} \nonumber \\
{\zeta _{y3}} &=& 2{i^2}{\alpha _i}^2\left( {{\alpha _{2\theta }} + {\beta _{2\theta }}} \right) \nonumber \\
{\zeta _{y4}} &=& 4{i^2}{\alpha _i}^2\theta {\eta _{2\theta }} \nonumber \\
{\zeta _{z1}} &=& 2{\left( {{\alpha _{2i}} + {\beta _{2i}}} \right)^2}\left( {\theta  + \delta \theta } \right)\left( {{\alpha _\theta } + {\beta _\theta }} \right) \nonumber \\
{\zeta _{z2}} &=& 4i\left( {\theta  + \delta \theta } \right)\left( {{\alpha _\theta } + {\beta _\theta }} \right){\eta _{2i}} \nonumber \\
{\zeta _{z3}} &=& 2i{\alpha _{2i}}\left( {{\alpha _\theta } + {\beta _\theta }} \right) \nonumber \\
{\zeta _{z4}} &=& 2i{\alpha _{2i}}\theta {\eta _\theta } \nonumber
\end{eqnarray}
Further putting SDC form of the nonlinear equation of motion of SFF in Hill's Frame and SDC form of the $J_2$ perturbation model we can write the total SDC model of the plant and perturbing forces, we can write the system equation of motion as follows \cite{Park}.
\begin{equation}
\mathbf{\dot X} = A_{J2}\left(\mathbf X \right)\mathbf X + B\mathbf U
\end{equation}
\section{Results and Discussions}
The Result section are divided in two parts, one is infinite time formulation results and another finite time formulation results. Each set of presented results have further ratification as, results of $SDC1$ and $SDC2$ model of plant dynamics.
\subsection{Infinite time SDRE Results}
Infinite time SDRE solution procedure is illustrated in section \ref{Infinite_Time_SDRE}. Since SDRE uses SDC formulation of nonlinear equation of motion, hence a larger base-line length formation and  eccentric chief satellite orbit is considered to test the capability of the controller. Table \ref{Initial_condition_SDRE}  lists the initial and final relative parameters of the deputy satellite. The simulation step size is selected $\triangle t =1sec$. The simulation is stopped once the tracking error becomes smaller than pre-selected tolerance value. The control weight $(R)$ is selected based on the a selection process where weight is varied from $R=10^8I_{3 \times 3}$ to $R=10^{11}I_{3 \times 3}$. Higher weight on control $(R)$ translates in  lesser control effort and higher settling time and lower $R$ translates in higher control and lesser settling time. Formation plots for $R=10^9I_{3 \times 3}$ and $R=10^{11}I_{3 \times 3}$value are given in this section. The settling time and control effort value for all control weight $(R)$ values are given in Table \ref{COMP_R}
\begin{table}[h!]
\caption{\small{Deputy Satellite Initial condition for Infinite time SDRE solution}}\label{Initial_condition_SDRE}
\begin{center}\begin{tabular}{|c|c|c|}
    \hline
    \textbf{Orbital} & \textbf{Initial Value}&\textbf{Final Value}\\
    \textbf{Parameters} &                     &\\
    \hline
    $\rho (km)$ & $5 km$& $25 (km)$\\
    \hline
    $\theta (deg) $& $45^0$ & $60^0$\\
    \hline
    $a (km)$& $0$ & $0$\\
    \hline
    $b (km)$& $0$ & $0$\\
    \hline
    $m$ (slope)& $1$ & $1.5$\\
    \hline
    $n$(slope)& $0$ & $0$\\
    \hline
\end{tabular}\end{center}
\end{table}
\begin{table}[h!]
\caption{\small{Chief Satellite Orbital Parameters, (SDRE)}}\label{ORB_PAR_SDRE}
\begin{center}\begin{tabular}{|c|c|}
    \hline
    \textbf{Orbital Parameters} & \textbf{Value}\\
    \hline
    \textbf{Semi-major axis} & $10000 km$ \\
    \hline
    \textbf{Eccentricity} & $0.15$ \\
    \hline
    \textbf{Orbit Inclination} & $0$ \\
    \hline
    \textbf{Argument of Perigee} & $0$ \\
    \hline
    \textbf{Longitude of ascending node} & $0$ \\
    \hline
    \textbf{Initial True Anomaly} & $10$ \\
    \hline
\end{tabular}\end{center}
\end{table}
\begin{figure}
\centering
  \includegraphics[width=5in]{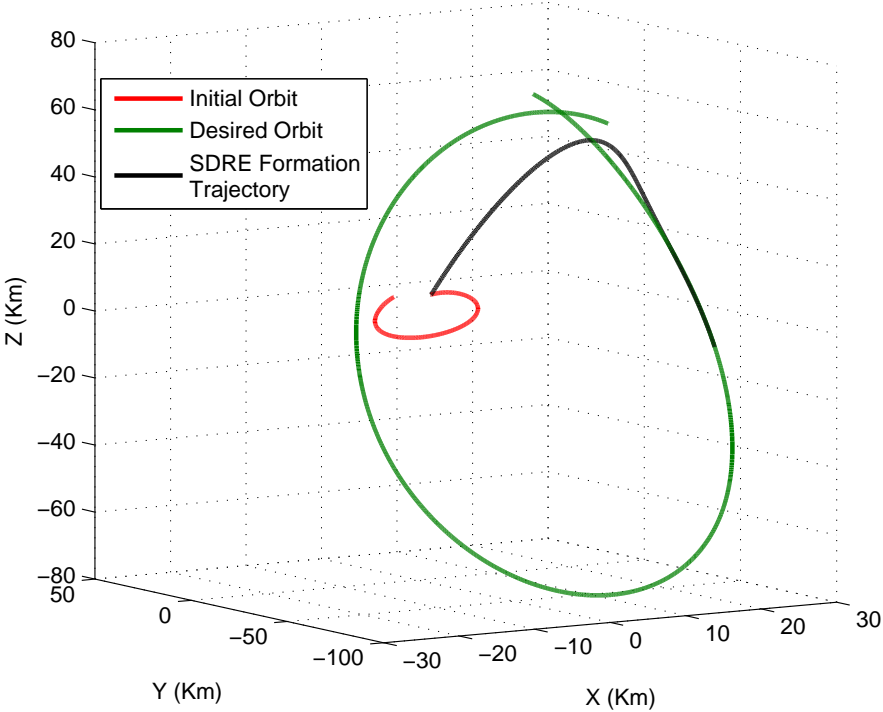}\\
  \caption{SDRE, Reconfiguration Trajectory}\label{SDRE_TRAJ}
\end{figure}
\begin{figure}
\centering
  \includegraphics[width=5in]{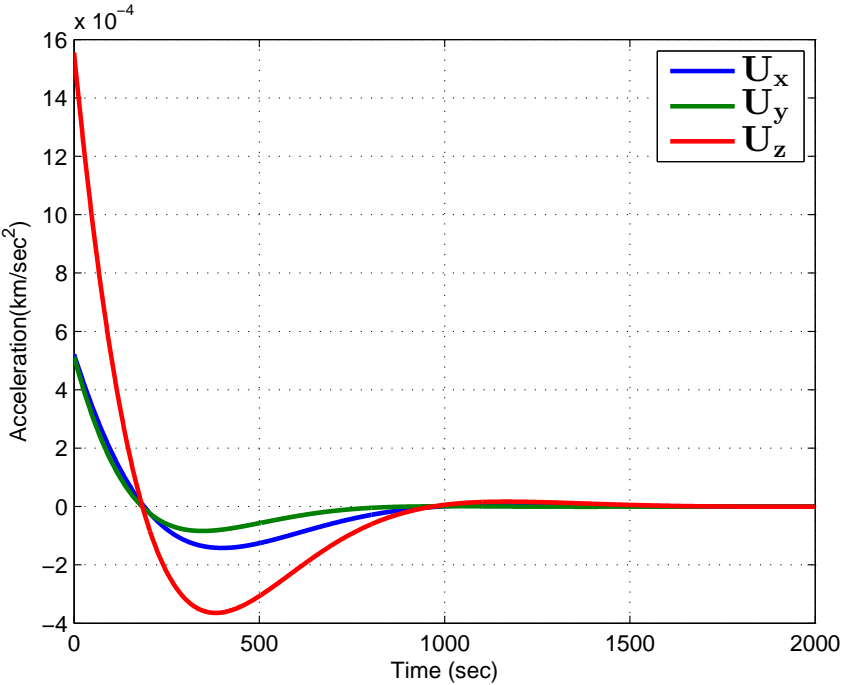}\\
  \caption{SDRE control $U$,  required for commanded formation reconfiguration}\label{SDRE_CNTRL}
\end{figure}
\begin{figure}
\centering
  \includegraphics[width=6in]{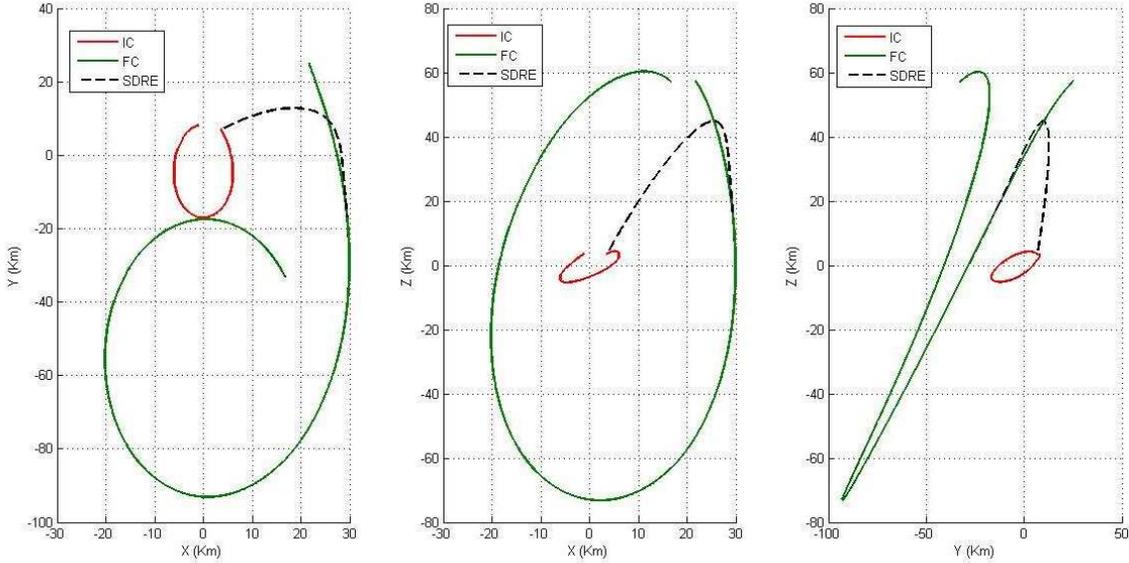}\\
  \caption{SDRE,Reconfiguration Trajectory in $XY$, $XZ$ and $YZ$ plane views.}\label{SDRE_TRAJ_VIEWS}
\end{figure}

\begin{figure}
\centering
  \includegraphics[width=5in]{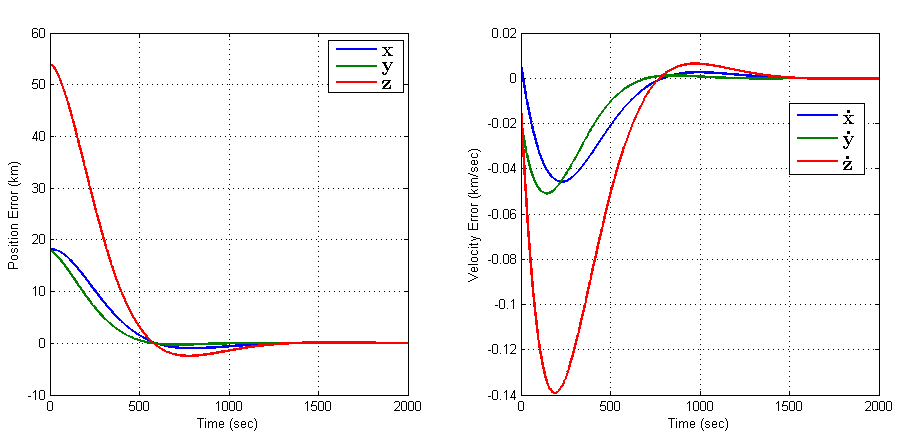}\\
  \caption{Position and Velocity Error plots for $\rho_{initial} = 5km$ and $\rho_{final} = 25km$}\label{POS_VEL_ERR_SDRE}
\end{figure}

\begin{figure}
\centering
  \includegraphics[width=5in]{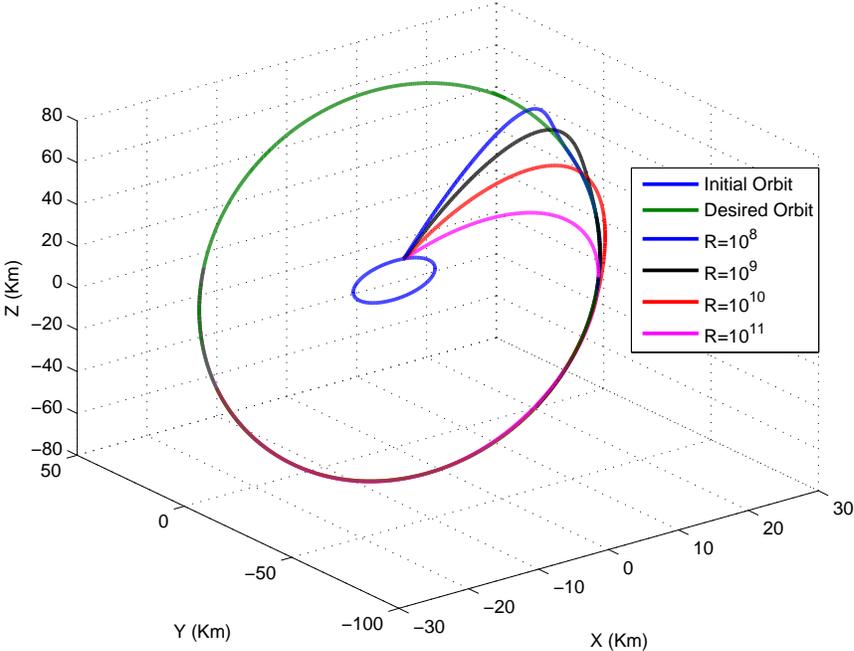}\\
  \caption{Formation Trajectory Circular Chief orbit $\rho_{initial} = 5km$ and $\rho_{final} = 25km$
  for $R=10^8I_{3 \times 3}$, $R=10^9I_{3 \times 3}$, $R=10^{10}I_{3 \times 3}$ and $R=10^{11}I_{3 \times 3}$}
  \label{R_TRAJ}
\end{figure}
\begin{figure}
\centering
  \includegraphics[width=5in]{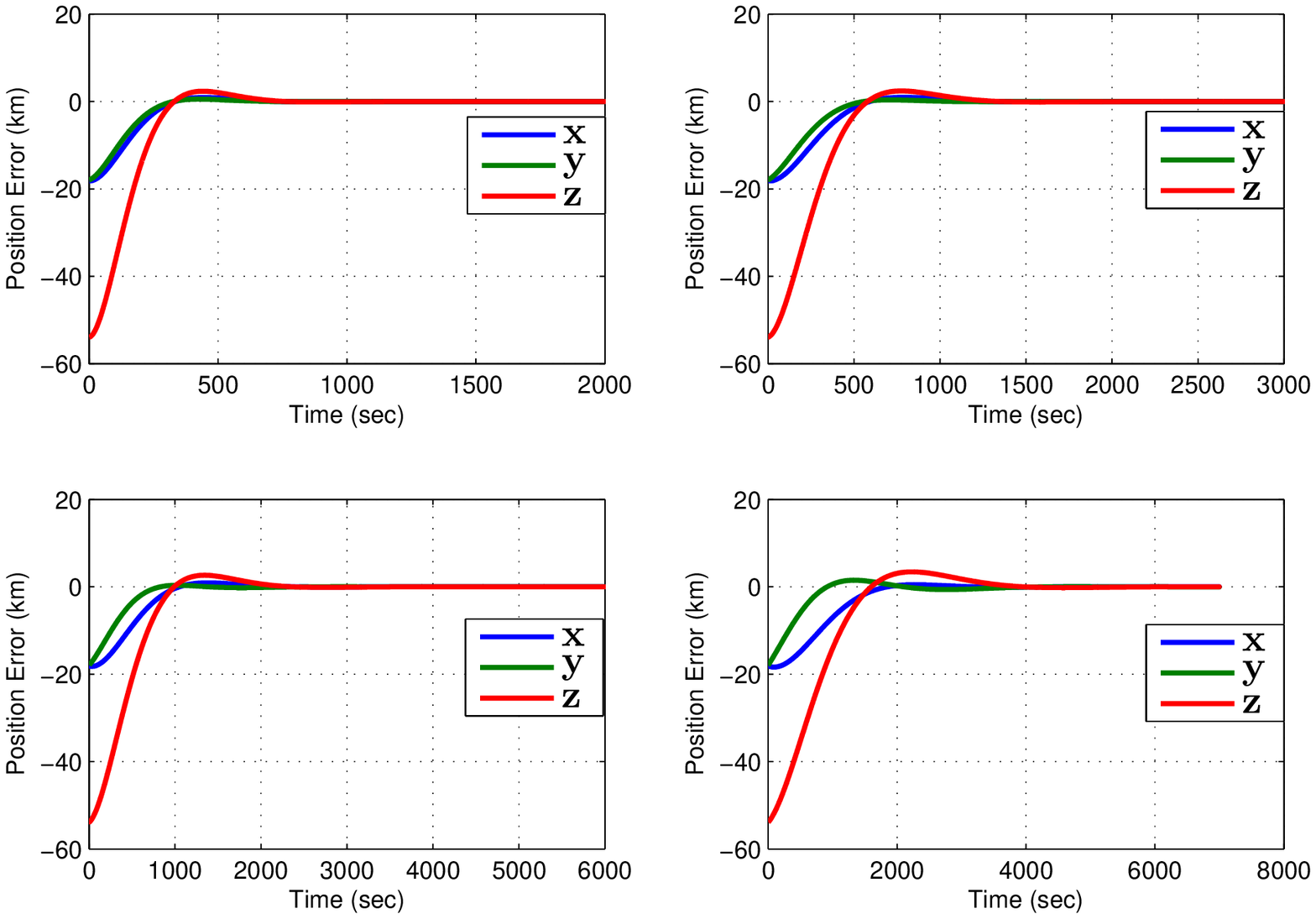}\\
  \caption{Position Error plots for $\rho_{initial} = 5km$ and $\rho_{final} = 25km$
  for $R=10^8I_{3 \times 3}$, $R=10^9I_{3 \times 3}$, $R=10^{10}I_{3 \times 3}$ and $R=10^{11}I_{3 \times 3}$ }\label{POS_ERR_R_PLOTS}
\end{figure}
The formation trajectory of deputy satellite in Hill's frame is shown in Figure \ref{SDRE_TRAJ}. The deputy satellite starts from the inner circle initial relative formation trajectory and is commanded to outer circular relative orbit. The orbit is also shown in $XY$, $XZ$ and $YZ$ planes in Figure \ref{SDRE_TRAJ_VIEWS} for better visualization of tracking of final desired trajectory in all three planes. Figure \ref{SDRE_CNTRL} gives control history for placing deputy satellite in the desired formation with respect to chief satellite. Figure \ref{R_TRAJ} is composite plots of formation trajectory for a circular chief satellite orbit with initial and commanded values being same as given in Table \ref{Initial_condition_SDRE} but for eccentricity of chief satellite is considered as zero. Figure \ref{R_TRAJ} shows the formation trajectory for different weight on control $R$. It is clear from the Figure \ref{R_TRAJ} that higher the $R$ value the control effort is less and there fore the trajectory more gradually and hence the time to terminal errors to be small and reach a steady state is also high Figure \ref{POS_ERR_R_PLOTS}. The following Table \ref{COMP_R} provides comparison of settling time and control effort for various $R$ values.
\begin{table}[h!]
\caption{\small{Control Effort and Settling time variation with $R$ values}}\label{COMP_R}
\begin{center}\begin{tabular}{|c|c|c|}
    \hline
    \textbf{Weight on} & \textbf{Settling time}&\textbf{Control effort}\\
    \textbf{Control(R)} &                     &$1e-02$\\
    \hline
    $10^8I_{3 \times 3}$ & $\approx1500 sec$& $4.19$\\
    \hline
    $10^9I_{3 \times 3}$& $\approx2000 sec$ & $4.12$\\
    \hline
    $10^{10}I_{3 \times 3}$& $\approx5000 sec$ & $3.58$\\
    \hline
    $10^{11}I_{3 \times 3}$& $\approx7500 sec$ & $3.02$\\
    \hline
\end{tabular}\end{center}
\end{table}

\begin{figure}
\centering
  \includegraphics[width=5in]{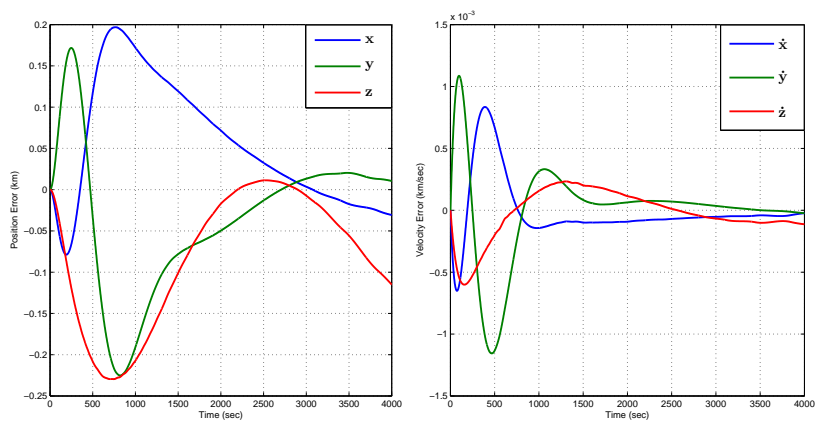}\\
  \caption{Relative Position and Velocity Error of SDC $Method:I$ and $Method:II$ for $\rho_{initial} = 5km$ and $\rho_{final} = 25km$ and $t_f = 4000 sec$ }\label{SDC_ERR}
\end{figure}

The above simulation for same initial and final condition is run for SDC formulation derived from $method:II$. The final state tracking error values for $method:I$ and $method:II$ are compared in Table \ref{STATE_ERROR_SDC} . It is clear from Table \ref{STATE_ERROR_SDC} that the state errors of $method:II$ are one order higher compared to SDC formulation by $method:I$. SDC formulation of system dynamics of SFF using $method:I$ retains the nonlinearity of the problem to maximum extent possible, where as the SDC formulation using $method:II$ is derived from the system equation of motion under assumption of circular chief satellite orbits. Hence the nonlinear behavior of the problem due to eccentric orbits are not captured well in this formulation and hence the error. This error will grow with higher eccentricity and larger semi-major axis chief satellite orbit problem. The relative error between SDC1 formulation and SDC2 formulation results is given in the Figure \ref{SDC_ERR}
\begin{table}[h!]
\caption{\small{SDRE trajectory State Errors for SDC formulation using $method:I$ and $method:II$}}\label{STATE_ERROR_SDC}
\begin{center}\begin{tabular}{|c|c|c|}
    \hline
    \textbf{State} & \textbf{SDC formulation}& \textbf{SDC formulation}\\
    \textbf{Error} & \textbf{$Method:I$} & \textbf{$Method:II$}  \\
    \hline
    $x(km)$    &     $0.0003$ &0.0085      \\
    \hline
    $\dot x (km/sec) $& $2.45 \times 10^{-6}$&$-5.3 \times 10^{-5}$ \\
    \hline
    $y (km)$& $-0.00096$&$0.002985$ \\
    \hline
    $\dot y (km/sec)$& $-1.47 \times 10^{-6}$&$2.95 \times 10^{-5}$ \\
    \hline
    $z (km)$& $-0.005931$&$-0.0011$ \\
    \hline
    $\dot z (km/sec)$& $9.312 \times 10^{-6}$&$-3.013 \times 10^{-5}$ \\
    \hline
\end{tabular}\end{center}
\end{table}
\subsection{Finite time SDRE Results}
Finite time SDRE solution is presented in the section \ref{Finite_Time_SDRE}. For the finite time SDRE simulation the weights on state is assumed to be zero that is $Q=0$ and weight on control $R$ is selected as $R = 10^9I_{3\times3}$. It is also to noted that finite time SDRE solution method achieve the final state has hard constraints. The simulation results are presented for both $SDC1$ and $SDC2$ models, and further a state error comparison is done for both SDC models. Out of two SDC formulation of nonlinear plant one with better terminal state error convergence is used for comparative method for MPSP results discussed in Chapter \ref{MPSP_Chapter}

The initial and desired relative parameters of the deputy satellite is given in Table \ref{Initial_condition_FSDRE}. The chief satellite orbital parameters considered are same as that for Infinite time SDRE solution case refer Table \ref{ORB_PAR_SDRE}, three cases are experimented that is $Case1 : e = 0$, $Case2 : e = 0.05$ and $Case3 : e = 0.15$ to illustrate the divergence of the solution of SDC2 formulation with increase in eccentricity of the chief satellite orbit. The final time $t_f$ for finite time SDRE simulation is selected from the settlings time for infinite time SDRE solution with $R = 10^9I_{3\times3}$ case from Table \ref{COMP_R} and simulation step size $\triangle t$ is selected as $1sec$

\begin{figure}
\centering
  \includegraphics[width=5in]{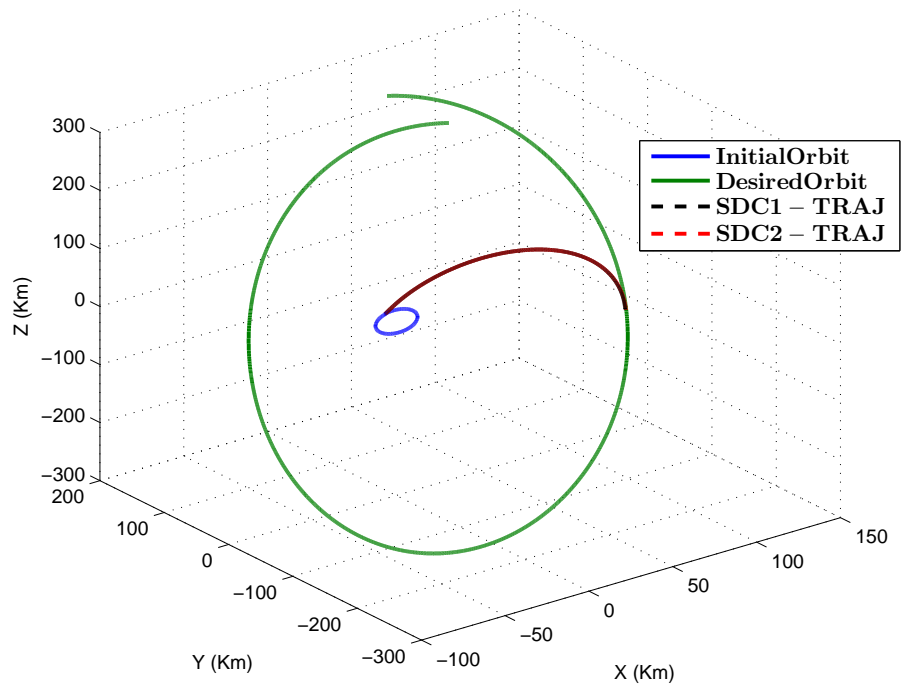}\\
  \caption{Finite time SDRE formation Trajectory for SDC $Method:I$ and $Method:II$ for $\rho_{initial} = 10km$ and $\rho_{final} = 100km$ and $e = 0$ }\label{SDRE_TRAJ_100_E0}
\end{figure}

\begin{figure}
\centering
  \includegraphics[width=5in]{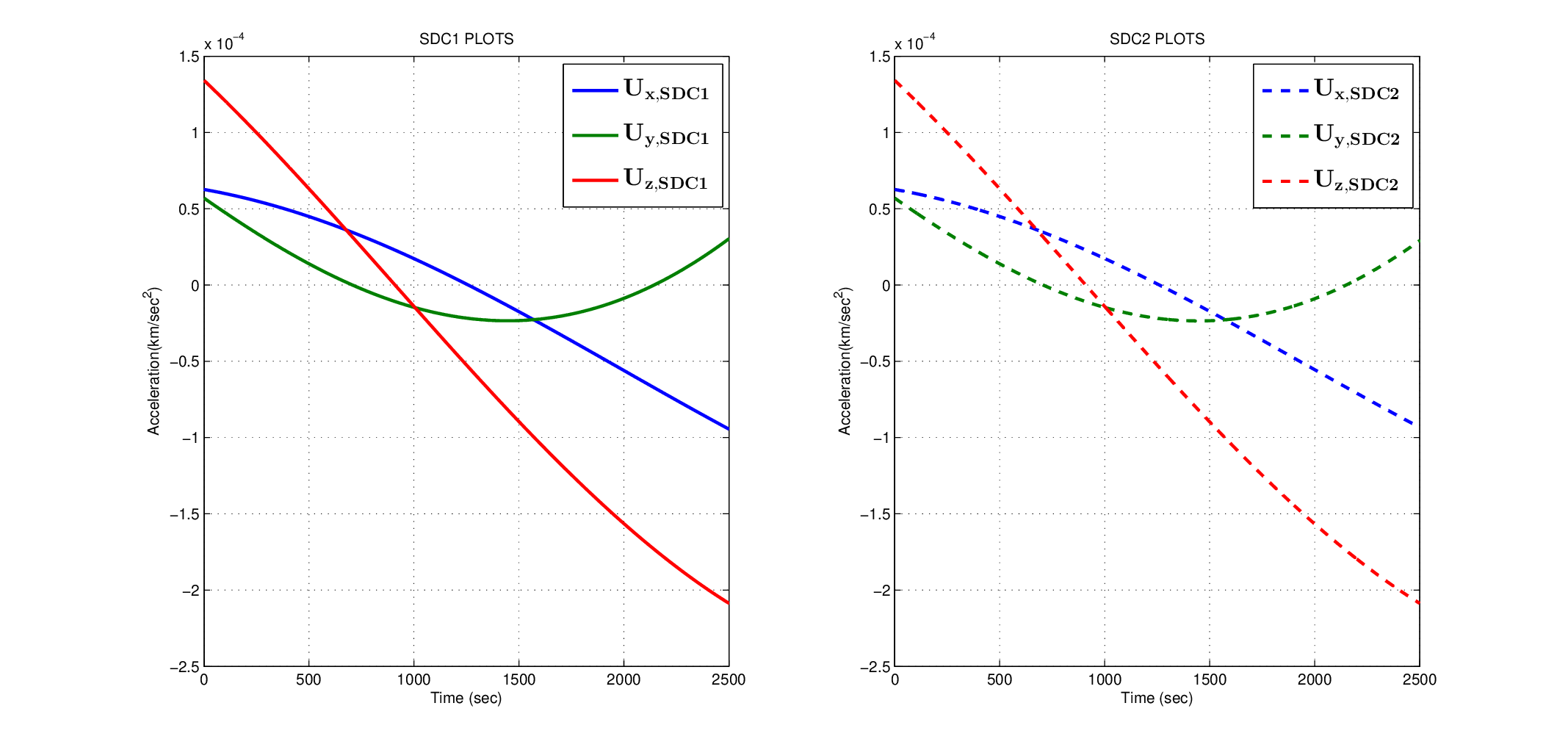}\\
  \caption{Control History for formation for SDC $Method:I$ and $Method:II$}\label{SDRE_CNTRL_100_E0}
\end{figure}

\begin{figure}
\centering
  \includegraphics[width=5in]{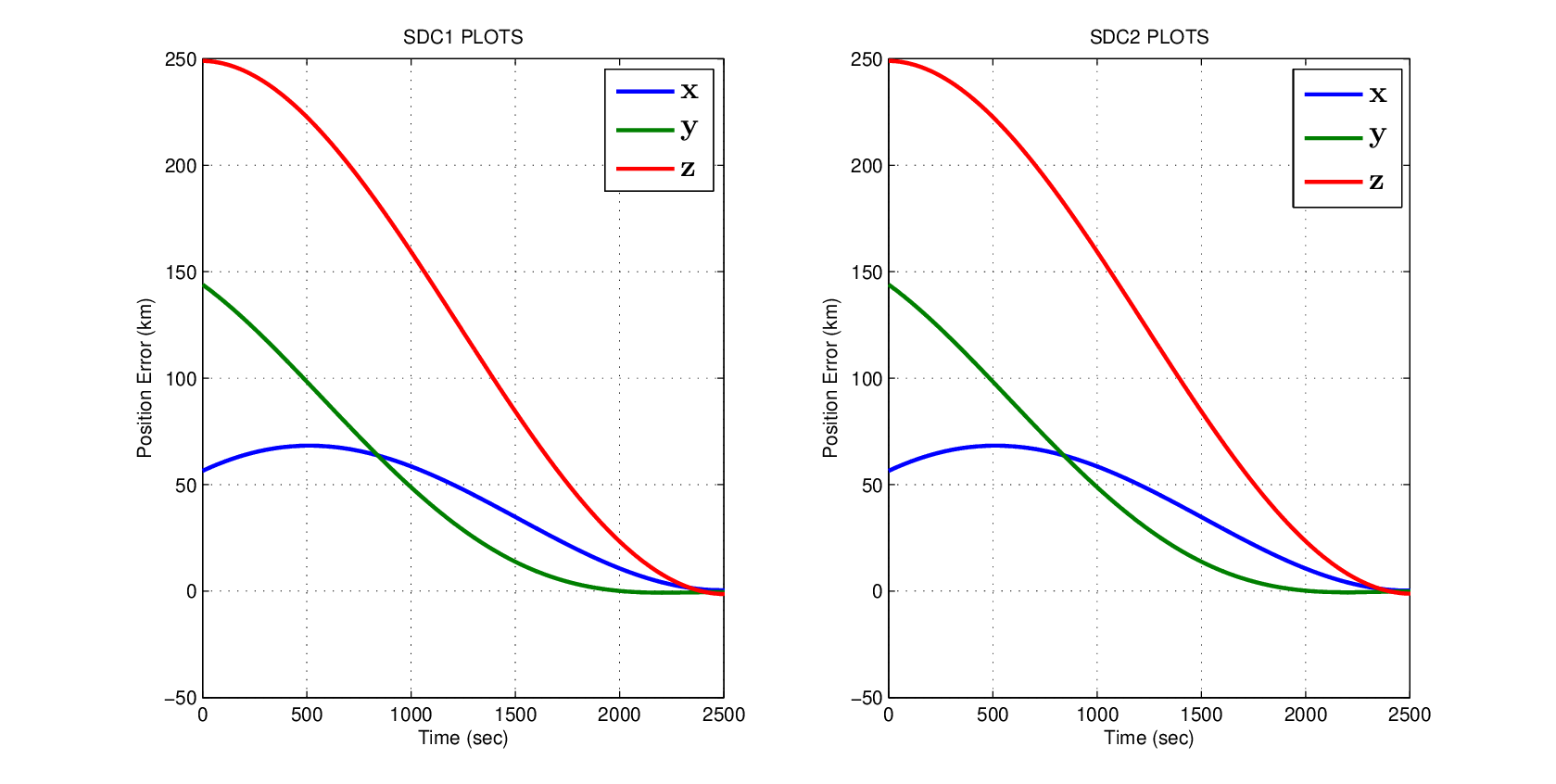}\\
  \caption{Formation Trajectory Position Error for SDC $Method:I$ and $Method:II$ for $\rho_{initial} = 10km$ and $\rho_{final} = 100km$ and $e = 0$ }\label{POS_ERR_E0}
\end{figure}

\begin{figure}
\centering
  \includegraphics[width=5in]{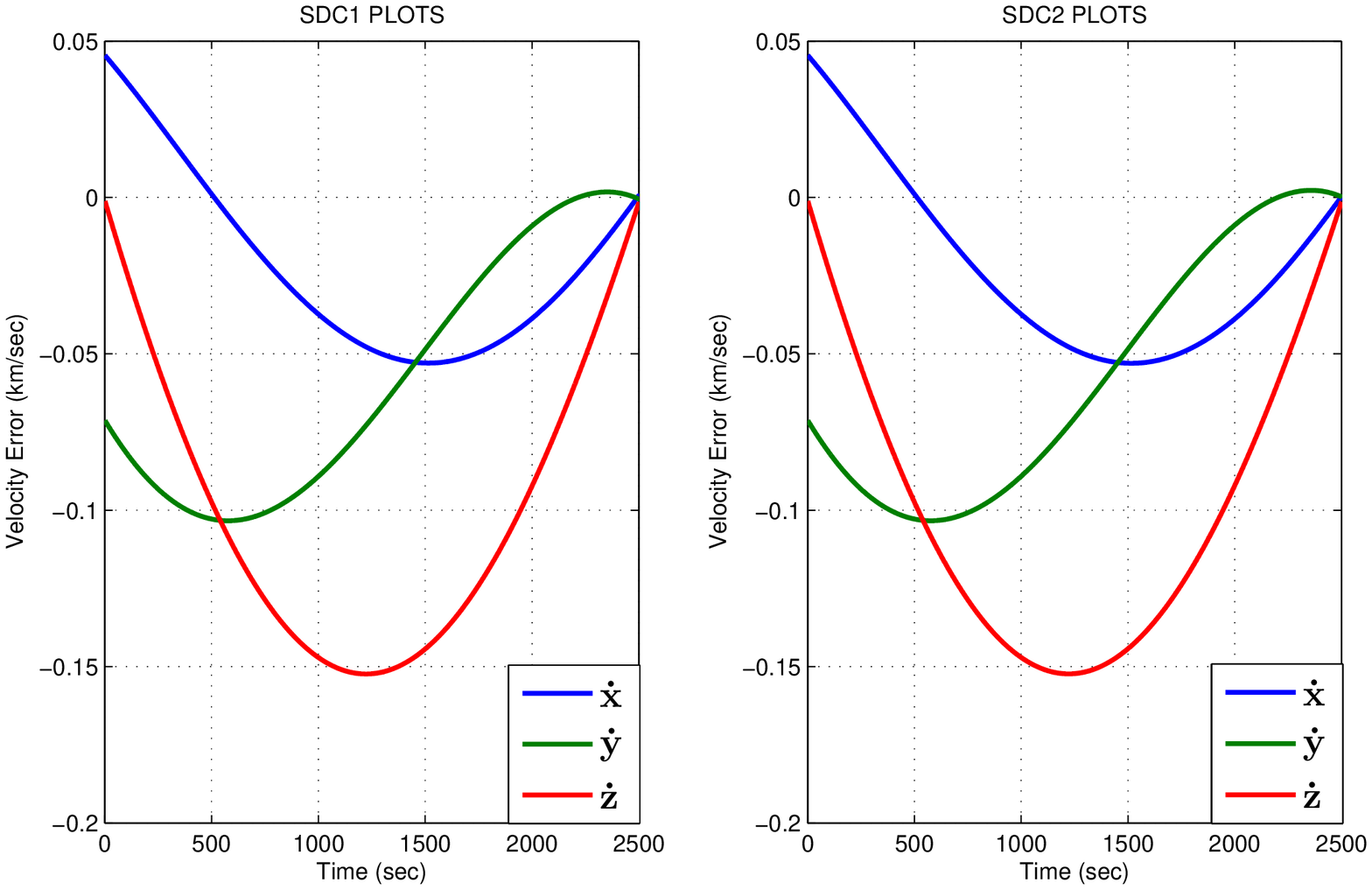}
  \caption{Formation Trajectory Velocity Error for SDC $Method:I$ and $Method:II$ for $\rho_{initial} = 10km$ and $\rho_{final} = 100km$ and $e = 0$ }\label{VEL_ERR E0}
\end{figure}

\begin{table}
\caption{\small{Deputy Satellite Initial condition for Finite time SDRE solution}}\label{Initial_condition_FSDRE}
\begin{center}\begin{tabular}{|c|c|c|}
    \hline
    \textbf{Orbital} & \textbf{Initial Value}&\textbf{Final Value}\\
    \textbf{Parameters} &                     &\\
    \hline
    $\rho (km)$ & $10 km$& $100 km$\\
    \hline
    $\theta (deg) $& $5^0$ & $35^0$\\
    \hline
    $a (km)$& $0$ & $0$\\
    \hline
    $b (km)$& $0$ & $0$\\
    \hline
    $m$ (slope)& $1$ & $1.5$\\
    \hline
    $n$(slope)& $0$ & $0$\\
    \hline
\end{tabular}\end{center}
\end{table}
Figure \ref{SDRE_TRAJ_100_E0} gives the details of the formation reconfiguration trajectory plot for case eccentricity $e=0$. The trajectory computed from both the SDC formulation $Method:I$ and $Method:II$ almost overlap each other and produce similar results for circular chief satellite orbit formation reconfiguration problem. The control history computed using SDC formulation $Method:I$ and $Method:II$ is presented in the figure \ref{SDRE_CNTRL_100_E0}. The position error and velocity error for both the methods are given in the plots \ref{POS_ERR_E0} and \ref{VEL_ERR E0} and terminal error in achieved final trajectory over commanded trajectory is given in the Table \ref{SDC_COMP_TABLE}

\begin{figure}
\centering
  \includegraphics[width=5in]{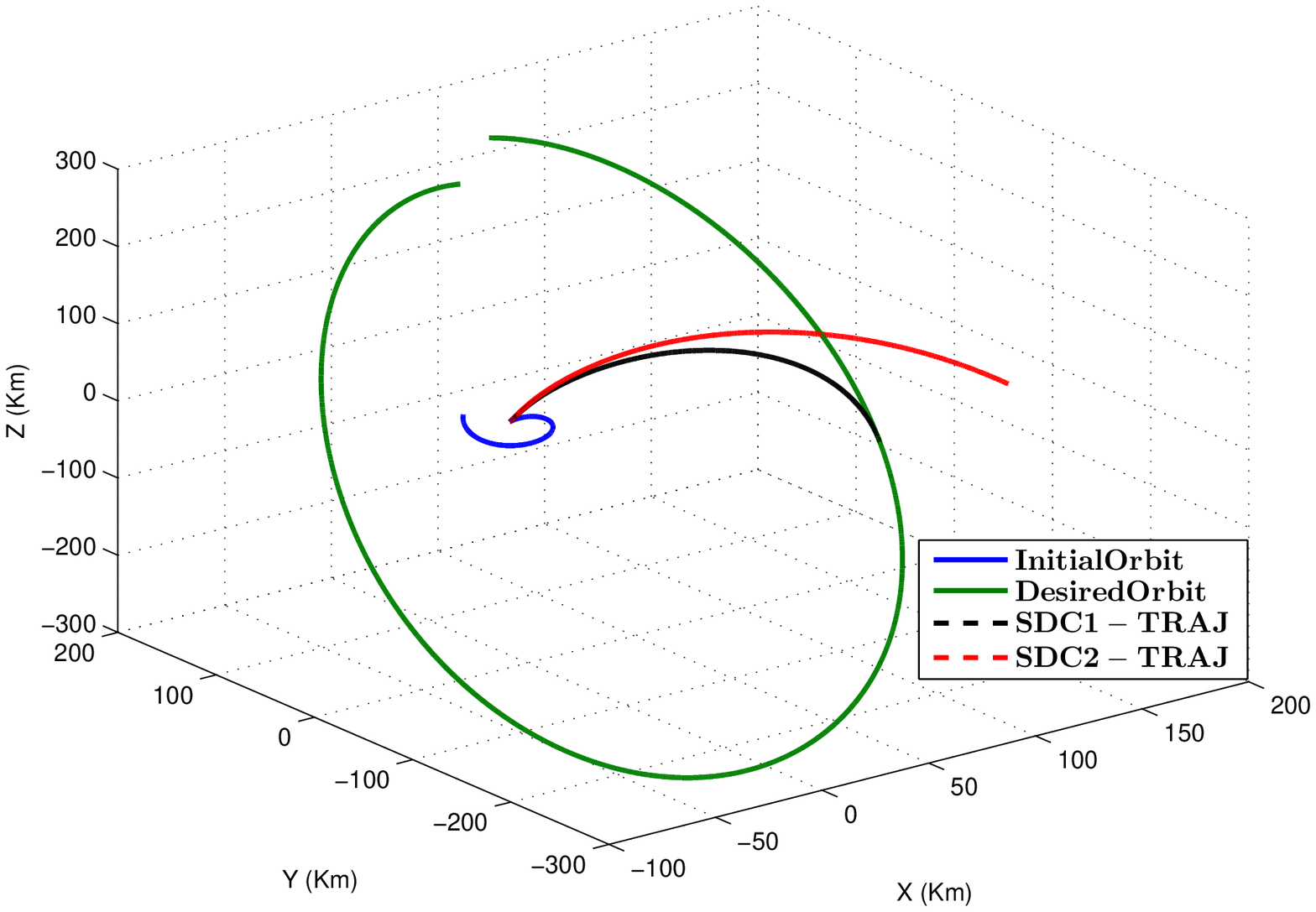}
  \caption{Finite time SDRE formation Trajectory for SDC $Method:I$ and $Method:II$ for $\rho_{initial} = 10km$ and $\rho_{final} = 100km$ and $e = 0.05$ }\label{SDRE_TRAJ_E005}
\end{figure}

\begin{figure}
\centering
  \includegraphics[width=6in]{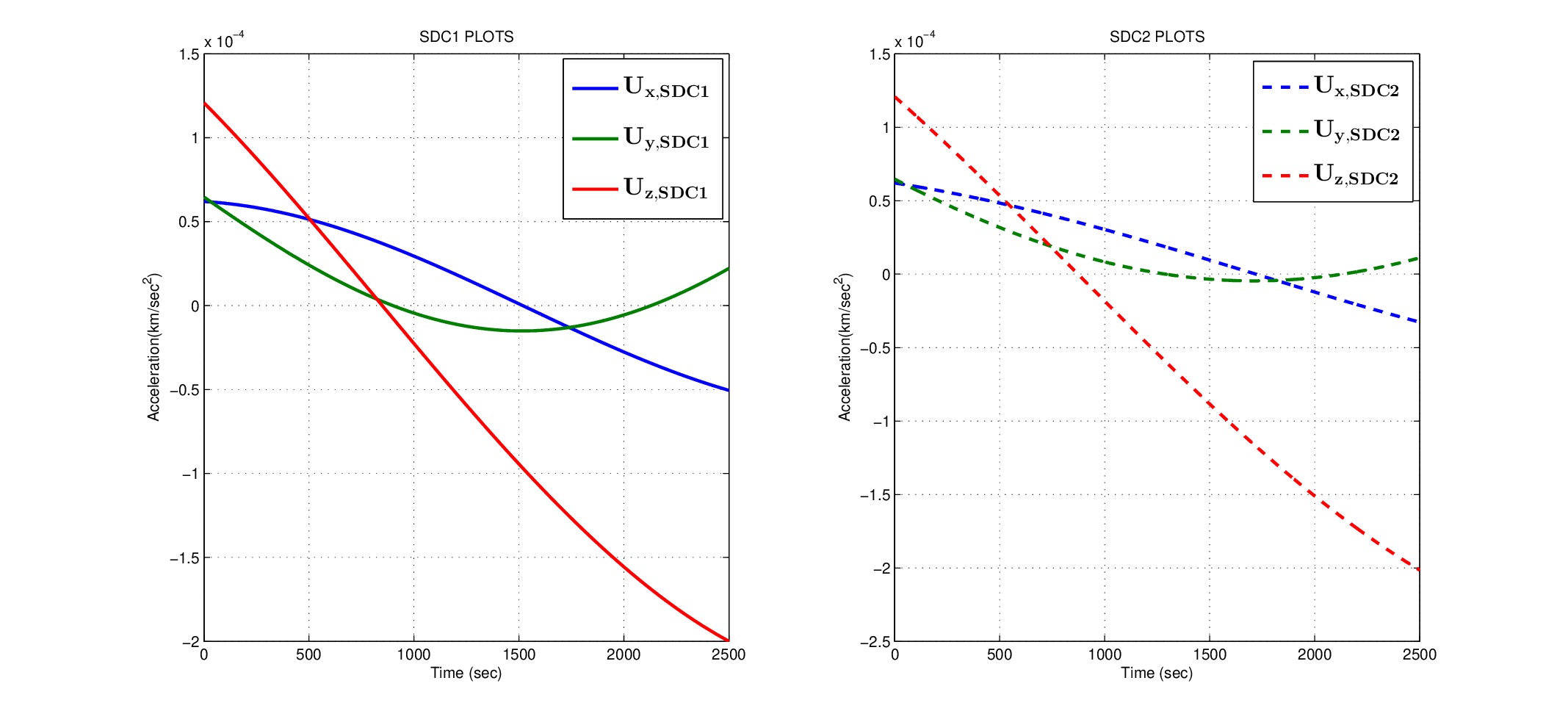}
  \caption{Control History for formation for SDC $Method:I$ and $Method:II$}\label{SDRE_CNTRL_E005}
\end{figure}

\begin{figure}
\centering
  \includegraphics[width=5in]{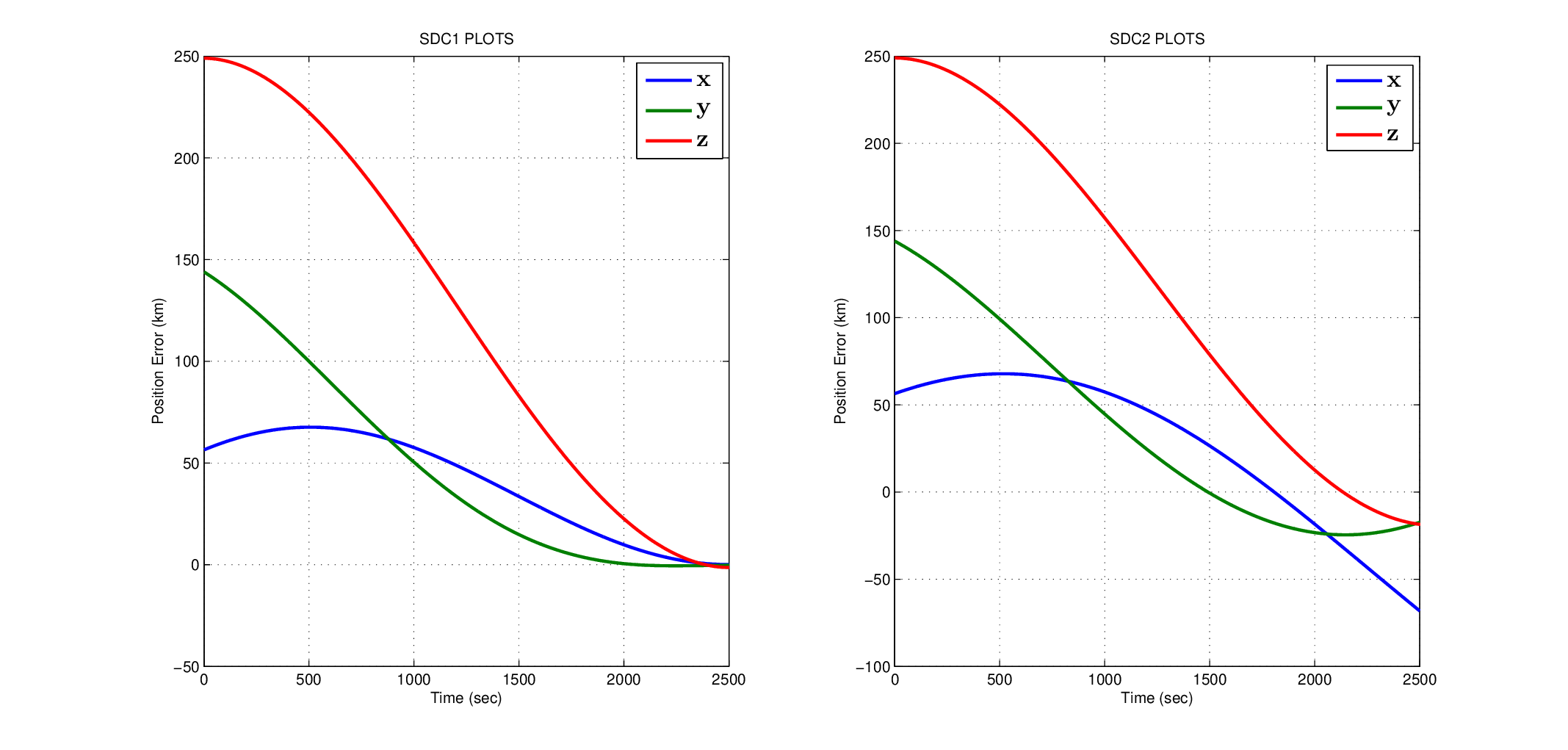}
  \caption{Formation Trajectory position error for SDC $Method:I$ and $Method:II$ for $\rho_{initial} = 10km$ and $\rho_{final} = 100km$ and $e = 0.05$ }\label{POS_ERR_E005}
\end{figure}

\begin{figure}
\centering
  \includegraphics[width=5in]{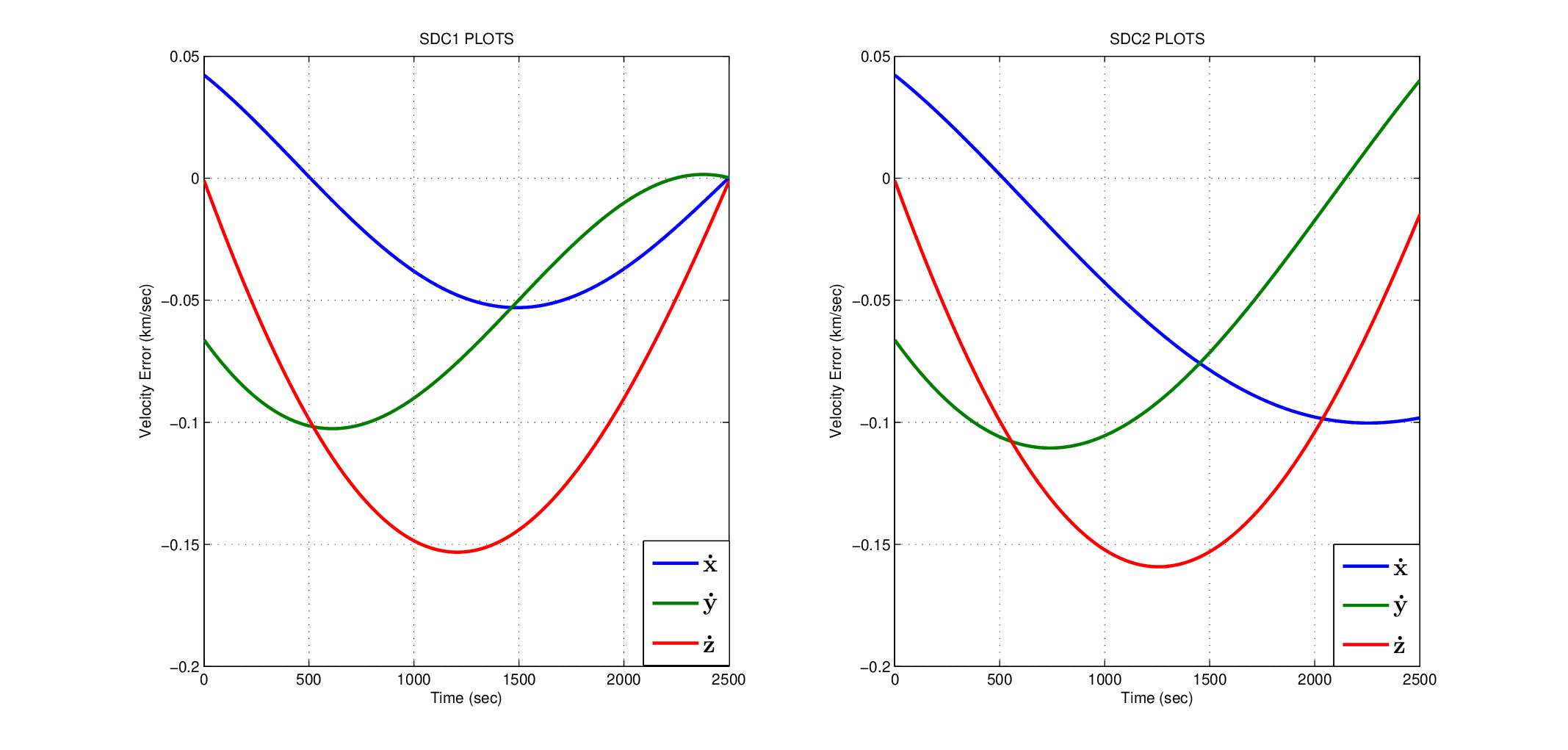}
  \caption{Formation Trajectory velocity error for SDC $Method:I$ and $Method:II$ for $\rho_{initial} = 10km$ and $\rho_{final} = 100km$ and $e = 0.05$ }\label{VEL_ERR_E005}
\end{figure}

Figure \ref{SDRE_TRAJ_E005} and \ref{SDRE_CNTRL_E005} gives plot of reconfiguration trajectory and associated control history for eccentric chief satellite case with $e=0.05$. Figure \ref{POS_ERR_E005} and \ref{VEL_ERR_E005} gives position and velocity error details with respect to the commanded trajectory values. The terminal error of reconfiguration trajectory is given in Table \ref{SDC_COMP_TABLE}


The simulation results of the case with eccentricity $e=0.15$ are similar to results of case $e=0.05$ where one can see formation trajectory computed by SDC2 method diverges much more and formation trajectory not converging to desired final orbit, where as the SDC1 method performs nominal and $\%\rho_{error} < 1\%$. The final state errors for SDC1 and SDC2 this simulation results are given in Table \ref{SDC_COMP_TABLE}.

\begin{table}[!htbp]
\centering
\caption{SDC1 and SDC2 formulation Terminal State error comparison for Case $e=0$, $e=0.05$ and $e=0.15$}\label{SDC_COMP_TABLE}
\begin{tabular}{|c|c|c|c|c|c|c|}
\hline
\textbf{State Error} &  \multicolumn{2}{c|}{\textbf e=0} & \multicolumn{2}{c|}{\textbf e=0.05}&\multicolumn{2}{c|}{\textbf e=0.15}\\
\hline
{}   & \textbf{SDC-1}   & \textbf{SDC-2}    & \textbf{SDC-1}   & \textbf{SDC-2} & \textbf{SDC-1} & \textbf{SDC-2}\\
\hline
$x(km)$  &  0.1798 & 0.3901   & 0.08  & -68.15 & -1.178 & 246.8\\
\hline
$\dot x (km/sec) $  &  0.0006& 0.00022   & 0.00019  & -0.0982 & -0.0008 & -0.0041\\
\hline
$y(km)$  &  -0.1314  &  -0.477   & -0.273  & -17.35 & 1.693 & -160\\
\hline
$\dot y (km/sec)$  &  0.00026  &  -0.00051   & 0.000205  & 0.04 & 0.005036 & -0.4\\
\hline
$z(km)$  &  -1.226  &  -1.411   & -1.389  & -18.51 & 0.80613 & -28.43\\
\hline
$\dot z(km)$  &  -.0015  &  -0.0017   & -0.0014  & -0.01502 & 0.0007 & -0.005\\
\hline
\end{tabular}
\end{table}
\section{Summary and Conclusions}
This chapter primarily introduced suboptimal SDRE control details.Inifnite time and Finite time SDRE formulation and solution procedure details are introduced in \ref{Infinite_Time_SDRE} and \ref{Finite_Time_SDRE}. SDRE control requires the nonlinear equation of motion of satellite formation flying problem to be written in the state dependent co-efficient form. In section details of SDC modeling methods are introduced, two SDC form are discussed. The results section displayed the versatility of the SDRE method to handle satellite formation problem involving high eccentricity and larger base-length formation separation problems.Infinite time and finite time solutions were discussed. A comparative study was made between SDC-1($Method:I)$ and SDC-2 ($Method:II)$ formulation solutions in both infinite time and finite time solution domain and it was found that SDC1 solutions converged to desired trajectory within the prescribed tolerance band of $<1\%$ error, where as SDC2 formulation failed to achieve the similar accuracy and solution diverged as eccentricity of chief satellite orbit was increased. It is inferred from the above exercise that SDC-1 formulation retains the nonlinearity of the problem to maximum extent possible and hence finite time solution of SDC-1 system model of SFF is chosen as comparative method for MPSP and G-MPSP in further section.

\cleardoublepage
\chapter{Model Predictive Static Programming}
\label{MPSP_Chapter}
This technique has been inspired from the philosophies of Model Predictive Control (MPC) \cite{rossiter} and Approximate Dynamic
Programming (ADP) \cite{Werbos}. MPSP technique caters for control synthesis for class of finite time horizon optimal control problem.Here in this chapter the mathematical details of MPSP method is presented.
Model predictive static programming method considers the general nonlinear system in discrete form. The discrete representation of state and output equations are given as follows,
\begin{eqnarray}
X_{k+1}& = & F_k(X_k, U_k) \label{discrete_sys_state}\\
Y_k & = & h(X_k)\label{discrete_sys_state_out}
\end{eqnarray}
where $X \in \Re^n, \ U \in \Re^m, \ Y \in \Re^p$ and $k = 1, 2,\dots, N$ are the time steps. The primary objective is to come up
with a suitable control history $U_k, \ k = 1, 2, \dots, N-1$(starting with a suitable guess), so that the output at the final
time step $Y_N$ goes to a desired value $Y_N^\ast$, i.e. $Y_N \rightarrow Y_N^\ast$. In addition, the aim is to achieve this
task with minimum control effort.

The MPSP method needs the initial guess control history. The guess control history can be any control values $U_0$ for grid points from $1,2 \ldots N$. This guess control is not expected to satisfy the objective of achieving zero terminal error. MPSP technique gives the technique to improve upon this initial guess control history by computing the control variable error history which needs to be subtracted from the previous control to compute the new improvised control history. The final objective is evaluated with the new control values applied at each grid point. If the convergence of the output vector $\left(Y_N\right)$ at final grid point is not sufficiently close enough to the desired value $\left(Y_{N}^*\right)$, then further iteration are carried out to refine the control history until the objective $\left( Y_N \to Y_{N}^* \right)$  is met. The control history update technique presented in MPSP frame work is a close form expression, and hence the evaluation of the same requires lesser computational requirements and hence is apt candidate for online implementation.

To proceed with the mathematical details, first the error in the output is defined as $ \triangle Y_N = Y_N - Y_N^*$. Next, using Taylor series expansion, $Y_N$ is expanded about $Y_N^\ast$ as follows
\begin{eqnarray} \label{taylor_expansion}
Y_N   = Y_N^* + \left[\frac{\partial Y_N}{\partial X_N}\right]\
dX_N + HOT
\end{eqnarray}
where $HOT$ contains the `higher order terms'. From (\ref{taylor_expansion}), neglecting HOT the error in the output
can be written as
 \begin{eqnarray}
 \label{error_out}
\triangle Y_N  \cong dY_N = \left[\frac{\partial Y_N}{\partial
X_N}\right]\ dX_N
\end{eqnarray}
However from (\ref{discrete_sys_state}), one can write the error in state at time step $(k+1)$ as
\begin{eqnarray}\label{dx_k+1}
dX_{k+1} = \left[\frac{\partial F_k}{\partial X_k}\right] \ dX_k +
\left[\frac{\partial F_k}{\partial U_k}\right] \ dU_k
\end{eqnarray}
where $dX_k$ and $dU_k$ are the error of state and control at time step $k$ respectively. Expanding $dX_N$ as in (\ref{dx_k+1}) for
$k=N-1$, and similarly for $dX_{N-1}$ for $k=N-2$ and so on, one can carry out the necessary algebra and continue until $k=1$.
Finally taking the help of (\ref{error_out}) one can write
\begin{eqnarray}\label{error_zero}
dY_N &=& A \ dX_1 + B_1  dU_1 + \cdots + B_{N-1} dU_{N-1}
\end{eqnarray}
where,
\begin{equation}
A \buildrel \Delta \over = \left[ {\frac{{\partial {Y_N}}}{{\partial {X_N}}}} \right]\left[ {\frac{{\partial {F_{N - 1}}}}{{\partial {X_{N - 1}}}}} \right] \cdots \left[ {\frac{{\partial {F_1}}}{{\partial {X_1}}}} \right]\nonumber \\
\end{equation}
\begin{equation}
\label{b_coeff}
 {B_k} \buildrel \Delta \over = \left[
{\frac{{\partial {Y_N}}}{{\partial {X_N}}}} \right]\left[
{\frac{{\partial {F_{N - 1}}}}{{\partial {X_{N - 1}}}}} \right]
\cdots \left[ {\frac{{\partial {F_{k + 1}}}}{{\partial {X_{k +
1}}}}} \right]\left[ {\frac{{\partial {F_k}}}{{\partial {U_k}}}}
\right]
\end{equation}
Since the initial condition is specified, there is no error in the first term. This means $dX_1 = 0$ and hence \ref{error_zero}
reduces to
\begin{eqnarray}\label{constra}
dY_N  = \sum_{k =1}^{N-1} B_{k}dU_{k}
\end{eqnarray}
\quad At this point, it can be pointed out that if one evaluates each of the $B_{k}, \ k = 1,\ldots,(N-1)$ as in (\ref{b_coeff}),
it will be a computationally intensive tasks (especially  when $N$ is high). However, it is possible to compute them
recursively \cite{Padhi:2008Paper}, \cite{Harshal} for details. Next, the idea is to minimize the following objective (cost) function
\begin{eqnarray} \label{cost_MPSP}
J = \frac{1}{2} \sum_{k =1}^{N-1}  (U^0_{k} - dU_{k})^T
R_{k}(U^0_{k} - dU_{k})
\end{eqnarray}
where $\ U_{k}^0$, $k = 1,  \dots ,(N-1)$ is the previous control history solution and $dU_{k}$ is the corresponding error in the
control history. The cost function in (\ref{cost_MPSP}) needs to be minimized subjected to the constraint in (\ref{constra}), where
$R_{k} > 0$ (a positive definite matrix) is the weighting matrix, which needs to be chosen judiciously by the control designer.
Equations (\ref{constra}) and  (\ref{cost_MPSP}) formulate an appropriate constrained static optimization problem. Hence, using
optimization theory \cite{bryson_book}, and carrying out the necessary algebra \cite{Padhi:2008Paper}\cite{Harshal}, the updated control at
time step $ k = 1,2,\ldots,(N-1)$ is given by
\begin{eqnarray}
\label{MPSP_CNTRL}
U_{k}\ =  \ U_{k}^0 - dU_{k} \ = \ \ R^{-1}_{k}B^T_{k}
A^{-1}_\lambda \left(dY_N - b_\lambda \right) \label{sol_u}
\end{eqnarray}
where,
\begin{eqnarray}
\begin{array}{*{20}{c}}
{A_\lambda } \buildrel \Delta \over = \left[ { - \sum\limits_{k =
1}^{N - 1} {{B_k}R_k^{ - 1}B_k^T} } \right],\quad {b_\lambda }
\buildrel \Delta \over = \left[ {\sum\limits_{k = 1}^{N - 1}
{{B_k}U_k^0} } \right]\nonumber
\end{array}
\end{eqnarray}
\quad In addition to the recursive computation of sensitivity matrices, it is clear that the updated control history solution in
(\ref{sol_u}) is a {\it closed form solution}, and hence, control solution can be updated with very minimal computational
requirement. We also mention that the relative magnitude of the control input at various time steps can be adjusted by properly
adjusting the weight matrixes $R_k$, $k = 1, \dots ,(N-1)$
associated with the cost function. For further details on MPSP one
can refer \cite{Padhi:2008Paper,Harshal}.

\section{Problem Formulation in MPSP Framework}
\label{MPSP_FORMULATION}
MPSP formulation needs the nonlinear equation of motion \ref{EQN_SFF_FUNC_FORM} to be written in discrete form. Euler discretization method is used for writing the nonlinear equation of motion in discrete form. Euler method is a first-order numerical procedure for solving ordinary differential equations (ODEs) with a given initial value. It is the most basic explicit method for numerical integration of ordinary differential. The discrete form the equation \ref{EQN_SFF_FUNC_FORM} can be obtained as follows,\\
\begin{eqnarray}
\label{Euler_EQN1}
\mathbf{\dot X} &=& f(\mathbf X,\mathbf U)\\
\label{Euler_EQN2}
\frac{{\mathbf{X_{k + 1}} - \mathbf{X_k}}}{{\Delta t}} &=& \mathbf f(\mathbf X{}_k,{\mathbf U_k})\\
\label{Euler_EQN3}
\mathbf{X_{k + 1}} &=& \mathbf{X_k} + \Delta t\left( {\mathbf f(\mathbf X{}_k,{\mathbf U_k})} \right)\\
\mathbf{X_{k + 1}} &=& {\mathbf F(\mathbf X{}_k,{\mathbf U_k})}
\end{eqnarray}
Where $\mathbf F(\mathbf X{}_k,{\mathbf U_k}) = \mathbf{X_k} + \Delta t\left( {\mathbf f(\mathbf X{}_k,{\mathbf U_k})} \right)$\\
Using the above technique the discretized form of equation of motion of relative dynamics of deputy satellite with respect to chief satellite can be written as follows,
\begin{eqnarray}
{x_{1k + 1}} &=& {x_{1k}} + \Delta t({x_{2k}})\\
{x_{2k + 1}} &=& {x_{2k}} + \Delta t\left( {2\dot \nu {x_{4k}} + \ddot \nu {x_{3k}} + {{\dot \nu }^2}{x_{1k}} - \frac{\mu }{{{\gamma _k}}}{x_{1k}} - \frac{\mu }{{{\gamma _k}}}{r_c} + \frac{\mu }{{r_c^2}} + {U_1} + a_{J2,k}} \right)\\
{x_{3k + 1}} &=& {x_{3k}} + \Delta t({x_{4k}})\\
{x_{4k + 1}} &=& {x_{4k}} + \Delta t\left( { - 2\dot \nu {x_{2k}} - \ddot \nu {x_{1k}} + {{\dot \nu }^2}{x_{3k}} - \frac{\mu }{{{\gamma _k}}}{x_{3k}} + {U_2} + a_{J2,k}} \right)\\
{x_{5k + 1}} &=& {x_{5k}} + \Delta t({x_{6k}})\\
{x_{6k + 1}} &=& {x_{6k}} + \Delta t\left( { - \frac{\mu }{{{\gamma _k}}}{x_{5k}} + {U_3} + a_{J2,k}} \right)
\end{eqnarray}
and the discrete form of system output is written as,
\begin{equation}
\mathbf Y_N = \mathbf X_N
\end{equation}
where $k = 1,2,3 \ldots N $ are time steps.
How ever Euler integration method is used for discretization of the system dynamics, a more accurate and reliable numerical integration technique Forth order Runge-Kutta method is used to simulate the system dynamics further in time using the control values computed from MPSP method.

The objective of the problem statement is to form the formation or to reconfigure the formation flying of satellites to the desired orbit. The Deputy satellite is initially in a orbit around the earth with initial formation separation
of $\rho_{initial}$. It is desired to place the deputy satellite in new formation with spatial
separation of $\rho_{final}$. The objective of the problem is to minimize the control effort required to reach the new orbit, and at the same time, deputy satellite should execute the reformation with minimum terminal state error. Mathematically we can put the problem objectives as follows. The main problem objective is to minimize the terminal position errors i.e. ${\left[ {\begin{array}{*{20}{c}}
{{x_1}}&{{x_3}}&{{x_5}}
\end{array}} \right]^T} \to {\left[ {\begin{array}{*{20}{c}}
{x_1^*}&{x_3^*}&{x_5^*}
\end{array}} \right]^T}$ at $t = {t_f}$ . However, since the velocity components should also match with the desired orbital parameters, one can also impose
${\left[ {\begin{array}{*{20}{c}} {{x_2}}&{{x_4}}&{{x_6}}
\end{array}} \right]^T} \to {\left[ {\begin{array}{*{20}{c}}
{x_2^*}&{x_4^*}&{x_6^*}
\end{array}} \right]^T}$ at $t = {t_f}$.
The error in the output ``$d\mathbf{Y_N}$`` is evaluated as follows
\begin{equation}
\label{eqn:23} d\mathbf{Y_N} = \mathbf{Y_N} - \mathbf{Y_N^*}
\end{equation}
where $\mathbf{Y_N^*}$ is the desired state vector.\\
\indent Aim is to compute the control command $U_k$, where $k = 1,
\ldots,(N - 1)$ so that $d\mathbf{Y_N} \to 0$. To achieve this objective, the coefficients $B_1$ to $B_{N-1}$ are evaluated using
\ref{b_coeff}. Finally the control command is updated using \ref{sol_u}.
The partial derivative of $F(\mathbf X_k,\mathbf U_k)$ and $\mathbf Y_N$ required to compute the sensitive matrices $B_k$ are
\begin{eqnarray}
\frac{{\partial \mathbf F\left( {\mathbf {X_{k,}}\mathbf {U_k}} \right)}}{{\partial {\mathbf X_k}}} &=& {I_{6 \times 6}} + \Delta t\left[ {\frac{{\partial \mathbf {f_k}}}{{\partial {\mathbf X_k}}}} \right]\\
\frac{{\partial \mathbf F\left( {\mathbf {X_{k,}}\mathbf {U_k}} \right)}}{{\partial
\mathbf {U_k}}} &=& \triangle t\left[ {\begin{array}{*{20}{c}}
0&0&0\\
1&0&0\\
0&0&0\\
0&1&0\\
0&0&0\\
0&0&1
\end{array}} \right]\\
\frac{{\partial {\mathbf Y_N}}}{{\partial {\mathbf X_N}}} &=& {I_{6 \times 6}}
\end{eqnarray}
The component of the partial derivative term $\left(\frac{{\partial {F_k}}}{{\partial {\mathbf X_k}}}\right)$ are given as follows,
\begin{eqnarray}
\frac{{\partial {f_1}}}{{\partial {x_{2k}}}} &=& 1\\
\frac{{\partial {f_2}}}{{\partial {x_{1k}}}} &=& {{\dot \nu_k }^2}- \mu \left[ {\frac{{\gamma  - 3{x_{1k}}\gamma
_k^{\frac{1}{2}}\left( {{r_{ck}} + {x_{1k}}} \right)}}{{\gamma_k^2}}} \right] + 3\mu {r_{ck}}\gamma _k^{ - \frac{5}{2}}\left( {{r_{ck}} + {x_{1k}}} \right)\\
\frac{{\partial {f_2}}}{{\partial {x_{3k}}}} &=& \ddot \nu_k  + 3\mu \gamma _k^{ - \frac{5}{2}}{x_{3k}}\left( {{r_{ck}} + {x_{1k}}} \right)\\
\frac{{\partial {f_2}}}{{\partial {x_{4k}}}} &=& 2\dot \nu_k \\
\frac{{\partial {f_2}}}{{\partial {x_{5k}}}} &=& 3\mu \left( {{r_{ck}} + {x_{1k}}} \right)\gamma _k^{ - \frac{5}{2}}{x_{5k}}\\
\frac{{\partial {f_3}}}{{\partial {x_{4k}}}} &=& 1\\
\frac{{\partial {f_4}}}{{\partial {x_{1k}}}} &=&  - \ddot \nu_k  - 3\mu {x_{3k}}\gamma _k^{ - \frac{5}{2}}\left( {{r_{ck}} + {x_{1k}}} \right)\\
\frac{{\partial {f_4}}}{{\partial {x_{2k}}}} &=&  - 2\dot \nu_k \\
\frac{{\partial {f_4}}}{{\partial {x_{3k}}}} &=& {{\dot \nu_k }^2} + \mu \left[ {\frac{{\gamma  - 3x_{3k}^2\gamma _k^{\frac{1}{2}}}}{{\gamma _k^2}}} \right]\\
\frac{{\partial {f_4}}}{{\partial {x_{5k}}}} &=&  - 3\mu {x_{3k}}{x_{5k}}\gamma _k^{ - \frac{5}{2}}\\
\frac{{\partial {f_5}}}{{\partial {x_{6k}}}} &=& 1\\
\frac{{\partial {f_6}}}{{\partial {x_{1k}}}} &=& 3\mu {x_{5k}}\left( {{r_{ck}} + {x_{1k}}} \right)\gamma _k^{ - \frac{5}{2}}\\
\frac{{\partial {f_6}}}{{\partial {x_{3k}}}} &=& 3\mu {x_{3k}}{x_{5k}}\gamma _k^{ - \frac{5}{2}}\\
\frac{{\partial {f_6}}}{{\partial {x_{5k}}}} &=&  - \mu \left[
{\frac{{{\gamma _k} - 3{x_{5k}}\gamma _k^{\frac{1}{2}}}}{{\gamma
_k^2}}} \right]
\end{eqnarray}
The component of the partial derivative of $J_2$ perturbation term, $\left(\frac{{\partial {J_{2,X_k}}}}{{\partial {\mathbf X_k}}}\right)$ are given as follows,
\begin{eqnarray}
\frac{{\partial {J_{2x}}}}{{\partial {x_1}}} &=&  - \frac{3}{2}\mu {J_2}R_e^2\left[ {\frac{{4{x_1}}}{{{{\left( {{r_c} + \rho } \right)}^5}\sqrt \rho  }} - \frac{{4{x_1}}}{{{{\left( {{r_c} + \rho } \right)}^5}\sqrt \rho  }}\left( {3{{\sin }^2}\left( {i + \delta i} \right){{\sin }^2}\left( {\theta  + \delta \theta } \right)} \right)} \right. \nonumber \\
 &+& \frac{3}{{{{\left( {{r_c} + \rho } \right)}^4}}}\left\{ {2\sin \left( {i + \delta i} \right)\cos \left( {i  + \delta i } \right)\Sigma _{31}^{ - 1}{{\sin }^2}\left( {\theta  + \delta \theta } \right) + 2{{\sin }^2}\left( {i + \delta i} \right)} \right.\nonumber \\
&&\left. {\left. {\sin \left( {\theta  + \delta \theta } \right)\cos \left( {\theta  + \delta \theta } \right)\Sigma _{21}^{ - 1}} \right\}} \right]\\
\frac{{\partial {J_{2x}}}}{{\partial {x_2}}} &=&  - \frac{3}{2}\mu {J_2}R_e^2\left[ {\frac{3}{{{{\left( {{r_c} + \rho } \right)}^4}}}\left\{ {2\sin \left( {i + \delta i} \right)\cos \left( {i + \delta i} \right)\Sigma _{32}^{ - 1}{{\sin }^2}\left( {\theta  + \delta \theta } \right)} \right.} \right.\nonumber \\
&&\left. {\left. { + 2{{\sin }^2}\left( {i + \delta i} \right)\sin \left( {\theta  + \delta \theta } \right)\cos \left( {\theta  + \delta \theta } \right)\Sigma _{22}^{ - 1}} \right\}} \right]\\
\frac{{\partial {J_{2x}}}}{{\partial {x_3}}} &=&  - \frac{3}{2}\mu {J_2}R_e^2\left[ {\frac{{4{x_3}}}{{{{\left( {{r_c} + \rho } \right)}^5}\sqrt \rho  }} - \frac{{4{x_3}}}{{{{\left( {{r_c} + \rho } \right)}^5}\sqrt \rho  }}\left( {3{{\sin }^2}\left( {i + \delta i} \right){{\sin }^2}\left( {\theta  + \delta \theta } \right)} \right)} \right. \nonumber  \\
 &+& \frac{3}{{{{\left( {{r_c} + \rho } \right)}^4}}}\left\{ {2\sin \left( {i + \delta i} \right)\cos \left( {i  + \delta i } \right)\Sigma _{33}^{ - 1}{{\sin }^2}\left( {\theta  + \delta \theta } \right) + 2{{\sin }^2}\left( {i + \delta i} \right)} \right.\nonumber \\
&&\left. {\left. {\sin \left( {\theta  + \delta \theta } \right)\cos \left( {\theta  + \delta \theta } \right)\Sigma _{23}^{ - 1}} \right\}} \right]\\
\frac{{\partial {J_{2x}}}}{{\partial {x_4}}} &=&  - \frac{3}{2}\mu {J_2}R_e^2\left[ {\frac{3}{{{{\left( {{r_c} + \rho } \right)}^4}}}\left\{ {2\sin \left( {i + \delta i} \right)\cos \left( {i + \delta i} \right)\Sigma _{34}^{ - 1}{{\sin }^2}\left( {\theta  + \delta \theta } \right)} \right.} \right.\nonumber \\
&&\left. {\left. { + 2{{\sin }^2}\left( {i + \delta i} \right)\sin \left( {\theta  + \delta \theta } \right)\cos \left( {\theta  + \delta \theta } \right)\Sigma _{24}^{ - 1}} \right\}} \right]\\
\frac{{\partial {J_{2x}}}}{{\partial {x_5}}} &=&  - \frac{3}{2}\mu {J_2}R_e^2\left[ {\frac{{4{x_5}}}{{{{\left( {{r_c} + \rho } \right)}^5}\sqrt \rho  }} - \frac{{4{x_5}}}{{{{\left( {{r_c} + \rho } \right)}^5}\sqrt \rho  }}\left( {3{{\sin }^2}\left( {i + \delta i} \right){{\sin }^2}\left( {\theta  + \delta \theta } \right)} \right)} \right.\nonumber \\
 &+& \frac{3}{{{{\left( {{r_c} + \rho } \right)}^4}}}\left\{ {2\sin \left( {i + \delta i} \right)\cos \left( {i  + \delta i } \right)\Sigma _{35}^{ - 1}{{\sin }^2}\left( {\theta  + \delta \theta } \right) + 2{{\sin }^2}\left( {i + \delta i} \right)} \right. \nonumber \\
&&\left. {\left. {\sin \left( {\theta  + \delta \theta } \right)\cos \left( {\theta  + \delta \theta } \right)\Sigma _{25}^{ - 1}} \right\}} \right]\\
\frac{{\partial {J_{2x}}}}{{\partial {x_6}}} &=&  - \frac{3}{2}\mu {J_2}R_e^2\left[ {\frac{3}{{{{\left( {{r_c} + \rho } \right)}^4}}}\left\{ {2\sin \left( {i + \delta i} \right)\cos \left( {i + \delta i} \right)\Sigma _{34}^{ - 1}{{\sin }^2}\left( {\theta  + \delta \theta } \right)} \right.} \right.\nonumber \\
&&\left. {\left. { + 2{{\sin }^2}\left( {i + \delta i} \right)\sin \left( {\theta  + \delta \theta } \right)\cos \left( {\theta  + \delta \theta } \right)\Sigma _{24}^{ - 1}} \right\}} \right]
\end{eqnarray}
Similarly the partial derivatives of $J_2$ perturbation component in $y$ and $z$ direction can be evaluated.
\subsection{Guess Control(LQR)}
The guess controller for MPSP SFF problem is obtained through LQR
solution approach. The infinite time horizon problem is considered
with linearized model.
\begin{equation}
\label{eqn:24} \mathbf{\dot X} = A \mathbf X + B \mathbf U
\end{equation}
System matrices $A$ and $B$ are defined in chapter \ref{LQR_chap}, cost function considered for LQR solution is as follows
\begin{equation}
J = \frac{1}{2}\int\limits_0^\infty  {\left( {\mathbf X_k^T{Q_{l}}{\mathbf X_k} +
\mathbf U_k^T{R_{l}}{\mathbf U_k}} \right)} dt
\end{equation}

For LQR guess solution the weight on state $Q_{l}=I_{6\times6}$ and $R_l=10^{9}\*I_{3\times3}$ are chosen.
Note that weights on control i.e. $R_k$ used to compute the control using equation \ref{MPSP_CNTRL} in MPSP
frame work, is selected as  $R_k=\triangle tR_l$, same as that used for guess solution method (LQR) multiplied by
 the time step. The state values at final time step $t_f$ from the desired commanded trajectory are used for evaluating
 state deviation $d \mathbf Y_N$ at final time.
\section{Results and Discussions}
\subsection{Results without $J_2$ perturbation}
\begin{figure}
  \centering
  \includegraphics[width=4.5in]{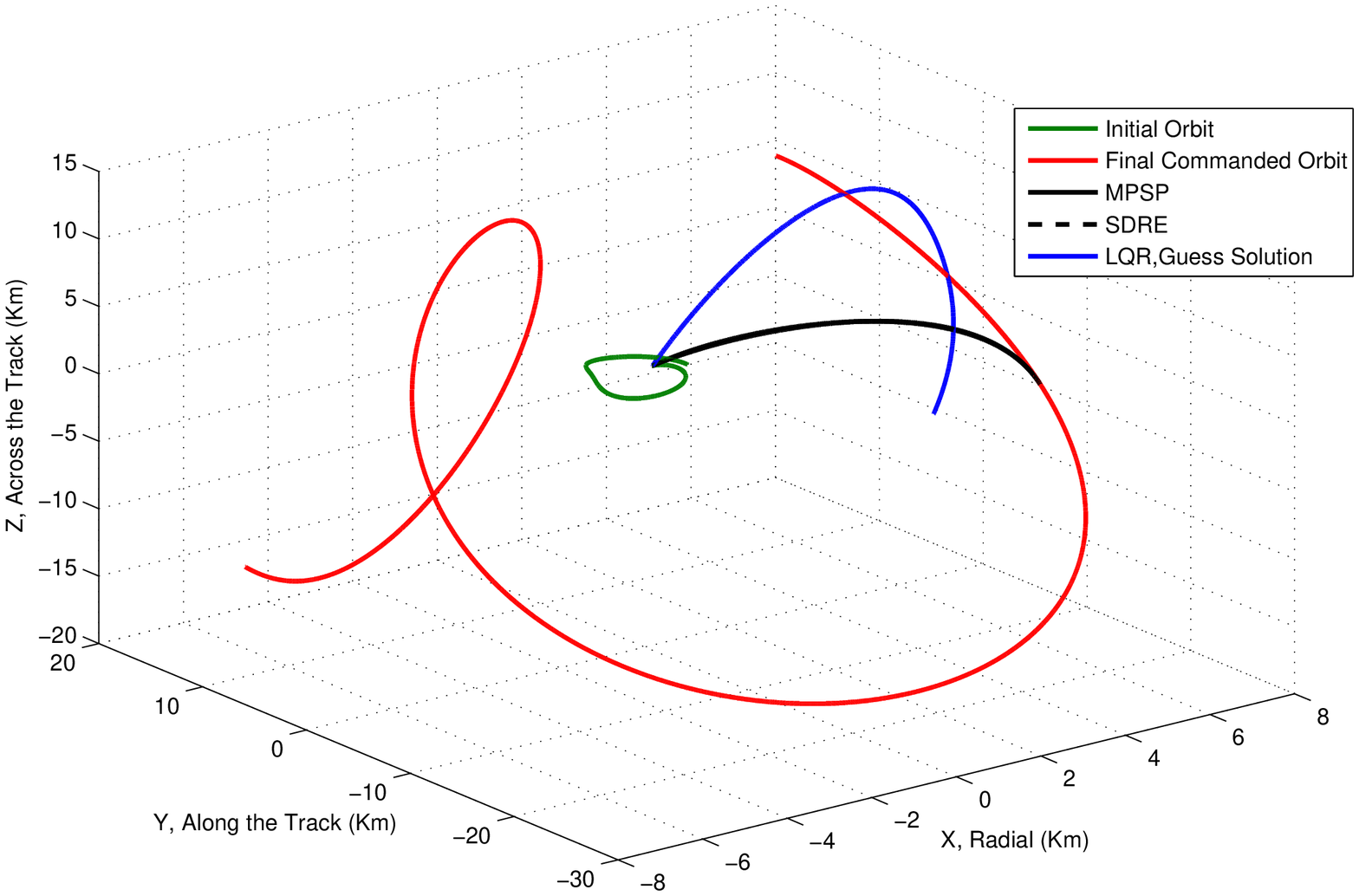}
  \caption{Satellite Orbit transfer trajectory plot,for guess LQR, MPSP and SDRE for eccentric chief satellite orbit}
  \label{TRAJ_ISO}
\end{figure}

In this section, two cases are studied (i) circular chief satellite orbit, $10,000 \hspace{1mm} km$ radius
vector (ii) Eccentric chief satellite orbit with eccentricity $0.15$ and $10,000 \hspace{1mm} km$ semi-major axis details are presented in Table \ref{ORB_PAR_MPSP}. Table \ref{Initial_condition_MPSP}  lists the initial and final relative parameters of the deputy satellite. The simulation step size is selected $\triangle t =1sec$. MPSP numerical simulation is stopped once the error criterion is met $\%{\rho_e}<0.5\%$ . The error criterion is specified in terms of percentage error over final desired base-line length formation commanded.
\begin{equation}
\%{\rho_e} = \frac{{\left( {{{\rho_f}} - {{\rho_d}}}
\right)}}{{{\rho _d}}} \times 100\
\end{equation}
\begin{itemize}
\item ${\rho_e}:$ Final reconfiguration base-line length error
\item ${\rho_f}:$ Final achieved\hspace{1 mm}${\rho}$
\item ${\rho_d}:$ Desired\hspace{1 mm}${\rho}$
\end{itemize}
Finite time State Dependent Ricatti (SDRE) solution presented in chapter \ref{SDRE} is used as comparative method for MPSP solution.
\begin{table}
\caption{\small{Deputy Satellite Initial condition for MPSP solution}}\label{Initial_condition_MPSP}
\begin{center}\begin{tabular}{|c|c|c|}
    \hline
    \textbf{Orbital} & \textbf{Initial Value}&\textbf{Final Value}\\
    \textbf{Parameters} &                     &\\
    \hline
    $\rho (km)$ & $0.5 km$& $5 km$\\
    \hline
    $\theta (deg) $& $45^0$ & $60^0$\\
    \hline
    $a (km)$& $0$ & $0$\\
    \hline
    $b (km)$& $0$ & $0$\\
    \hline
    $m$ (slope)& $1$ & $1.5$\\
    \hline
    $n$(slope)& $0$ & $0$\\
    \hline
\end{tabular}\end{center}
\end{table}
\begin{table}
\caption{\small{Chief Satellite Orbital Parameters, (MPSP)}}\label{ORB_PAR_MPSP}
\begin{center}\begin{tabular}{|c|c|}
    \hline
    \textbf{Orbital Parameters} & \textbf{Value}\\
    \hline
    \textbf{Semi-major axis} & $10000 km$ \\
    \hline
    \textbf{Eccentricity} & Case:1 $e=0$, Case:2 $e=0.15$ \\
    \hline
    \textbf{Orbit Inclination} & $0$ \\
    \hline
    \textbf{Argument of Perigee} & $0$ \\
    \hline
    \textbf{Longitude of ascending node} & $0$ \\
    \hline
    \textbf{Initial True Anomaly} & $10$ \\
    \hline
\end{tabular}\end{center}
\end{table}
\begin{figure}
  \centering
  \includegraphics[width=5in]{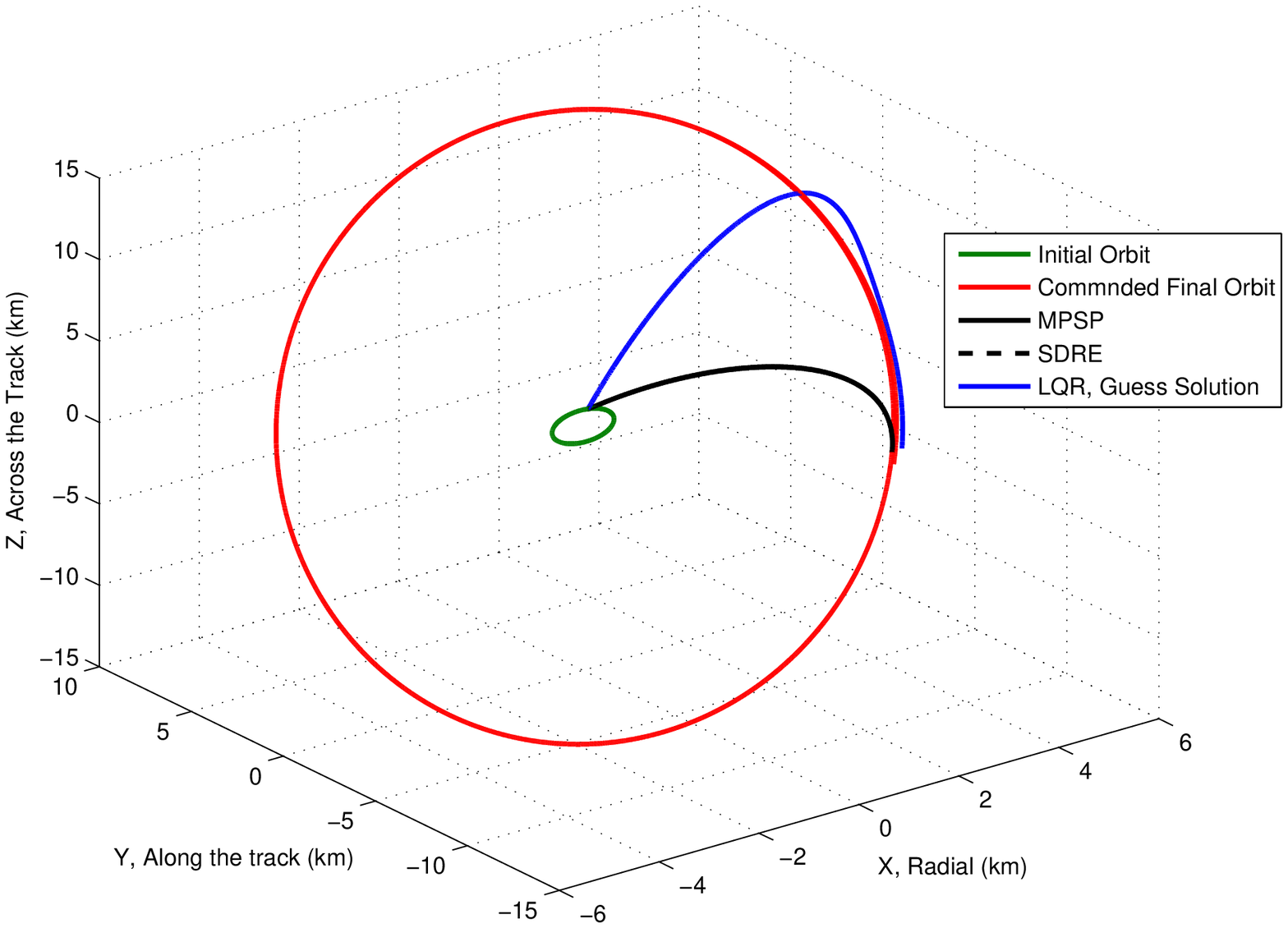}
  \caption{Satellite Orbit transfer trajectory plot,for guess LQR, MPSP and SDRE for Circular chief satellite orbit}
  \label{TRAJ_CIRCULAR}
\end{figure}
\begin{figure}
  \centering
  \includegraphics[width=4.5in]{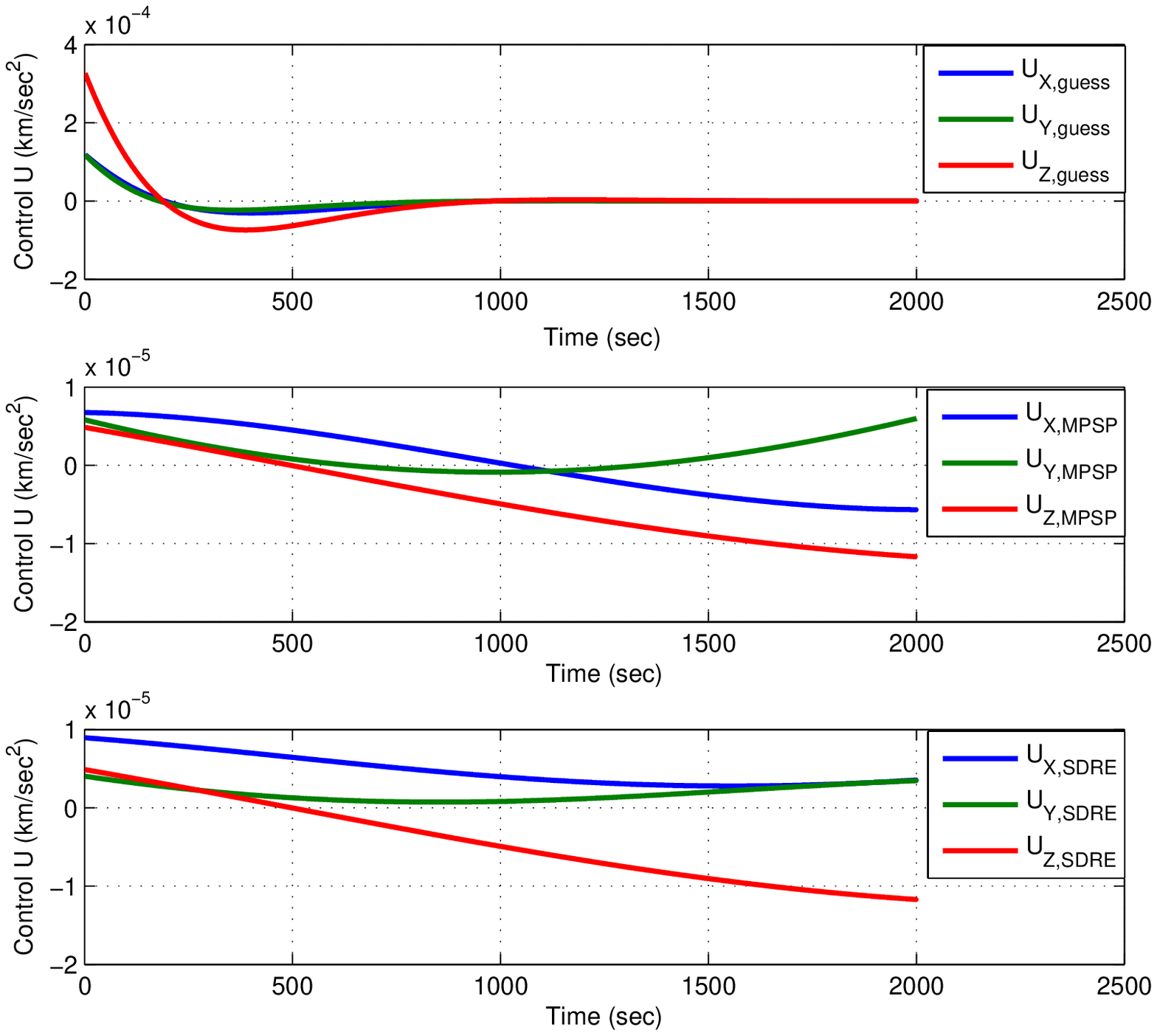}
  \caption{Control History for Guess control LQR and subsequent Updated MPSP controls and SDRE solution}
  \label{CNTRL}
\end{figure}
\begin{figure}
  \centering
  \includegraphics[width=6.7in]{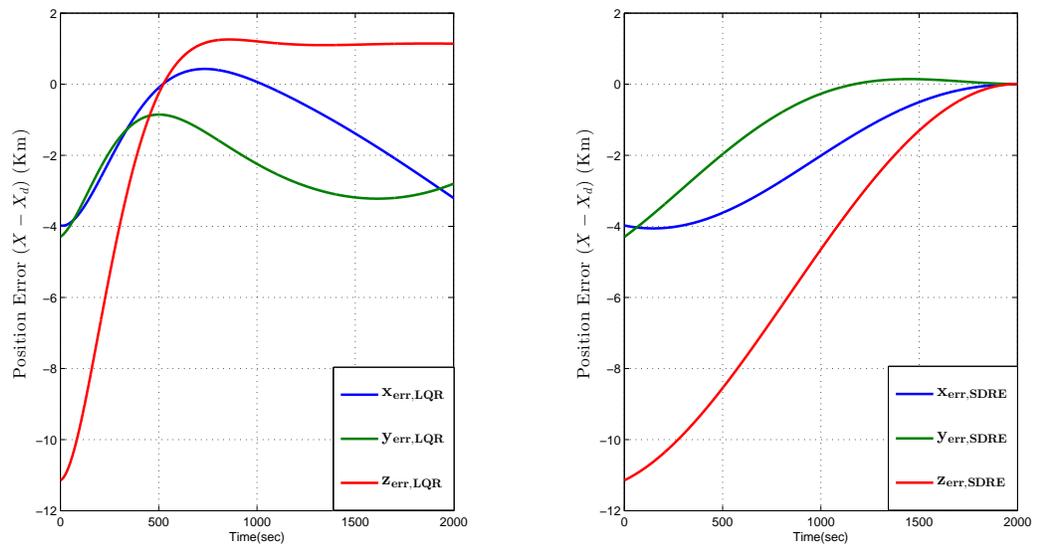}
  \caption{Position Error for Initial Guess solution LQR and SDRE}
  \label{POS_ERR_LQR_SDRE}
\end{figure}
\begin{figure}
  \centering
  \includegraphics[width=5in]{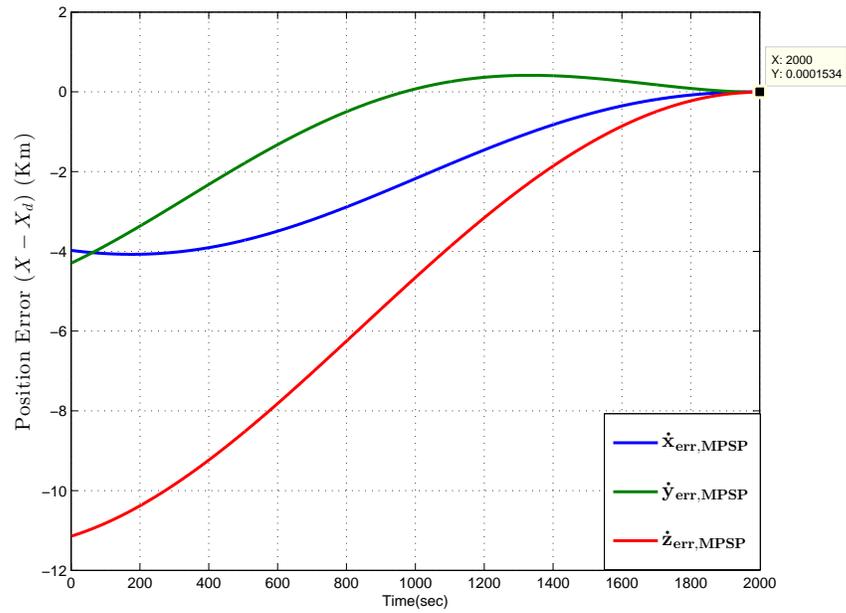}
  \caption{Position Error for MPSP final iteration (Iteration No. 10) }
  \label{POS_ERR_MPSP}
\end{figure}
\begin{figure}
  \centering
  \includegraphics[width=6.7in]{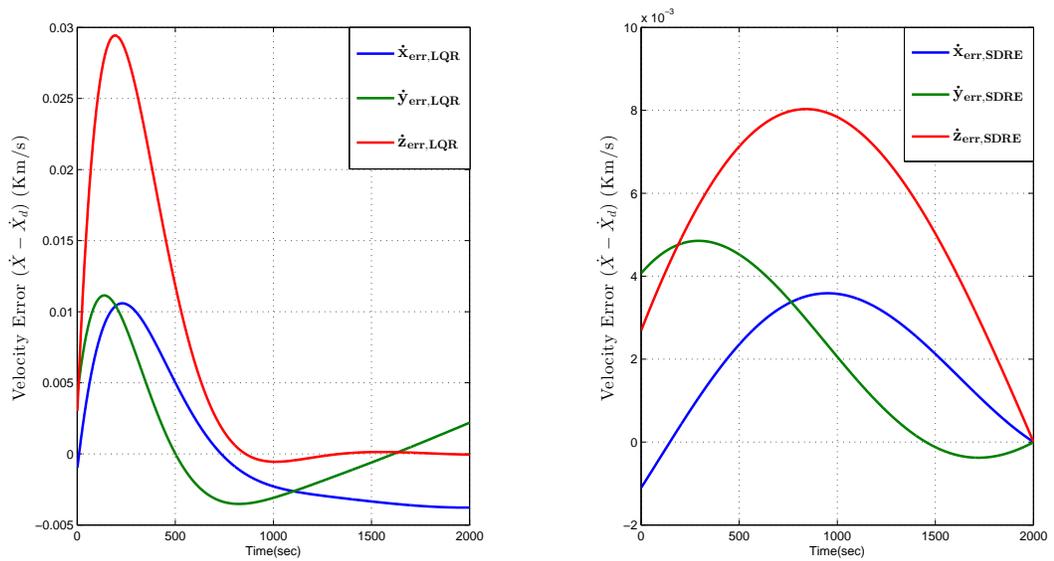}
  \caption{Velocity Error for Initial Guess solution LQR and SDRE}
  \label{VEL_ERR_LQR_SDRE}
\end{figure}
\begin{figure}
  \centering
  \includegraphics[width=5in]{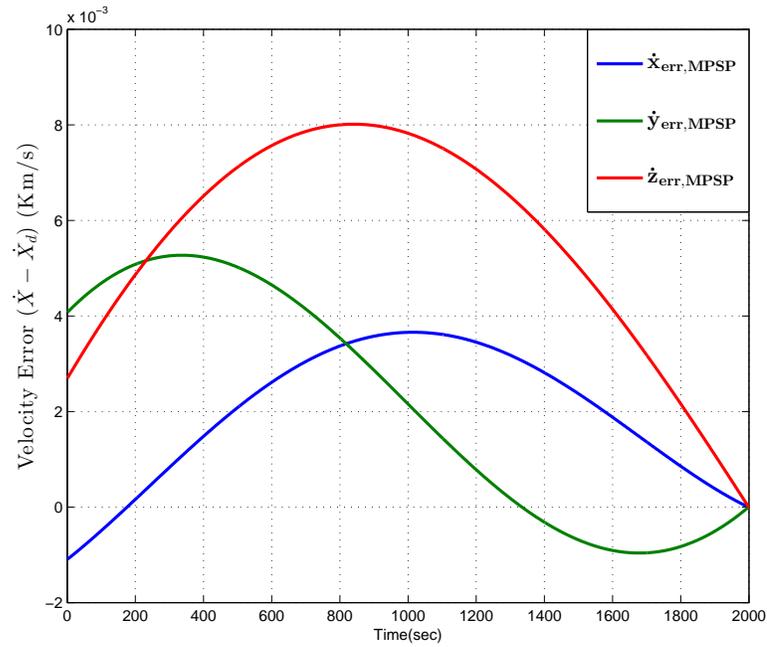}
  \caption{Velocity Error for MPSP final iteration (Iteration No. 10) }
  \label{VEL_ERR_MPSP}
\end{figure}
\begin{figure}
  \centering
  \includegraphics[width=4.5in]{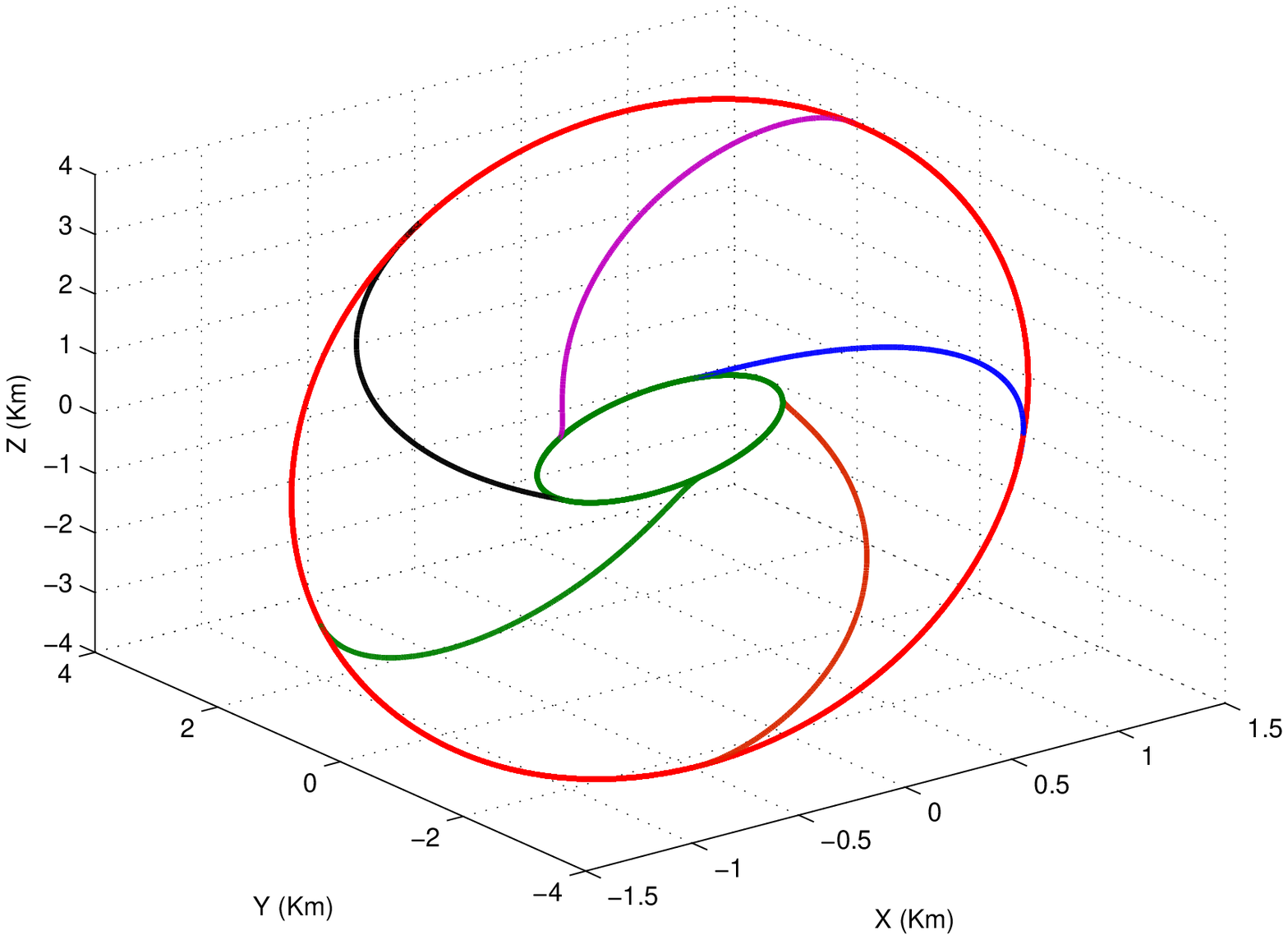}
  \caption{MPSP guidance Formation trajectory for different Initial Conditions and circular chief satellite orbit}
  \label{MPSP_IC}
\end{figure}\\
Figure \ref{TRAJ_CIRCULAR} and \ref{TRAJ_ISO} shows in 3D orbit transfer from the initial formation to new commanded formation trajectory for circular and eccentric chief satellite orbit respectively. MPSP trajectory is significantly different from the initial guess (LQR trajectory). MPSP solution tries to minimizes the control and achieve the final states as hard constraints. Ten iterations are carried out and corresponding state error for LQR, MPSP and SDRE solution are tabulated in Table \ref{table:1}. Figure \ref{CNTRL},  \ref{POS_ERR_LQR_SDRE}, \ref{POS_ERR_MPSP}, \ref{VEL_ERR_LQR_SDRE}, \ref{VEL_ERR_MPSP}and \ref{TRAJ_ISO} refers to eccentric chief satellite orbit results. From Table-\ref{table:1} it can be noticed that improvement in final accuracy in achieved states is great and this accuracy is achieved along with control minimization (Figure \ref{COST}). Figure \ref{CNTRL} shows the control plots for LQR, MPSP and SDRE methods. The total control effort (area under the curve in figure \ref{COST}) for
SDRE is $ 76.0195 \left( {\frac{{km}}{{{{\sec }^2}}}} \right)^2$ and MPSP is $ 69.0704\left( {\frac{{km}}{{{{\sec }^2}}}}
\right)^2$. The control effort required for MPSP method for placing the satellite in desired formation is significantly lesser
compared to SDRE. Figures  \ref{POS_ERR_LQR_SDRE}, \ref{POS_ERR_MPSP} and \ref{VEL_ERR_LQR_SDRE}, \ref{VEL_ERR_MPSP} presents the
error in position and velocity respectively for guess control LQR, SDRE and MPSP respectively. For a SFF problem achieving the final velocity states along with position states on the final orbit is very crucial. Since reaching the desired
position on the desired orbit does not suffice the formation requirement, to be on the orbit and maintain desired relative
separation the injection velocity at the desired orbit are to be met very closely. Note that rendezvous mission which are subset of formation flying where final separation distance is very small, maintaining tight tolerance on the final achieved relative
velocities is very critical for success of mission. Else over a period the satellite drifts away from the required formation
thereby needing to apply control repeatedly to maintain the formation. It can be seen that the velocity error for MPSP
trajectory converges very close to zero value (see Table \ref{table:1}). Figure \ref{MPSP_IC} shows the plot of MPSP
solution for different initial condition on the initial formation orbit for circular chief satellite orbit with initial separation of $0.5\hspace{1mm}km$ to commanded radial separation between and deputy satellite as $1.5\hspace{1mm}km$. For every different
initial conditions the MPSP solution converges satisfactorily to the desired orbit.

\begin{table}
\caption{LQR,SDRE and Iteration wise MPSP state error.(Position errors are in ``$km$`` and velocity in ``$\frac{km}{sec}$``)} 
\centering 
\begin{tabular}{|c|c|c|c|c|} 
\hline
\textbf {Error}  & \textbf {Initial} & \textbf {Itr\#1} & \textbf {Itr\#3} & \textbf {Itr\#5}\\ [0.5ex] 
\textbf {in States} &  \textbf {Guess(LQR)} &      &        &         \\
\hline 

$x$ & -3.1985 & -3.7777  &  -0.9491 & -0.0407   \\ 
\hline
$\dot{x}$ & -0.0038 & -0.0028 & -0.0020 &  -0.0001  \\
\hline
$y$ & -2.7933 & 3.3318 & 0.1499 & -0.0049 \\
\hline
$\dot{y}$ & 0.0022 & 0.0030 & 0.0007 & 0.0000297 \\
\hline
$z$ & 1.1496 & -0.0634 &  -0.0000250 & 0.0000012\\
\hline
$\dot{z}$ & -0.0001 & -0.0001 & 0.00000024 & 0.000000018 \\[1ex] 

\hline 
\textbf {Error}  &  \textbf {Itr\#7} & \textbf {Itr\#9}  & \textbf {Itr\#10}  & \textbf {SDRE}  \\ [0.5ex]
\textbf {in States}              &   & $(1e-3)$ & $(1e-4)$  & \\

\hline 

$x$ &  0.0017 & 0.2787  & -0.6386 & -0.0027   \\ 
\hline
$\dot{x}$ & 0.0000022 & 0.0006 &  -0.0015& 5.129e-06  \\
\hline
$y$ & -0.0009 & -0.0361 & -0.0150 & 0.00593 \\
\hline
$\dot{y}$ & -0.0000011 &  -0.0002 & 0.0004 &  -5.02e-06 \\
\hline
$z$ &  0.19e-06 & -0.41e-04 & 0.289e-05 & 0.00835 \\
\hline
$\dot{z}$ &  0.3e-09 & 0.16e-06 & 0.22e-06 & 7.625e-06 \\[1ex] 
\hline 
\end{tabular}
\label{table:1} 
\end{table}

\begin{table}
\caption{LQR,SDRE and MPSP state error(Eccentric chief satellite orbit solution, Final $\rho = 5\hspace{1mm}km$). (Position errors are in ``$km$`` and velocity in ``$\frac{km}{sec}$``)} 
\centering 
\begin{tabular}{|c|c|c|c|} 
\hline
\textbf{Error}  & \textbf{Initial} & \textbf{SDRE} & \textbf{MPSP} \\ [0.5ex] 
\textbf{in States} &  \textbf{Guess(LQR)} &      &     $(1e-4)$           \\
\hline 

$x$  & -3.7777  & -0.00267 & -0.6386    \\ 
\hline
$\dot{x}$ & -0.0028 & 5.129e-06& -0.0015     \\
\hline
$y$ & 3.3318 & 0.00593 & -0.0150   \\
\hline
$\dot{y}$ & 0.0030 & -5.02e-06& 0.00039   \\
\hline
$z$& -0.0634 & 0.00835 & 0.289e-05 \\
\hline
$\dot{z}$& -0.0001 & 7.625e-06 & 0.22e-06 \\ 

\hline 
\end{tabular}
\label{table:1} 
\end{table}

\begin{figure}
  \centering
  \includegraphics[width=4.5in]{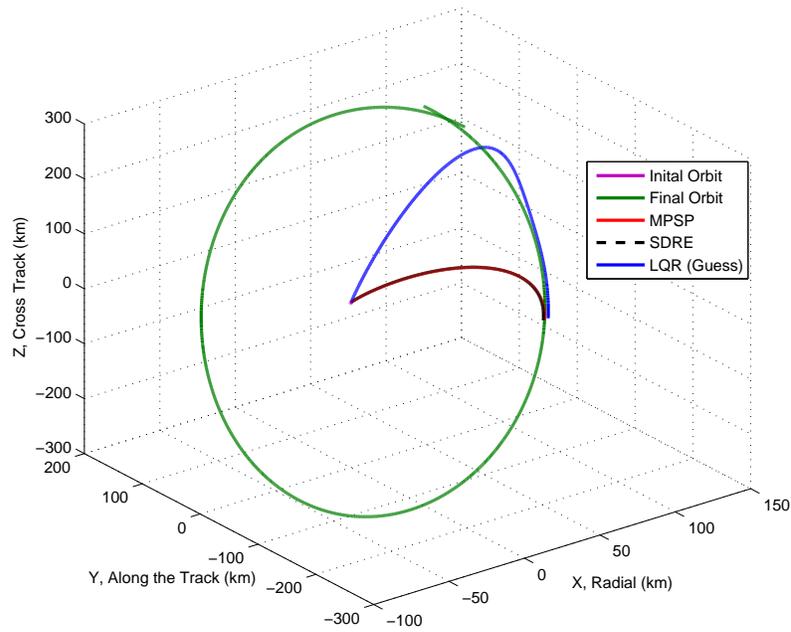}
  \caption{Orbital Trajectory for Formation Flight with $100km$ base length using LQR and MPSP guidance}
  \label{TRAJ_100}
\end{figure}
Figure \ref{TRAJ_100} shows the trajectory plot for formation flying with initial condition of $0.5\hspace{1mm}km$ base-line length ($\rho$) to $100\hspace{1mm}km$ spatial separation with rest of the orbital parameters being same as given in Table \ref{Initial_condition_MPSP} and circular chief satellite orbit. These set of final parameters with high separation trajectory is considered to demonstrate the accuracy of MPSP over Linearized dynamics solution.\\ The state errors for the LQR (guess solutions), MPSP and SDRE solutions are given in Table \ref{table:100Km_state}. The error in the final $\rho$ achieved
using LQR is $2\hspace{1mm}km$, SDRE is $1.4\hspace{1mm}km$ and MPSP value of $0.1\hspace{1mm}km$, the convergence criterion for
MPSP solution is meet with eight iteration $(\%\rho_e=0.1\%)$ and numerical simulation is stopped.
\begin{table}
\caption{LQR, MPSP and SDRE STATE ERROR ($\rho = 100km$)(Position errors are in ``$km$`` and velocity in ``$\frac{km}{sec}$``)} 
\centering 
\begin{tabular}{|c|c|c|c|} 
\hline
\textbf{Error}  &  \textbf{LQR} &  \textbf{SDRE} & \textbf{MPSP}\\
   \textbf{in States}             &                &   &\\

\hline 

$x$ & 26.166 &   0.813&-0.0263  \\
\hline
$\dot{x}$ & 0.0305 & 0.001384& -0.0003  \\
\hline
$y$ & 10.2913 & 0.217& -0.2691\\
\hline
$\dot{y}$ & -0.0151 & 0.000434&-0.0015\\
\hline
$z$ & 1.1628 & 0.1271&0.0000486  \\
\hline
$\dot{z}$ & 0.0009 & 0.0006177& 0.000000128 \\ 
\hline 
\end{tabular}
\label{table:100Km_state} 
\end{table}

\subsection{Results with $J_2$ perturbation effects considered along plant model}
I this section the simulation results presented involve the perturbation model for $J_2$ term. The $J_2$ disturbance term is exogenous to system and acts as additional component of acceleration in all three direction in Hill's frame along with applied control forces. The control value computed from MPSP technique with known model of $J_2$ perturbation account for this perturbing forces and effectively achieve the set objective of tracking a commanded trajectory with minimum terminal error. Similar to MPSP results with no $J_2$ effects the final time $t_f$ is selected as $2000sec$ and simulation time step is chosen as $\triangle t = 1 sec$.

To have a comparative study of MPSP technique capability to synthesize the controller under external perturbation forces, the initial and final relative parameters of the deputy satellite are considered are same as used for MPSP simulation results given in Table \ref{Initial_condition_MPSP}. The orbital parameter of chief satellite is given Table \ref{ORB_PAR_MPSP_J2} . The simulation step size is selected $\triangle t =1sec$. MPSP numerical simulation is stopped once the error criterion is met $\%{\rho_e}<0.5\%$
\begin{table}
\caption{\small{Chief Satellite Orbital Parameters, (MPSP) with $J_2$ perturbation model}}\label{ORB_PAR_MPSP_J2}
\begin{center}\begin{tabular}{|c|c|}
    \hline
    \textbf{Orbital Parameters} & \textbf{Value}\\
    \hline
    \textbf{Semi-major axis} & $10000 km$ \\
    \hline
    \textbf{Eccentricity} & $e=0.15$ \\
    \hline
    \textbf{Orbit Inclination} & $60$ \\
    \hline
    \textbf{Argument of Perigee} & $0$ \\
    \hline
    \textbf{Longitude of ascending node} & $0$ \\
    \hline
    \textbf{Initial True Anomaly} & $10$ \\
    \hline
\end{tabular}\end{center}
\end{table}
\begin{figure}
  \centering
  \includegraphics[width=4.5in]{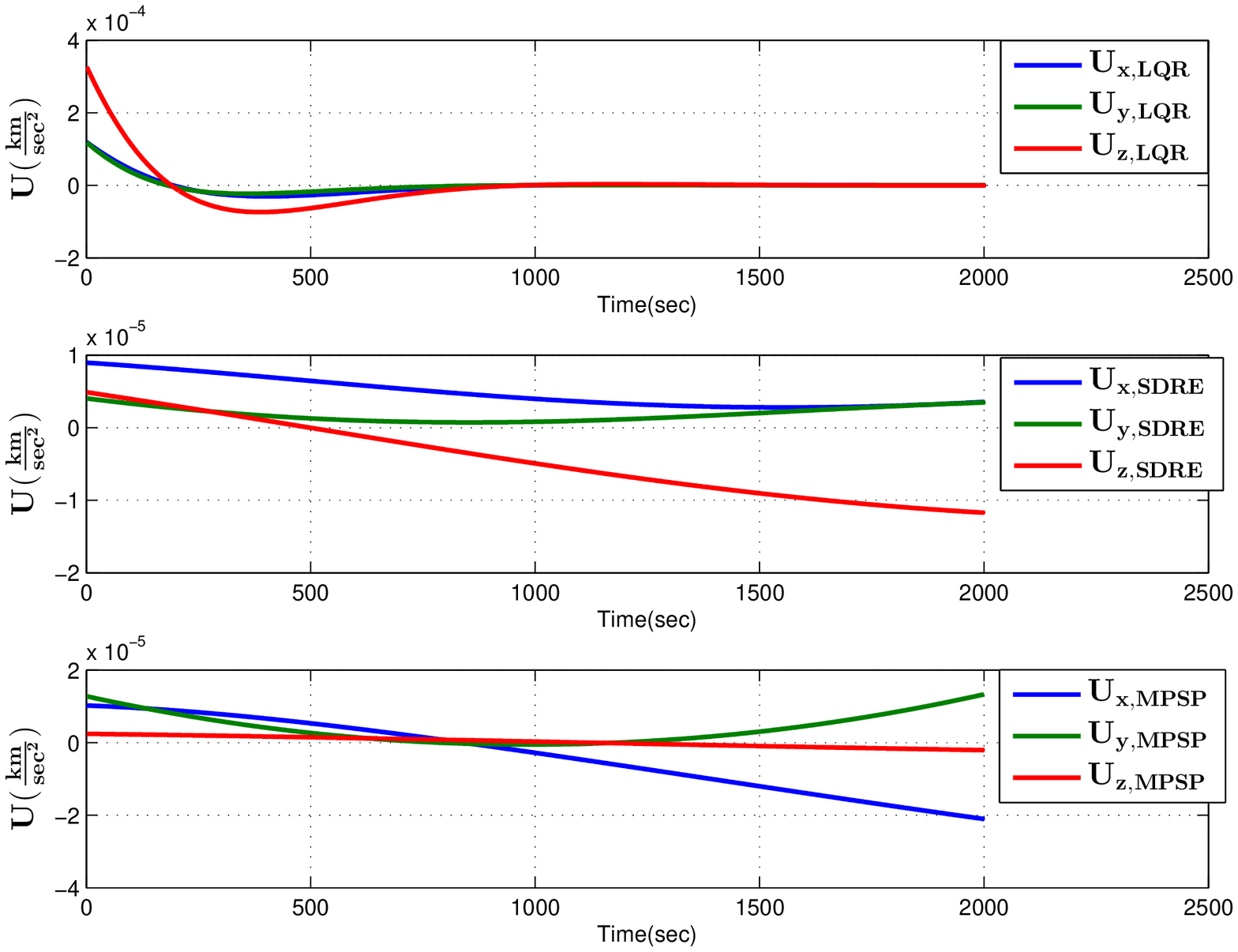}
  \caption{Control History for Guess control LQR and subsequent Updated MPSP controls and SDRE solution with $J_2$ perturbation}
  \label{CNTRL_J2}
\end{figure}
\begin{figure}
  \centering
  \includegraphics[width=6.7in]{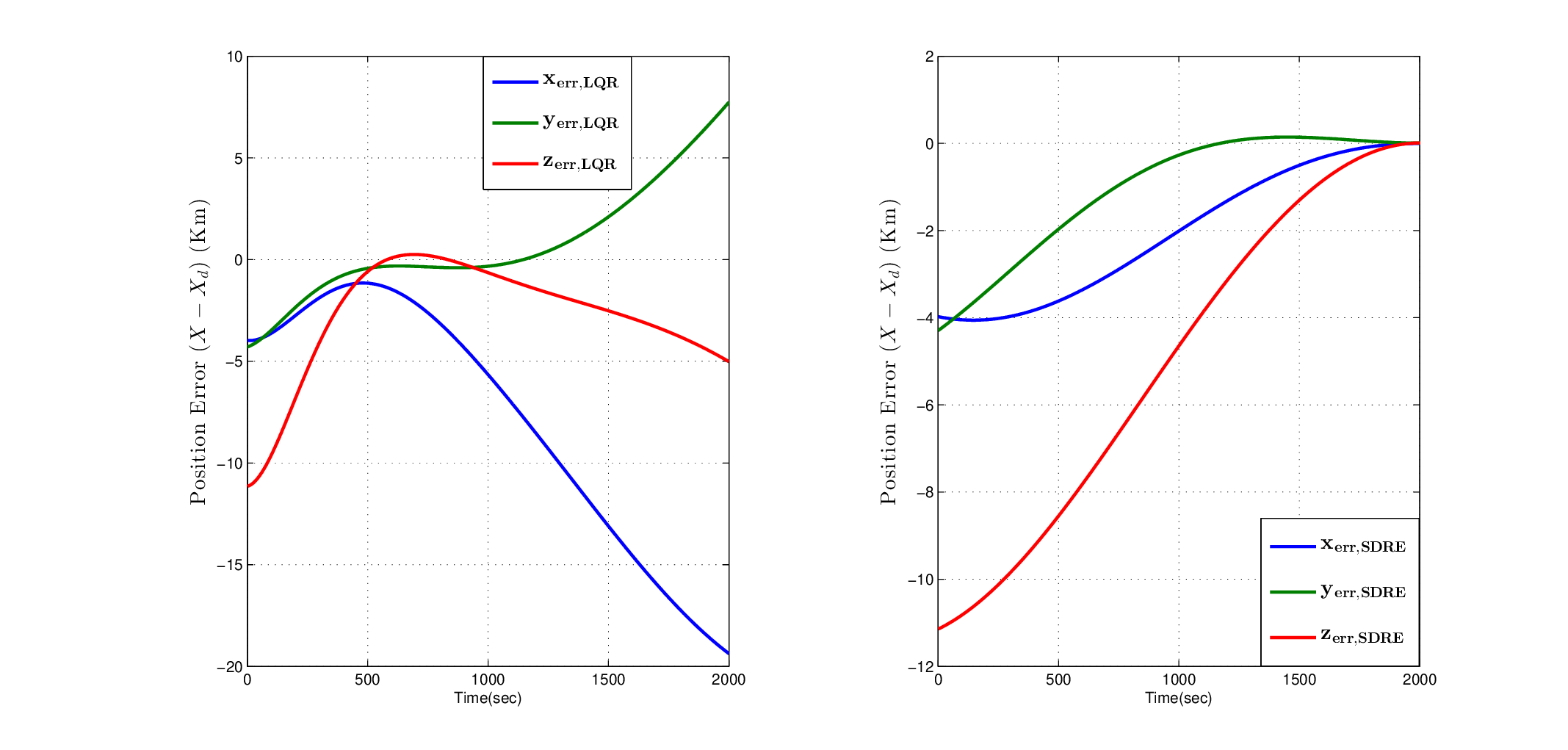}
  \caption{Position Error for Initial Guess solution LQR and SDRE with $J_2$ perturbation}
  \label{POS_ERR_LQR_SDRE_J2}
\end{figure}
\begin{figure}
  \centering
  \includegraphics[width=5in]{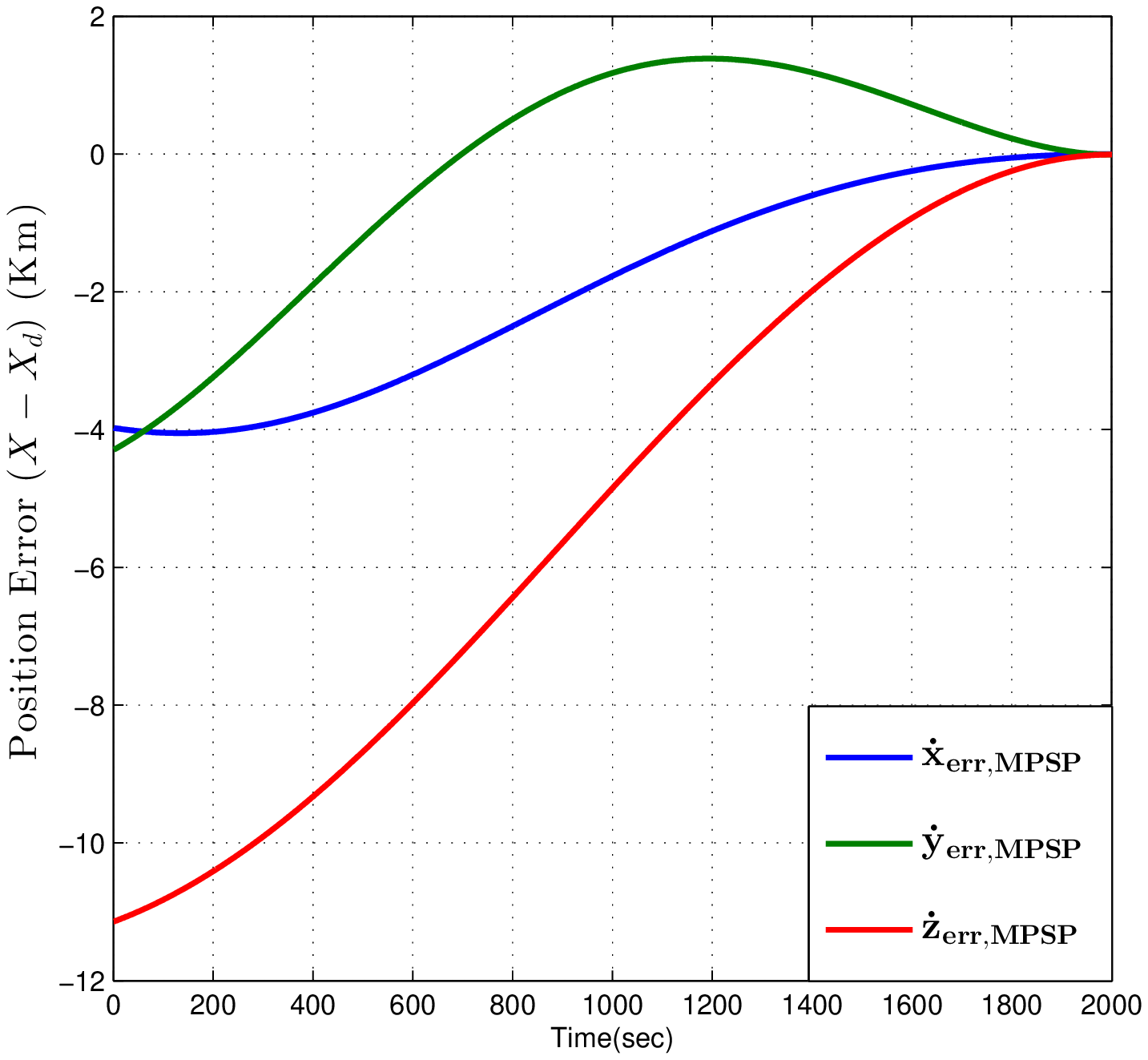}
  \caption{Position Error for MPSP final iteration with $J_2$ perturbation (Iteration No. 10) }
  \label{POS_ERR_MPSP_J2}
\end{figure}
\begin{figure}
  \centering
  \includegraphics[width=6.7in]{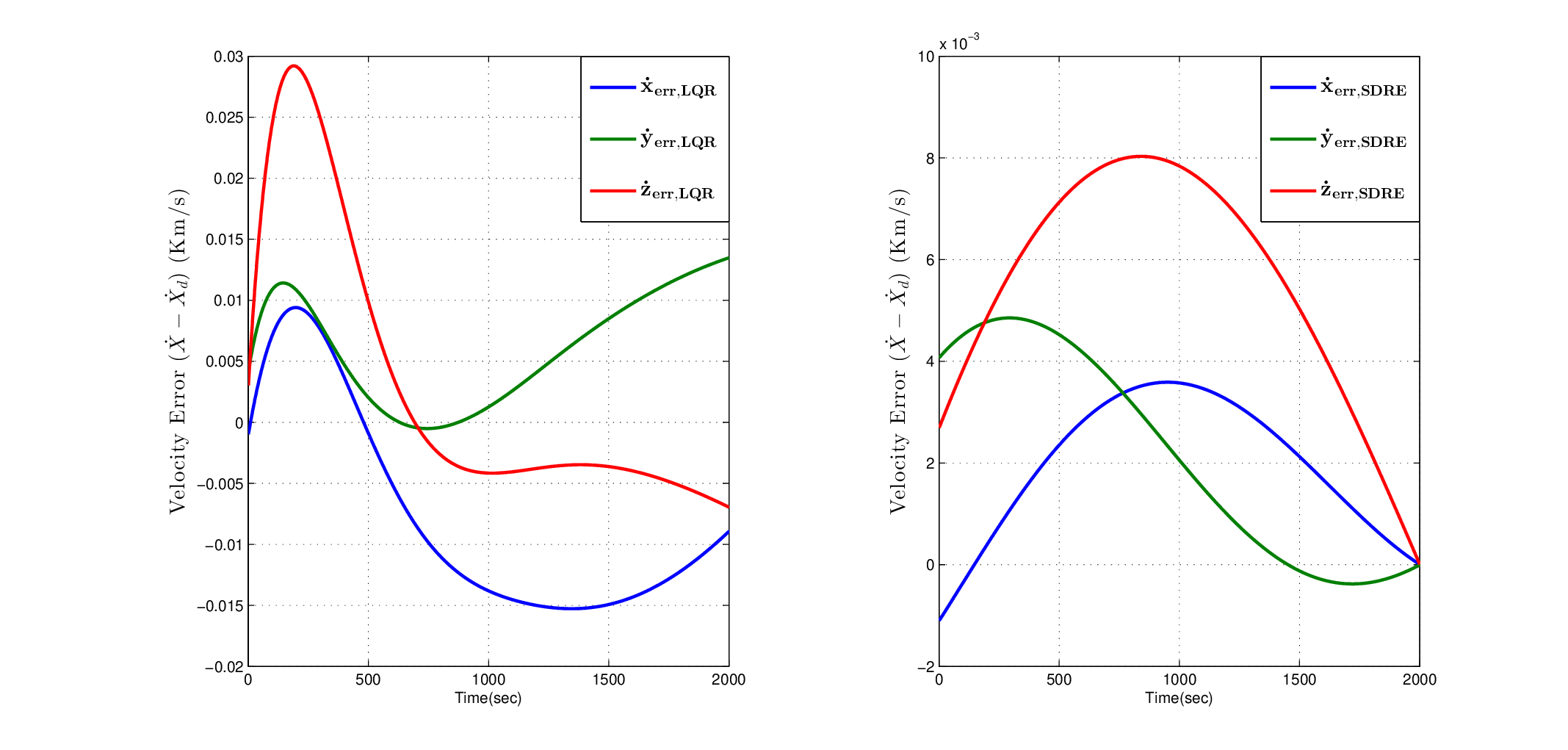}
  \caption{Velocity Error for Initial Guess solution LQR and SDRE with $J_2$ perturbation}
  \label{VEL_ERR_LQR_SDRE_J2}
\end{figure}
\begin{figure}
  \centering
  \includegraphics[width=5in]{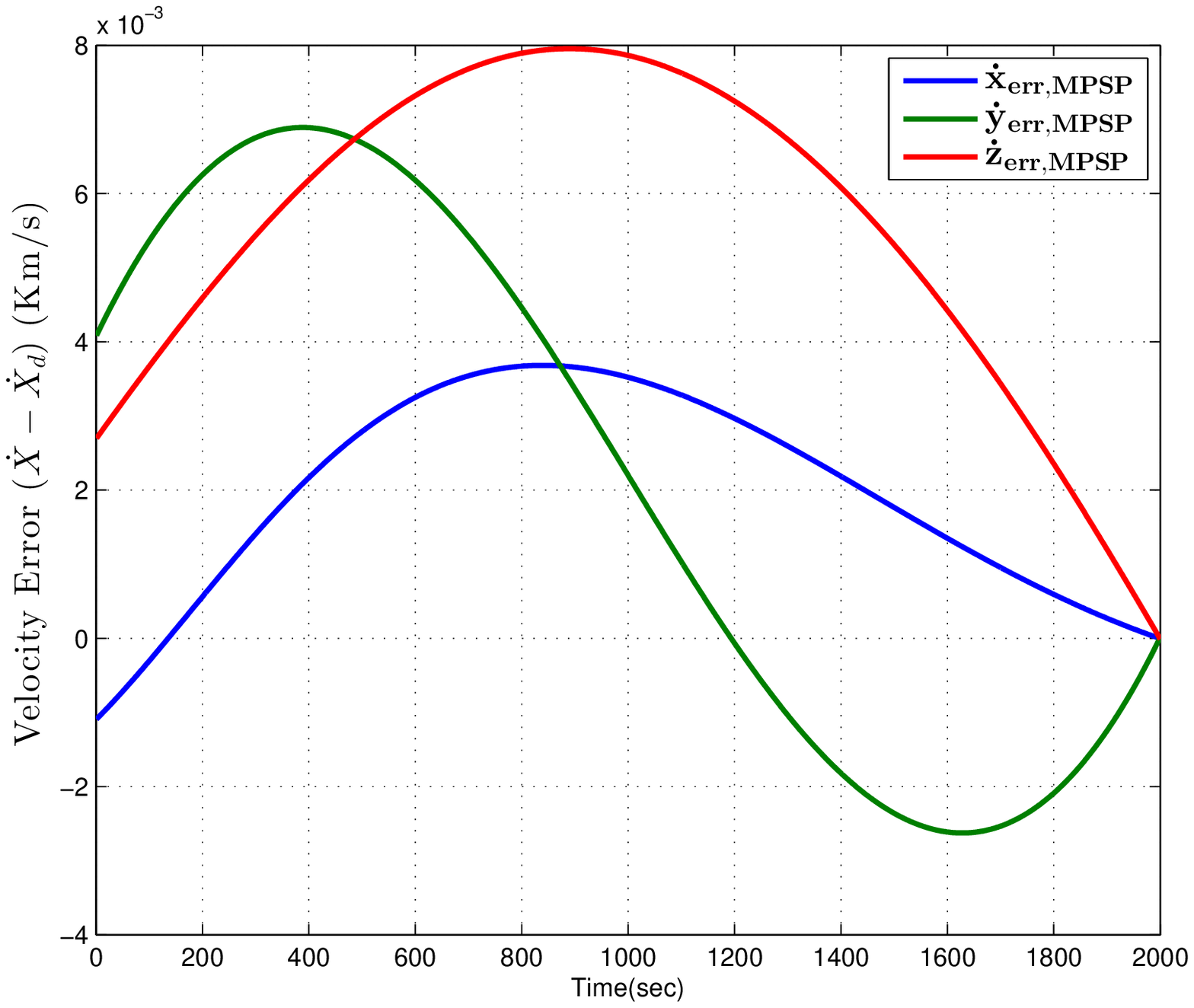}
  \caption{Velocity Error for MPSP final iteration with $J_2$ perturbation (Iteration No. 10) }
  \label{VEL_ERR_MPSP_J2}
\end{figure}
Figure \ref{POS_ERR_LQR_SDRE_J2}, \ref{POS_ERR_MPSP_J2} gives position error for guess control LQR, comparative method SDRE and MPSP method under the effects of $J_2$ perturbation. Figure \ref{CNTRL_J2} gives the control effort of guess control, SDRE and MPSP respectively. Table gives the details of terminal state errors for LQR, SDRE and MPSP techniques under effect of $J_2$ perturbation acceleration. Table illustrates the comparative behavior of MPSP and SDRE simulation results for reformation problem with and without $J_2$ perturbation. It can be seen that SDRE results diverge under consideration of $J_2$ perturbation effects where as MPSP method satisfactorily drives the terminal state to desired orbital states hence leading to close tracking of commanded orbit.
\begin{table}
\caption{LQR, MPSP and SDRE STATE ERROR ($\rho = 5km$ with $J_2$ perturbation)(Position errors are in ``$km$`` and velocity in ``$\frac{km}{sec}$``)} 
\centering 
\begin{tabular}{|c|c|c|c|} 
\hline
\textbf{Error}  &  \textbf{LQR} &  \textbf{SDRE} & \textbf{MPSP}\\
   \textbf{in States}             &                &   &\\

\hline 

$x$ & -19.361 & -0.7303&7.05e-04  \\
\hline
$\dot{x}$ & -0.008942 & -0.0014& 1.201e-06  \\
\hline
$y$ & 7.727& -0.2502& 3.45e-04\\
\hline
$\dot{y}$ & 5.02e-06 &  0.004&1.466e-07\\
\hline
$z$ & -5 & 0.5559&3.684e-04  \\
\hline
$\dot{z}$ & -0.006941 & 0.0007& 4.827e-07 \\ 
\hline 
\end{tabular}
\label{table:COMP_J2} 
\end{table}
\section{Summary and Conclusions}
In this section the details satellite formation flying using MPSP control is presented. The MPSP theory is introduced in initial section of this chapter. The SFF problem is defined in MPSP frame work. The simulation are carried out with and without the effects of $J_2$ perturbation model. IN both the case it is found out that MPSP solution results are superior compared to comparative SDRE solution, and that MPSP in both situation could compute the controller such that the terminal state error is minimal and within the tolerance limit that is $\%\rho_e < 1\%$. In next chapter G-MPSP controller for SFF is introduced, and formation reconfiguration results are discussed.

\cleardoublepage
\chapter{Generalized Model Predictive Static Programming (G-MPSP)}
\label{GMPSP}
In this section, the theoretical details of the generalized model predictive static programming (G-MPSP) are presented. Note that a brief summary of MPSP theory has been presented in Appendix. For more details of MPSP theory, one can refer to recent
publications~\cite{Padhi:2008Paper,PDas,Harshal}. In the proposed design, a
general nonlinear systems in continuous time setting is considered with the following state dynamics and output equation:
\begin{equation} \label{Eqn:sys_dyn}
\dot X\left( t \right) = f\left( {X\left( t \right),U\left( t
\right)} \right)
\end{equation}
\begin{equation} \label{Eqn:sys_output}
Y\left( t \right) = h\left( {X\left( t \right)} \right)
\end{equation}
where, $X \in \Re^n$, $\ U \in \Re^m$ and $\ Y \in \Re^p$. The primary objective is to obtain a suitable control history $U(t)$
so that the output $Y(t_f)$ at the fixed final time $t_f$ goes to a desired value $Y^*(t_f)$, i.e. $Y(t_f) \rightarrow Y^*(t_f)$. It is also required that this task is achieved with minimum control effort.
For the technique presented here, one needs to start from a ``guess history'' of the control solution. With the application of
such a guess history, obviously the objective is not expected to be met, and hence, there is a need to improve this solution. In
this section, we present a way to compute an error history of the control variable, which needs to be subtracted from the previous history to get an improved control history. This iteration continues until the objective is met, i.e., until $Y(t_f)
\rightarrow Y^*(t_f)$. Note that the technique presented here produces an update in control history in a closed form thereby
reducing the computational load substantially as well as making it computationally very efficient. Next, the mathematical details of the G-MPSP design are presented.

Let the error in output at the final time $t_f$ be given as
follows:
\begin{equation}\label{Eqn:del_Y_tf}
\begin{array}{l}
 \delta Y\left( {X\left( {t_f } \right)} \right) = \left[ {Y\left( {t_f } \right) - Y^ *  \left( {t_f } \right)} \right]
 \end{array}
\end{equation}

Multiplying both sides of \ref{Eqn:sys_dyn} by a matrix
$W(t)$ produces
\begin{equation}\label{Eqn:Multiply_W}
W \left( t \right)\dot X = W \left( t \right)f\left( {X\left( t
\right),U\left( t \right)} \right)\mathrm{,}
\end{equation}
where, the computation of the matrix $W\left(t \right) \in \Re^{p
\times n}$ is described later in this section.

The following is obtained by integrating both sides of
\ref{Eqn:Multiply_W} from $t_0$ to $t_f$ as
\begin{equation}\label{Eqn:Integral_Multiply_W}
\begin{array}{l}
\int_{t_0 }^{t_f } {\left[ {W \left( t \right)\dot X\left( t
\right)} \right]} dt = \int_{t_0 }^{t_f } {\left[ {W \left( t
\right)f\left( {X\left( t \right),U\left( t \right)} \right)}
\right]} dt\mathrm{.}
 \end{array}
\end{equation}

Next, adding the quantity $Y\left( {X\left( {t_f } \right)}
\right)$ to both sides of \ref{Eqn:Integral_Multiply_W} and
using algebraic manipulation, the following is obtained as
\begin{equation}\label{Eqn:Yf_add}
\begin{array}{l}
Y\left( {X\left( {t_f } \right)} \right) = Y\left( {X\left( {t_f }
\right)} \right) + \int_{t_0 }^{t_f } {\left[ {W \left( t
\right)f\left( {X\left( t \right),U\left( t \right)} \right)}
\right]} dt - \int_{t_0 }^{t_f } {\left[ {W \left( t \right)\dot
X\left( t \right)} \right]} dt\mathrm{.}
 \end{array}
\end{equation}

Considering the last term of the right hand side of
\ref{Eqn:Yf_add} and integrating by parts produces
\begin{equation}\label{Eqn:Integrating_by_parts_last}
\begin{array}{l}
 \int_{t_0 }^{t_f } {\left[ {W \left( t \right)\dot X\left( t \right)} \right]} dt \\
  = \left[ {W \left( t \right)X\left( t \right)} \right]_{t_0 }^{t_f }  - \int_{t_0 }^{t_f } {\left[ {\left( {\frac{{dW\left( t \right)}}{{dt}}} \right) X\left( t \right)} \right]} dt \\
  = \left[ {W \left( {t_f } \right)X\left( {t_f } \right) - W \left( {t_0 } \right)X\left( {t_0 } \right)} \right] - \int_{t_0 }^{t_f } {\left[ {\dot W \left( t \right)X\left( t \right)} \right]} dt \mathrm{.}\\
 \end{array}
\end{equation}

Substituting \ref{Eqn:Integrating_by_parts_last} in
\ref{Eqn:Yf_add} leads to the following
\begin{equation}\label{Eqn:Yf_add_Integrating_by_parts}
\begin{array}{l}
 Y\left( {X\left( {t_f } \right)} \right) = Y\left( {X\left( {t_f } \right)} \right) - \left[ {W \left( {t_f } \right)X\left( {t_f } \right)} \right] + \left[ {W \left( {t_0 } \right)X\left( {t_0 } \right)} \right] \\
 \quad \quad \quad \quad \quad \quad + \int_{t_0 }^{t_f } {\left[ {W \left( t \right)f\left( {X\left( t \right),U\left( t \right)} \right) + \dot W \left( t \right)X\left( t \right)} \right]} dt \mathrm{.}\\
 \end{array}
\end{equation}

The following expression is obtained by considering the variation
of the both sides of \ref{Eqn:Yf_add_Integrating_by_parts}
as
\begin{equation} \label{Eqn:del_Y_tf_variation}
\begin{array}{l}
 \delta Y\left( {X\left( {t_f } \right)} \right) = \left[ {\left( {\frac{{\partial Y\left( {X\left( t \right)} \right)}}{{\partial X\left( t \right)}} - W \left( t \right)} \right)\delta X\left( t \right)} \right]_{t = t_f }  + \left[ {W \left( {t_0 } \right)\delta X\left( {t_0 } \right)} \right] \\
 \quad \quad \quad \quad \quad + \int_{t_0 }^{t_f } {\left[ {\left( {W \left( t \right)\frac{{\partial f\left( {X\left( t \right),U\left( t \right)} \right)}}{{\partial X\left( t \right)}} + \dot W \left( t \right)} \right)\delta X\left( t \right)} \right.} \left. { + \left( {W \left( t \right)\frac{{\partial f\left( {X\left( t \right),U\left( t \right)} \right)}}{{\partial U\left( t \right)}}} \right)\delta U\left( t \right)} \right]dt \\
 \end{array}
\end{equation}

Next, it is desired to determine the variations $\delta Y\left(
{X\left( {t_f } \right)} \right)$ produced by the given $\delta
U\left( t \right)$. The idea is to choose the $W(t)$ in a way that
causes the coefficients of $\delta X\left( t \right)$ in the above
equation to vanish. The following is thus in order:
\begin{equation}\label{Eqn:Wdot_update}
\dot W\left( t \right) =  - W\left( t \right) \left(
{\frac{{\partial f\left( {X\left( t \right),U\left( t \right)}
\right)}}{{\partial X\left( t \right)}}} \right)\mathrm{,}
\end{equation}
\begin{equation}\label{Eqn:Wdot_tf}
W\left( {t_f } \right) = \frac{{\partial Y\left( {X\left( {t_f }
\right)} \right)}}{{\partial X\left( {t_f } \right)}}\mathrm{.}
\end{equation}

There is no error in initial condition because the specified
initial condition is a known entity. Hence, the expression $\delta
X\left( {t_0 } \right) = 0$ holds true. Furthermore, substituting
\ref{Eqn:Wdot_update} and \ref{Eqn:Wdot_tf} into
\ref{Eqn:del_Y_tf_variation} produces
\begin{equation}\label{Eqn:del_Y_tf_deriv}
\begin{array}{l}
 \delta Y\left( {X\left( {t_f } \right)} \right) = \int_{t_0 }^{t_f } {\left[ {B_c\left( t \right)\delta U\left( t \right)} \right]} dt \mathrm{,}\\
 \end{array}
\end{equation}
where,
\begin{equation}\label{Eqn:Bc_deriv}
B_c\left( t \right) = W \left( t \right)\frac{{\partial f\left(
{X\left( t \right),U\left( t \right)} \right)}}{{\partial U\left(
t \right)}}\mathrm{.}
\end{equation}

Let the following performance index be considered for optimal
control formulation:
\begin{equation}\label{Eqn:J_fun_min}
J = \frac{1}{2}\int_{t_0 }^{t_f } {\left[ {\left( {U^0 \left( t
\right) - \delta U\left( t \right)} \right)^T R\left( t
\right)\left( {U^0 \left( t \right) - \delta U\left( t \right)}
\right)} \right]dt}
\end{equation}
where, $U^0 \left( t \right)$ is the previous control history
solution. The cost function in \ref{Eqn:J_fun_min} needs to
be minimized subjected to the constraint in
\ref{Eqn:del_Y_tf_deriv}, where the positive definite
weighting matrix $R\left( t \right)> 0$ needs to be chosen
judiciously by the control designer. The selection of such a cost
function is motivated by the fact that it is desired to find an
$l_2$-norm minimizing control history, since $(U^0 \left( t
\right) - \delta U\left( t \right))$ is the updated control value
at time $t$.

Equations~(\ref{Eqn:del_Y_tf_deriv}) and (\ref{Eqn:J_fun_min})
formulate an approximate constrained static optimization problem.
Using the static optimization theory~\cite{bryson_book}, the
augmented cost function is given by
\begin{equation}\label{Eqn:aug_J_fun_c}
\begin{array}{l}
 \bar J = \frac{1}{2}\int_{t_0 }^{t_f } {\left[ {\left( {U^0 \left( t \right) - \delta U\left( t \right)} \right)^T R\left( t \right)\left( {U^0 \left( t \right) - \delta U\left( t \right)} \right)} \right]dt}
+ \lambda ^T \left[ {\delta Y\left( {t_f } \right) - \int_{t_0 }^{t_f } {\left[ {B_c\left( t \right)\delta U\left( t \right)} \right]dt} } \right] \\
 \end{array}
\end{equation}
where, $\lambda$ is the Lagrange multiplier.

Consider next the first variation of \ref{Eqn:aug_J_fun_c}
is given by the expression
\begin{equation}\label{Eqn:del_aug_Jc}
\begin{array}{l}
 \delta \bar J =  - \int_{t_0 }^{t_f } {\left[ {\left\{ {R\left( t \right)\left( {U^0 \left( t \right) - \delta U\left( t \right)} \right) + \left( {B_c\left( t \right)} \right)^T \lambda } \right\}\delta \left( {\delta U\left( t \right)} \right)} \right]dt}  \mathrm{,}\\
 \end{array}
\end{equation}
from which it is clear that a minimum of $\bar J$ occurs if the
following expression holds true:
\begin{equation}\label{Eqn:delUc_derv}
\delta U\left( t \right) = \left( {R\left( t \right)} \right)^{ -
1} \left( {B_c\left( t \right)} \right)^T \lambda  + U^0 \left( t
\right)
\end{equation}

Substituting \ref{Eqn:delUc_derv} into
\ref{Eqn:del_Y_tf_deriv} leads to
\begin{equation}\label{Eqn:del_Y_tf_mod}
\begin{array}{l}
 \delta Y\left( {t_f } \right) = \int_{t_0 }^{t_f } {B_c\left( t \right)\left( {\left( {R\left( t \right)} \right)^{ - 1} \left( {B_c\left( t \right)} \right)^T \lambda  + U^0 \left( t \right)} \right)dt}  \\
 \quad \quad \quad  = A_\lambda  \lambda  + b_\lambda  \mathrm{,} \\
 \end{array}
\end{equation}
where,
\begin{equation}\label{Eqn:A_lamda_c}
A_\lambda   \buildrel \Delta \over = \left[ {\int_{t_0 }^{t_f }
{\left[ {B_c\left( t \right)\left( {R\left( t \right)} \right)^{ -
1} B_c^T \left( t \right)} \right]dt} } \right]\mathrm{,}
\end{equation}
and
\begin{equation}\label{Eqn:b_lamda_c}
b_\lambda   \buildrel \Delta \over = \left[ {\int_{t_0 }^{t_f }
{\left[ {B_c\left( t \right)U^0 \left( t \right)} \right]dt} }
\right]\mathrm{.}
\end{equation}

Assuming that $A_\lambda$ is a non-singular matrix, the following
expression is obtained from \ref{Eqn:del_Y_tf_mod} as
\begin{equation} \label{Eqn:lamda_c}
\lambda  = \left( {A_\lambda  } \right)^{ - 1} \left[ {\delta
Y\left( {t_f } \right) - b_\lambda  } \right]\mathrm{,}
\end{equation}
substituting which into \ref{Eqn:delUc_derv} produces
\begin{equation}\label{Eqn:delUc_derv_mod2}
\delta U\left( t \right) = \left( {R\left( t \right)} \right)^{ -
1} \left( {B_c\left( t \right)} \right)^T \left[ {\left(
{A_\lambda } \right)^{ - 1} \left[ {\delta Y\left( {t_f } \right)
- b_\lambda } \right]} \right] + U^0\left( t \right)\mathrm{.}
\end{equation}

Hence, the updated control is given by
\begin{equation}\label{Eqn:total_Uc}
U\left( t \right) = U^0 \left( t \right) - \delta U\left( t
\right) =  - \left( {R\left( t \right)} \right)^{ - 1} \left(
{B_c\left( t \right)} \right)^T \left[ {\left( {A_\lambda  }
\right)^{ - 1} \left[ {\delta Y\left( {t_f } \right) - b_\lambda }
\right]} \right]
\end{equation}
It is clear from \ref{Eqn:total_Uc} that the updated control
history solution in \ref{Eqn:total_Uc} is a {\it closed form
solution}. In this approach, the idea is to convert the dynamic
optimization problem into a constrained static optimization
problem and then to compute a closed form control history update
for a class of finite-horizon problems. Furthermore, the necessary
error coefficients in \ref{Eqn:Bc_deriv} are computed
recursively using \ref{Eqn:Wdot_tf} and
\ref{Eqn:Wdot_update}. Overall it leads to a very fast
computation of the control history update, and hence, is a
computationally very efficient technique.

At this point, we would like to point out that the process needs
to be repeated in an iterative manner. Concepts such as output
convergence to terminate the algorithm and iteration
unfolding \cite{McHenry:1979Paper} (where the control history is
updated only a finite number of times in a particular time step)
can also be incorporated to enhance the computational efficiency
further (at the cost of sub-optimality of the solution).

We observe the following points related to the proposed G-MPSP:
\begin{enumerate}
    \item In this G-MPSP formulation, the discretization of the system
dynamics is not required, which is required for the MPSP.

    \item In this G-MPSP, any higher-order of the
integration technique (e.g. forth-order Runge-Kutta
scheme) can be used for computing
recursively the sensitivity matrices (see
(\ref{Eqn:Wdot_update})).

    \item In this G-MPSP, it can be observed from \ref{Eqn:del_Y_tf_deriv} that the error
in output at final time $t_f$ defined by $\delta Y(X(t_f))$ given
in \ref{Eqn:del_Y_tf_deriv} is derived using the first order
terms of the corresponding Taylor's series expansion of the
continuous time optimal control formulation. While, in the MPSP,
the error in output at final time step $k=N$ defined by $\Delta
Y_N$ given in \ref{error_out} and finally in
\ref{constra} is derived using the first order
terms of the corresponding Taylor's series expansion on the
discretized version of the dynamics \ref{Eqn:sys_dyn} which
amounts to two approximations, namely, one given by a numerical
discretization method (e.g. Euler's discretization
scheme) and the second given by Taylor's
series expansion. Thus, the G-MPSP formulation needs only one
approximation due to its continuous time formulation.

\end{enumerate}
{\section{G-MPSP Implementation Algorithm}\label{Section:AlgoSteps}
The G-MPSP technique is an iterative algorithm which starts from a
guess history and continues until the desired accuracy in the
terminal error of the output $Y(t_f)$ is achieved. The following
algorithmic steps
are performed in every guidance cycle once the dynamics and the guess history are defined:}%

\begin{enumerate}
    \item\label{G-MPSP:Step1} Initialize the
previous control history $U^0(t)$ as the guess control history
with some guess such that the guess trajectories are not very far
from the desired trajectories.

    \item\label{G-MPSP:Step2} Define the present state as $t = t_0$ and the desired state as
    $t = t_f$.

    \item\label{G-MPSP:Step3} Propagate the system dynamics given by \ref{Eqn:sys_dyn} using
    $U^0(t)$ until to get the final state of the system dynamics $X(t_f)$
    and compute output $Y(t_f)$. Required output $Y^*(t_f)$ is known. Therefore, $\delta Y(t_f)$ can
    also be computed.

    \item\label{G-MPSP:Step4} If either element of $\delta Y(t_f)$ is more than the desired limit,
then go to next step. If not, stop. Use this converged solution
for guidance which will be used as the guidance command. This step
represents the end of prediction mode.

    \item\label{G-MPSP:Step5} Compute the update matrix
$W(t)$ at each time step $t$ using a numerical integration scheme
(e.g. either the forward Euler or any other more accurate method
such as the forth order Runge-Kutta
(RK4)) using \ref{Eqn:Wdot_update}
and the final value \ref{Eqn:Wdot_tf}.

    \item\label{G-MPSP:Step6} Use the matrix $W(t)$
to compute the matrix $B_c(t)$ using \ref{Eqn:Bc_deriv}.

    \item\label{G-MPSP:Step7} Once $B_c(t)$ is
    computed, $A_{\lambda}$ and $b_{\lambda}$ can be computed using \ref{Eqn:A_lamda_c}
and (\ref{Eqn:b_lamda_c}) respectively.

    \item\label{G-MPSP:Step8} Compute $\delta U(t)$ and new
control $U(t)$ using equation (\ref{Eqn:delUc_derv_mod2}) and
(\ref{Eqn:total_Uc}) respectively. Prepare for the next iteration
by assigning $U^0(t) = U(t)$ and go to step \ref{G-MPSP:Step3}.
This step represents the end of correction mode.
\end{enumerate}
\section{Results and Discussions}
G-MPSP is generalized form of MPSP formulation which eliminates the necessity of using discretized form of system equation. Similar problem formulation as in MPSP (Refer section: \ref{MPSP_FORMULATION}) is used for G-MPSP solution. The problem objective is that deputy satellite states should track the desired relative orbit state with minimum terminal error that is $\mathbf X \to \mathbf {X_d}$ (in G-MPSP formulation frame work, $\delta \mathbf{Y(t_f)} \to 0$) and minimization of control effort.
A finite time SDRE solution presented in the section \ref{Finite_Time_SDRE} with SDC1 system model is used as comparative method for G-MPSP results. Similar to MPSP simulation infinite time LQR solution is used as initial guess history for G-MPSP algorithm.

Simulation results for satellite formation reconfiguration problem with $J_2$ perturbation model is presented in this section. The initial and desired orbital parameters for the deputy satellite and orbital parameters of the chief satellite are presented in the Table \ref{Initial_condition_GMPSP} and \ref{ORB_PAR_GMPSP} respectively.

\begin{figure}
  \centering
  \includegraphics[width=4.5in]{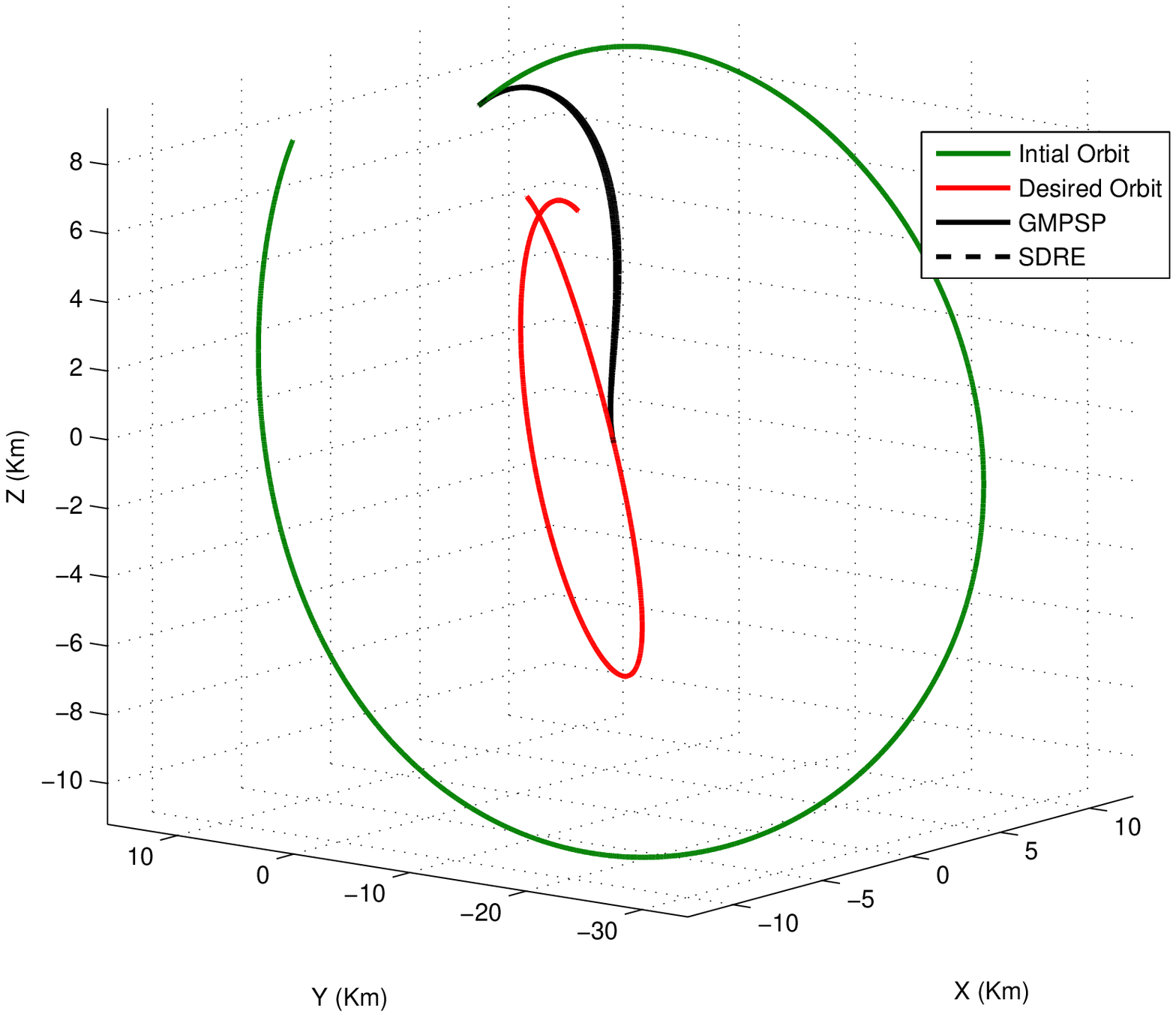}
  \caption{Satellite Orbit transfer trajectory plot,for guess LQR, G-MPSP and SDRE for Circular chief satellite orbit}
  \label{GMPSP_TRAJ}
\end{figure}

\begin{figure}
  \centering
  \includegraphics[width=4in]{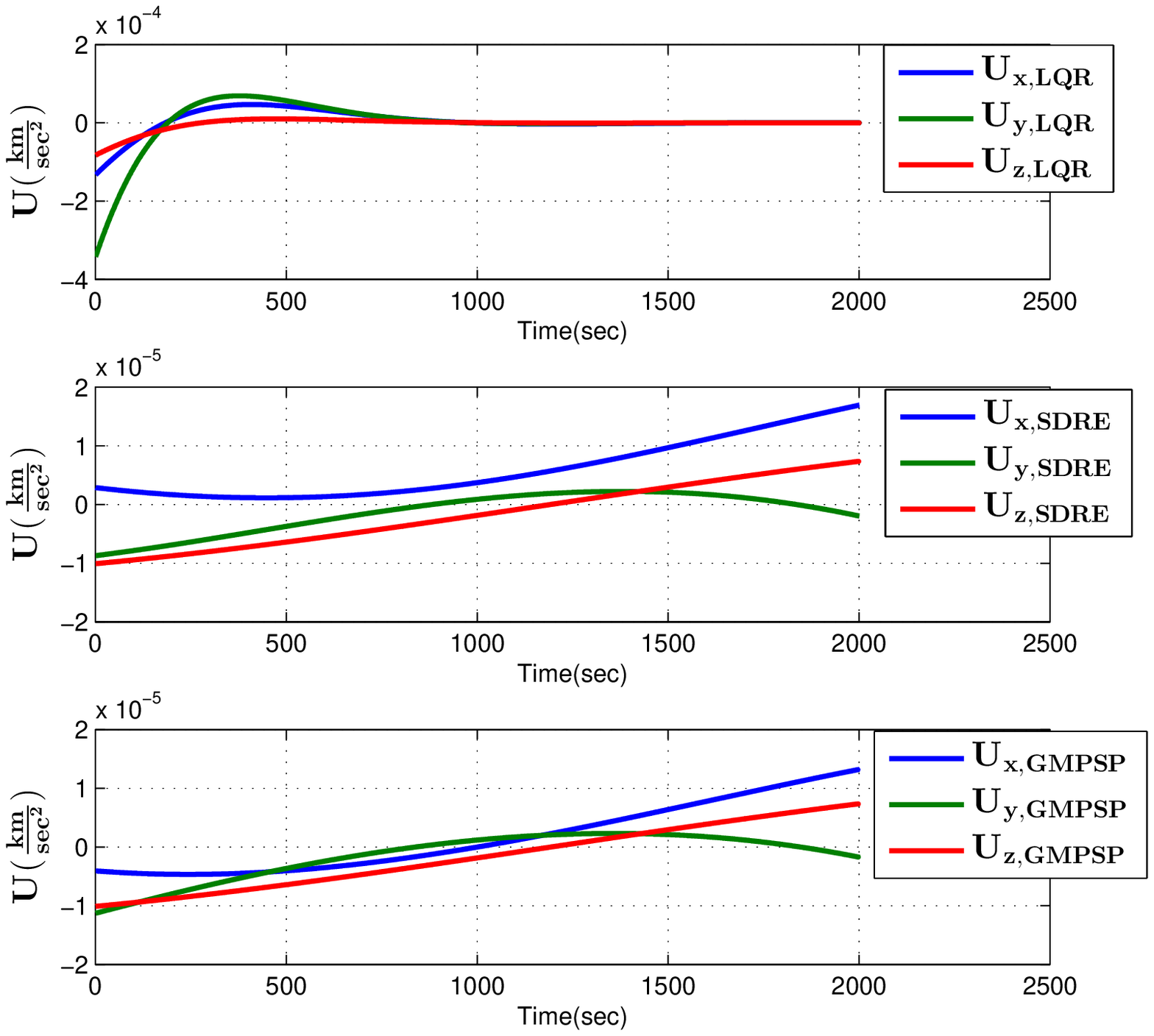}
  \caption{Control History for Guess control LQR and subsequent Updated G-MPSP controls and SDRE solution with $J_2$ perturbation}
  \label{CNTRL_GMPSP}
\end{figure}

\begin{figure}
  \centering
  \includegraphics[width=6.5in]{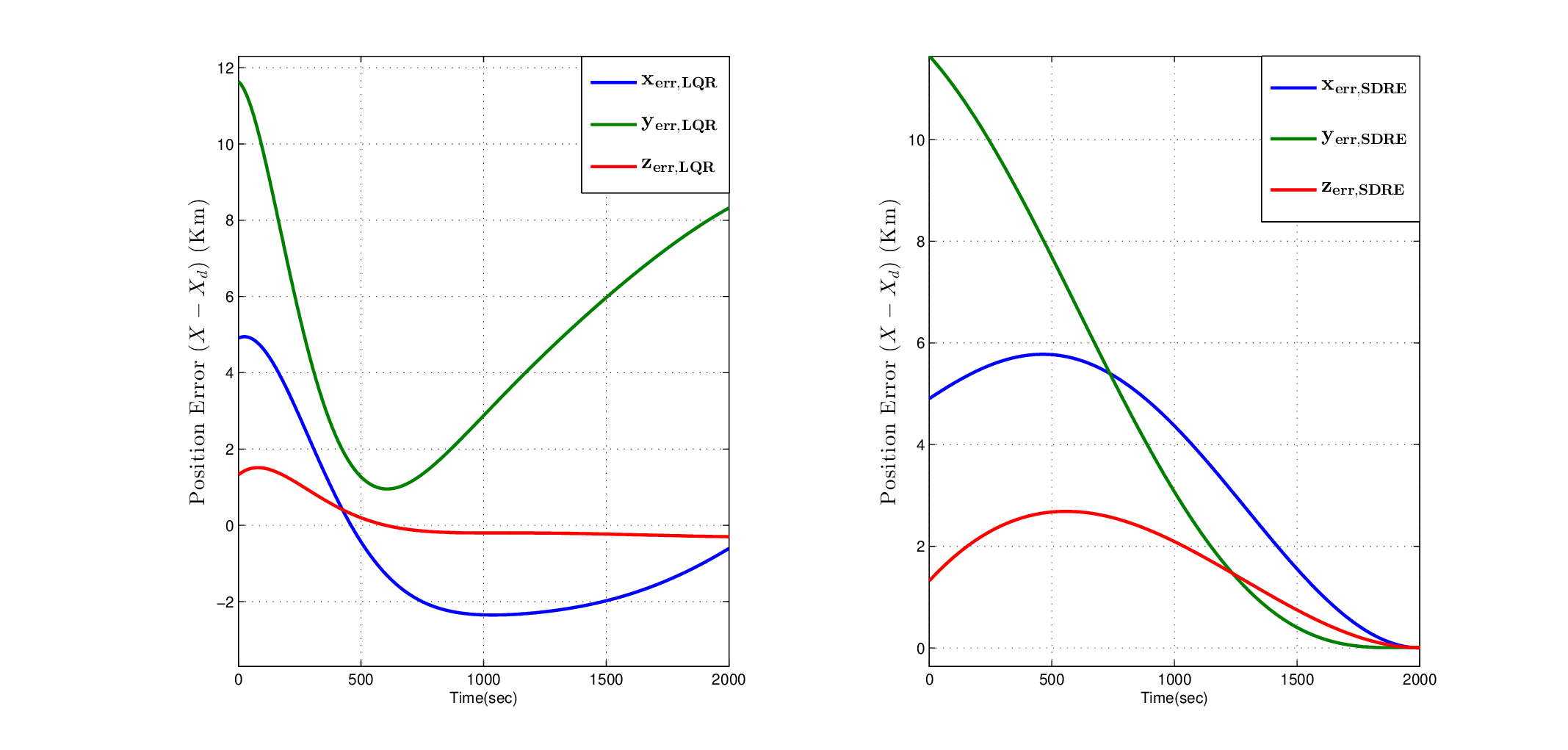}
  \caption{Position Error for Initial Guess solution LQR and SDRE with $J_2$ perturbation}
  \label{POS_ERR_LQR_SDRE_J2_GMPSP}
\end{figure}

\begin{figure}
  \centering
  \includegraphics[width=4.2in]{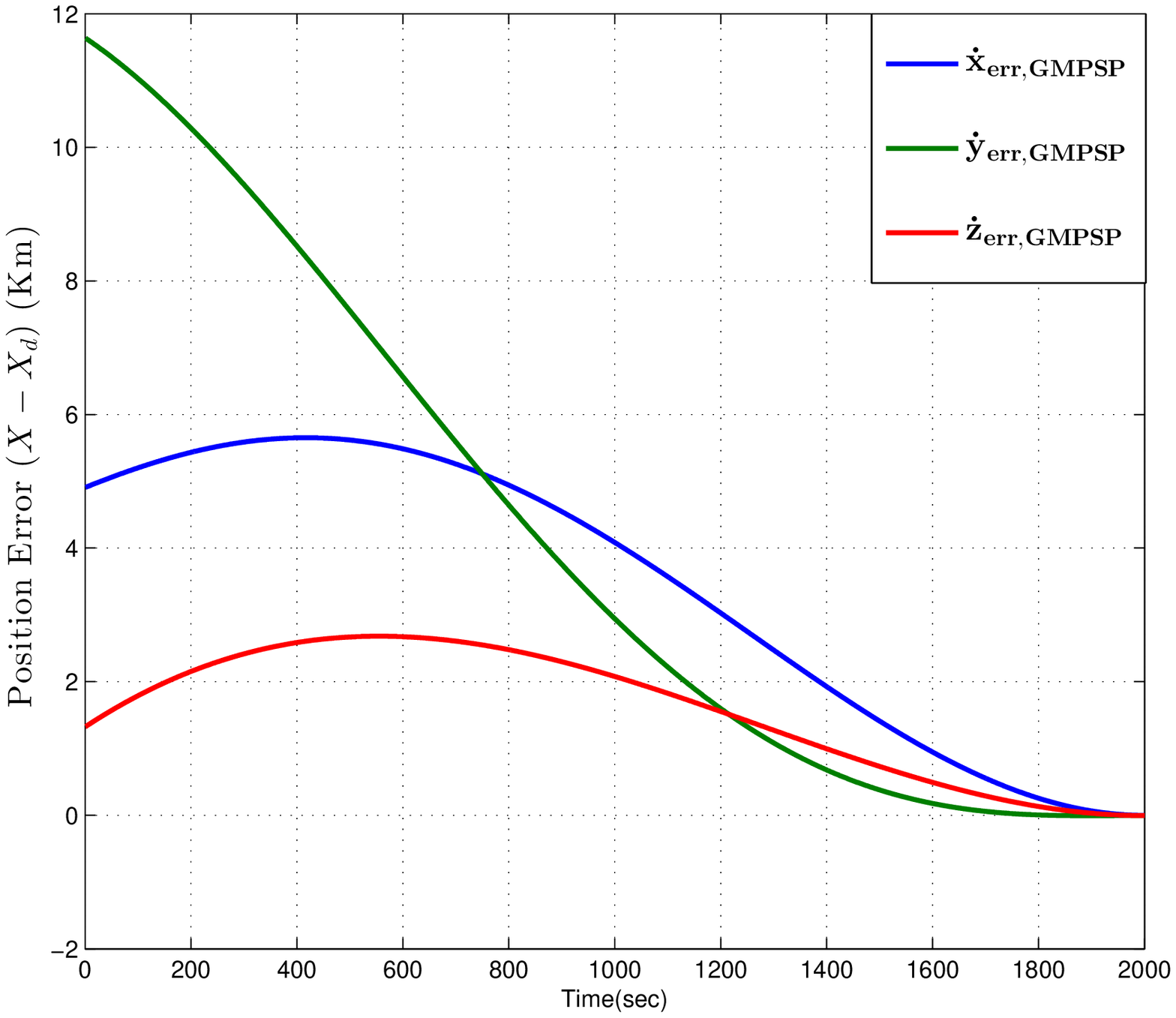}
  \caption{Position Error for G-MPSP final iteration with $J_2$ perturbation (Iteration No. 10) }
  \label{POS_ERR_MPSP_J2_GMPSP}
\end{figure}

\begin{figure}
  \centering
  \includegraphics[width=6.7in]{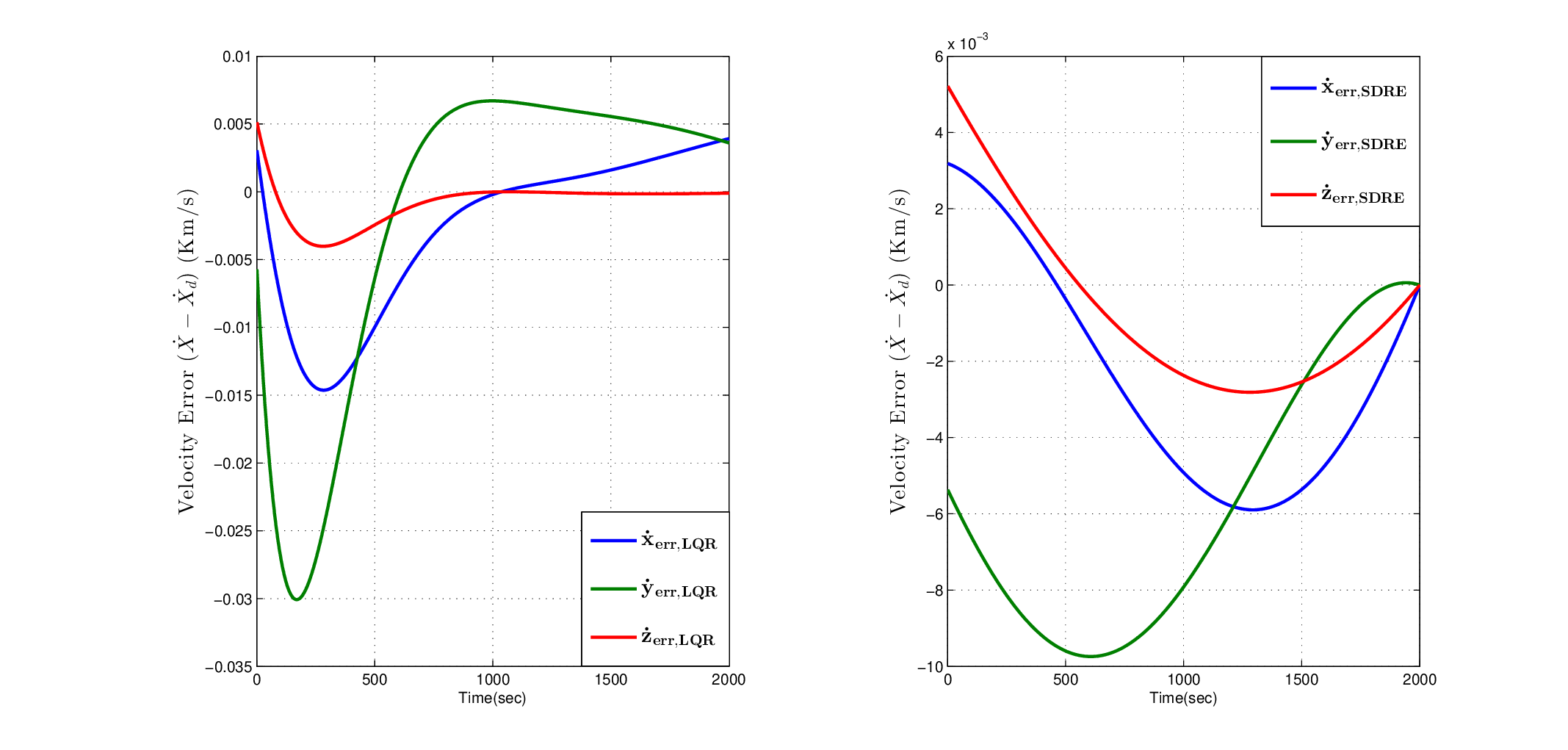}
  \caption{Velocity Error for Initial Guess solution LQR and SDRE with $J_2$ perturbation}
  \label{VEL_ERR_LQR_SDRE_J2_GMPSP}
\end{figure}

\begin{figure}
  \centering
  \includegraphics[width=4.2in]{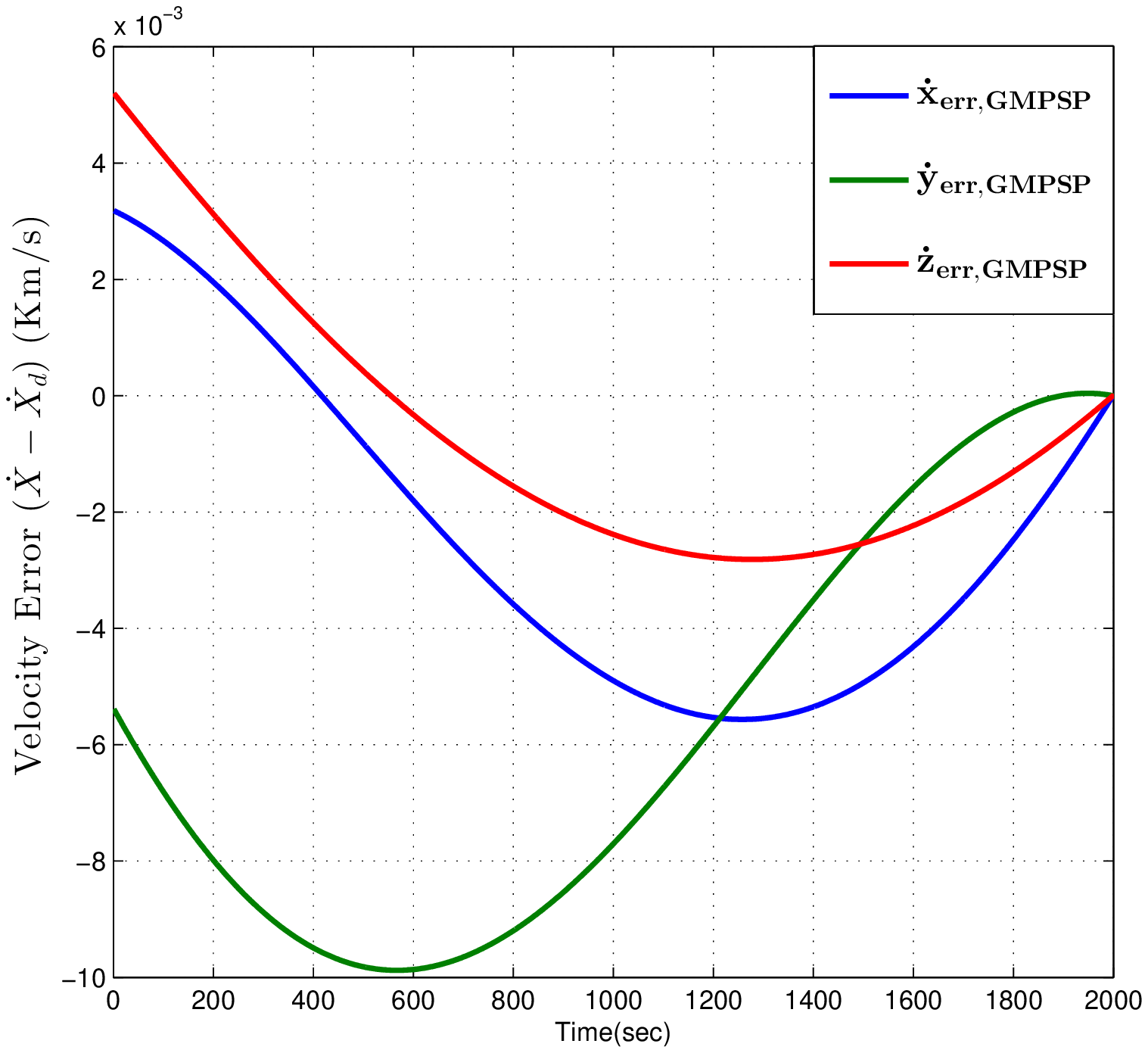}
  \caption{Velocity Error for G-MPSP final iteration with $J_2$ perturbation (Iteration No. 10)}
  \label{VEL_ERR_MPSP_J2_GMPSP}
\end{figure}

The Figure \ref{GMPSP_TRAJ} gives the detail of formation reconfiguration trajectory. The deputy satellite is initially in the relative orbit with $\rho = 10 km$, the satellite is commanded to move into a closer formation separation of $\rho = 2.5 km$, the solid black line and dotted black line in Figure \ref{GMPSP_TRAJ} shows the reconfiguration trajectory of the deputy satellite computed using G-MPSP and SDRE technique respectively. Figure \ref{CNTRL_GMPSP} shows the control history for guess control (LQR), G-MPSP updated control history for final iteration and SDRE control respectively. The position error and velocity error achieved by G-MPSP is significantly lesser than the initial guess control LQR state errors and SDRE state error. The position and velocity error history plot are given in Figure \ref{POS_ERR_LQR_SDRE_J2_GMPSP}, \ref{POS_ERR_MPSP_J2_GMPSP} and \ref{VEL_ERR_LQR_SDRE_J2_GMPSP}, \ref{VEL_ERR_MPSP_J2_GMPSP}. Table \ref{table:COMP_GMPSP} gives the details of the terminal error comparison of three method LQR, SDRE and G-MPSP. The iteration is stopped once the terminal state errors are within the tolerance limit of $\rho_e < 1\%$
\begin{table}
\caption{\small{Deputy Satellite Initial condition for G-MPSP solution}}\label{Initial_condition_GMPSP}
\begin{center}\begin{tabular}{|c|c|c|}
    \hline
    \textbf{Orbital} & \textbf{Initial Value}&\textbf{Final Value}\\
    \textbf{Parameters} &                     &\\
    \hline
    $\rho (km)$ & $10 km$& $2.5 km$\\
    \hline
    $\theta (deg) $& $45^0$ & $60^0$\\
    \hline
    $a (km)$& $0$ & $0$\\
    \hline
    $b (km)$& $0$ & $0$\\
    \hline
    $m$ (slope)& $1$ & $1.5$\\
    \hline
    $n$(slope)& $0$ & $0$\\
    \hline
\end{tabular}\end{center}
\end{table}
\begin{table}
\caption{\small{Chief Satellite Orbital Parameters, (G-MPSP)}}\label{ORB_PAR_GMPSP}
\begin{center}\begin{tabular}{|c|c|}
    \hline
    \textbf{Orbital Parameters} & \textbf{Value}\\
    \hline
    \textbf{Semi-major axis} & $10000 km$ \\
    \hline
    \textbf{Eccentricity} &  $e=0.1$ \\
    \hline
    \textbf{Orbit Inclination} & $60$ \\
    \hline
    \textbf{Argument of Perigee} & $0$ \\
    \hline
    \textbf{Longitude of ascending node} & $0$ \\
    \hline
    \textbf{Initial True Anomaly} & $10$ \\
    \hline
\end{tabular}\end{center}
\end{table}
\begin{table}
\caption{LQR, G-MPSP and SDRE STATE ERROR ($\rho_{final} = 2.5km$ with $J_2$ perturbation)(Position errors are in ``$km$`` and velocity in ``$\frac{km}{sec}$``)} 
\centering 
\begin{tabular}{|c|c|c|c|} 
\hline
\textbf{Error}  &  \textbf{LQR} &  \textbf{SDRE} & \textbf{G-MPSP}\\
   \textbf{in States}             &                &   &\\

\hline 

$x$ & 8.321 & 0.01204 & 0.0035  \\
\hline
$\dot{x}$ & 0.003919 & -1.129e-06 & -1.326e-06  \\
\hline
$y$ & -0.6062 & 0.00318 & -0.0002024\\
\hline
$\dot{y}$ &0.003616 &  -4.74e-06 & 1.289e-06\\
\hline
$z$ &-0.299& 0.006446 & -0.00428  \\
\hline
$\dot{z}$ &-9.95e-05 & -4.6e-06 & 8e-06 \\ 
\hline 
\end{tabular}
\label{table:COMP_GMPSP} 
\end{table}
\section{Summary and Conclusions}
In this chapter theoretical details of G-MPSP method is presented. The problem of satellite formation reconfiguration using G-MPSP suboptimal control is considered the problem formulation is similar to MPSP method. Simulation results pertaining to scenario of the reconfiguration where the chief and deputy satellite are brought into close formation from a larger base-line length separation is experimented. The results of the simulation are presented and it can be inferred that like MPSP method G-MPSP algorithm achieves the reconfiguration with very minimum terminal state errors. G-MPSP method has advantage over MPSP with no requirement of writing the system dynamics in discrete form. It is concluded that like MPSP method G-MPSP is successful in synthesizing the control for formation flying of satellites under perturbing effects of $J_2$ gravitational forces, and yet achieve fine tracking of the desired orbit. In light of these results MPSP and G-MPSP forms the most suited control logics which can be implemented in rendezvous mission where meeting close tolerance in formation is the key to success. In the next chapter a Robust control logic for SFF problem is presented.

\cleardoublepage
\chapter{Robust Satellite Formation Flying}
The need for nonlinear controller has become a necessity, since
present day missions demanding higher inter satellite separation
and eccentric chief satellite orbits. In such cases linear
controller fail to meet the mission objective to transfer or
maintain the deputy satellite in the desired orbit. Most common
controller used in small satellite mission owing to their limited
computation capability is LQR. LQR controller are not suitable for
mission involving eccentric orbits, higher formation distance and
external disturbance such as J2 perturbation. The novelty of this
work is neural network augmented LQR controller, where the neural
network approximates the nonlinearity of the plant (due to
eccentric chief satellite orbit and large baseline separation) and
also the exogenous disturbance terms due gravitational
perturbation of the oblate earth (J2 perturbation). A control term
is computed taking into consideration the approximated disturbance
terms, which along with the LQR control adds up to form the total
control term which is applied to meet the desired formation
mission objective. The key benefit of the idea is the small
satellite can continue to implement LQR controller, complectly
neglecting the nonlinear plant and $J2$ perturbation model but
augmenting with the proposed neural network structures ensures that it acts as a robust nonlinear optimal controller.

One set of networks, called as $NN_1$, is used for driving the LQR
controller towards the optimal control for the nonlinear system.
The other set of networks, called as $NN_2$, is used to capture
the unmodeled dynamics (including slowly-varying external
disturbance terms), thereby improving the plant model and helping
the $NN_1$ in the process. Both sets of neural networks are
trained online using `closed form expressions' and do not require
any iterative process. This technique is subsequently applied to the
challenging problem of satellite formation flying. Simulation
studies show that the presented control synthesis approach is able
to ensure close formation flying catering for large initial
separation, high eccentricity orbits, uncertain semi-major axis of
chief satellite and $J2$ gravitational effects, which is usually
considered as an exogenous perturbation.
\section{Generic Problem Formulation}
This section gives the generic problem formulation for class of
problem of the form,
\[\mathbf {\dot X} = f(\mathbf X) + B \mathbf U + d'(\mathbf X)\]where $d'(\mathbf X)$ denotes the
disturbance external to the system. We can re-write the above
equation as
\begin{equation}
\label{GENERIC}
 \mathbf \dot X = A\mathbf X+B\mathbf U+d(\mathbf X)
\end{equation}
where, $d(\mathbf X) = (f(\mathbf X) - A\mathbf X + d'(\mathbf X))$.\\
Here $d(\mathbf X)\in{\Re}^n$ is the total uncertainty term in the system.
The control synthesis to system of the form \ref{GENERIC} is
explained in following section (Refer \ref{Dynamic
Re-optimization}). Aim of the controller is to minimize the state
deviation with minimum control effort, cost function considered
for this purpose is as follows.
\begin{equation}
\label{COST_FUNCTION}
 J = \frac{1}{2}\int\limits_0^\infty  {\left(
{(\mathbf X-\mathbf X_d)^T{Q}{(\mathbf X-\mathbf X_d)} + \mathbf U^T{R}{\mathbf U}} \right)} dt
\end{equation}
where, $\mathbf X_d$ is state vector for desired final orbit.
\section{Control Synthesis Structure}
\label{Dynamic Re-optimization} \quad In this section the
methodology used for online optimization of LQR controller is
described. The collective uncertainty due to omitted algebraic
terms in linearization process and external disturbance is written
in terms of a lumped up term which is denoted as un-modeled
dynamics in the rest of the chapter. SFF falls to class of problem
where control $\mathbf U$ is not associated with system state $\mathbf X$, i.e. $B$
is constant matrix. Hence it is assumed that the unknown function
is dependent on state alone and is not a function of the control.
\subsection{Basic Philosophy}
\quad This section explains the philosophy behind the working of
online optimization method of LQR controller. There are a total of
two neural networks along with the LQR controller involved in the
method.
\begin{enumerate}
\item LQR controller: LQR controller operates on the linear plant
model. For input of state $X_k$ the LQR controller block gives the
co-state value $\lambda_{1,k+1}$ \item $NN_{1}$: These networks
approximate the additional costate required based on the
information given by the online training algorithm. The networks
used are Radial Basis Function Neural Networks (RBFNN). \item
$NN_{2}$: These networks approximate the un-modeled dynamics
which is crucial for online training and the weights are used for
computation of the partial derivative of the un-modeled dynamics
with respect to $X$ . They use the channel wise error information
in the state for training. These are single layer networks with
basis as the terms in plant dynamics, details in
\ref{Sim_Studies}.
\end{enumerate}
The net costate for $k^{th}$ time step is given by
\begin{eqnarray}\label{eq:COSTATEBRK}
    \lambda_{k} = \lambda_{1,k}+\lambda_{2,k}
\end{eqnarray}
where, $\lambda_{1,k}$ is the output from LQR and $\lambda_{2,k}$
is the output of $NN_{1}$ (RBF network). The control is evaluated
using $\lambda_{k}$ through optimal control equation.
\subsection{LQR Controller}
 The LQR operates on linearized plant model given as follows,
\begin{equation}
\label{eqn:24} \mathbf \dot X = A \mathbf X + B \mathbf  U
\end{equation}
where, system matrices $A$ and $B$ are given in chapter \ref{LQR_chap}
Cost function considered for LQR solution is given in
\ref{COST_FUNCTION}. Co-state is evaluated using $\mathbf \lambda =
P(\mathbf  X-\mathbf X_d)$ where $P$ is Ricatti coefficient obtained from solution of Ricatti equation \ref{Ricatti_EQN}
\begin{equation}
\label{Ricatti_EQN}
PA{\rm{  +  }}{A^T}P{\rm{ }} + {\rm{ }}Q{\rm{ }} - {\rm{ }}PB{R^{ - 1}}{B^T}P{\rm{  = 0}}
\end{equation}
Optimal control $\mathbf U$ is evaluated using $\mathbf U = -R^{-1}B^{T}P(\mathbf X-\mathbf X_d)$
(Refer \cite{bryson_book}). $P$ is constant matrix evaluated off-line
and stored as gain value to be used online.
\subsection{NN1 network synthesis and weight update rule}
\quad $NN1$ is a Radial basis function network consisting of input
layer, output layer and single hidden or intermediate layer.
Gaussian function is selected as the basis for the network. The
response of RBF NN1 network is given as $\lambda_{2,k+1} =
W_{c}^T\phi_c(\mathbf X_k)$ where $W_{c}$ are weights and $\phi_c(\mathbf X_k)$
is basis function (Gaussian function). Weight update rule for NN1
network in derived from the error cost function minimization. A
cost function of form
\begin{equation}
J_{NN1} = \frac{1}{2}{\left( {W_c^{*T}{\phi _c}({\mathbf X_k}) -
W_c^T{\phi _c}({\mathbf X_k})} \right)^T}\left( {W_c^{*T}{\phi _c}({\mathbf X_k})
- W_c^T{\phi _c}({\mathbf X_k})} \right)\nonumber
\end{equation}
\begin{equation}
+ \frac{1}{2}{\left( {W_c - {W_p}} \right)^T}{R_1}\left( {W_c -
{W_p}} \right)\nonumber
\end{equation} is formed. The term $W_p$ is the stored
weight from previous iteration. Term $W_c^{*T}{\phi _c}({\mathbf X_k}) =
\mathbf {\lambda_{2,k+1}^t}$ are target values for NN1 network.
Differentiating the above equation with respect to $W_c$ and
equating to zero we obtain the expression for weight update rule
for NN1 network.
\begin{equation}
\label{NN2_Weight} {W_c} = \frac{{\left( {W_c^{*T}{\phi
_c}({\mathbf X_k})\phi _c^T({\mathbf X_k}) + W_p^T{R_1}} \right)}}{{\left( {\phi
_c^T({\mathbf X_k}){\phi _c}({\mathbf X_k}) + {R_1}I} \right)}}
\end{equation}
$R_1$ is the weight on error $W_c-W_p$ and $I$ is identity matrix
of size $(\phi _c^T({\mathbf X_k}){\phi _c}({\mathbf X_k}))$
\subsection{NN2 Network synthesis and weight update rule}
$NN2$ neural network is designed to capture the un-modeled
dynamics of the plant, and help $NN1$ network to come up with the
extra co-state term needed addition to LQR co-state value. Control
evaluated from total value of the co-state caters to the perturbed
system model.\\
\quad The actual plant model can be written as follows.
\begin{eqnarray}\label{eq:ACTPLNT}
    \mathbf \dot X = A\mathbf X+B\mathbf U+d(\mathbf X),\>\>\mathbf X(0)=\mathbf X_0
\end{eqnarray}
Here $d(\mathbf X)\in{\Re}^n$ is the un-modeled dynamics term. A Virtual
plant whose states are $\mathbf X^a$ is created and the dynamics of the
virtual plant is given as
\begin{eqnarray}\label{eq:VRTPLNT}
    \mathbf \dot X^a = A\mathbf X+B\mathbf U+\hat d(\mathbf X)+K_{\tau}(\mathbf X-\mathbf X^a)
\end{eqnarray}
The $\hat d(\mathbf X)$ is an approximation of the actual function $d(\mathbf X)$
and $K_{\tau}$ is a Hurwitz matrix which contains the desired time
constants, it is desired that virtual plant should track the
actual plant. Error term can be defined as
\begin{eqnarray}\label{eq:ERROR}
    E = \mathbf X-\mathbf X^a
\end{eqnarray}
The error dynamics can be obtained by differentiating the above
equation with time and substituting~\ref{eq:ACTPLNT} and
~\ref{eq:VRTPLNT}
\begin{eqnarray}\label{eq:ERRORDOT}
    \dot E &=& \mathbf \dot X-\mathbf \dot X^a \nonumber\\
    \dot E &=& d(\mathbf X) -\hat d(\mathbf X)-K_{\tau}E
\end{eqnarray}
Error is decomposed into individual channels as $e_i =
x_i-x^a_i$. The $i^{th}$ channel error dynamics is given as
\begin{eqnarray}\label{eq:ERRORDOTI}
    \dot e_i &=& \dot x_i-\dot x^a_i, \>\>i=1,2,\ldots,n\nonumber\\
    \dot e_i &=& d_i(\mathbf X) -\hat d_i(\mathbf X)-k_{\tau_i}e_i
\end{eqnarray}
Single layer neural network with nonlinear basis functions is
chosen to approximates the un-modeled dynamics $d_i(X)$ in the
$i^{th}$ channel.
\begin{eqnarray}\label{eq:DHAT}
    \hat d_i(\mathbf X) = \hat{W_i}^T \Phi_i(\mathbf X),\>\>W_i\in \Re^p
\end{eqnarray}
where, $\hat W_i$ are the weights and $\Phi_i(\mathbf X)$ are the basis.
Lets consider there exists an ideal approximator for the unknown
function which approximates $d_i(\mathbf X)$ with an ideal approximation
error $\epsilon_i$ for the chosen basis $\Phi_i(\mathbf X)$.
\begin{eqnarray}\label{eq:D}
    d_i(\mathbf X) = {W_i}^T \Phi_i(\mathbf X)+\epsilon_i
\end{eqnarray}
The weights $W_i$ are the ideal weights which are unknown. Channel
wise error dynamics can be written as
\begin{eqnarray}\label{eq:ERRORDOTIW}
    \dot e_i &=& {W_i}^T \Phi_i(\mathbf X)+\epsilon_i-\hat{W_i}^T \Phi_i(\mathbf X)-k_{\tau_i}e_i
\end{eqnarray}
The error in weights of the $i^{th}$ approximating network is
defined as
\begin{eqnarray}\label{eq:ERRORW}
    \tilde{W_i} &=& {W_i}-\hat{W_i}\\
    \dot{\tilde{W_i}} &=&
    -\dot{\hat{W_i}},\;\;\;\;{{W_i}}=constant
\end{eqnarray}
Aim is that weights of the approximating networks $\hat{W_i}$
should approach the ideal weights ${W_i}$ asymptotically, i.e.,
\begin{eqnarray}
    \tilde{W_i} \rightarrow 0\;\;\; as \;\;\;t\rightarrow
    \infty\nonumber
\end{eqnarray}
The un-modeled information is stored in terms of the weights of
the approximating networks. A Lyapunov approach is discussed in
the next subsection for updating $\hat{W_i}$ online.

\subsection{Lyapunov Analysis and Weight Update Rule}
\quad The choice of Lyapunov function candidate is a important
part of any Lyapunov analysis \cite{Slotine}. There are three quantities whose
asymptotic stability are to be guaranteed,
\begin{enumerate}
\item $e_i$, the $i^{th}$ channel error

\item $\tilde{W}_i$, the error in $i^{th}$ network weights

\item $\left[\frac{\partial d_i(X)}{\partial X} - \frac{\partial
\hat{d_i}(X)}{\partial X}\right]$, the error in $i^{th}$ unknown
function partial derivative
\end{enumerate}
The positive definite Lyapunov function candidate is
 \begin{eqnarray}\label{eq:LYPN1}
 V_i(e_i, \tilde{W_i})&=& \beta_i \frac{e^2_i}{2} + \frac{\tilde{{W_i}^T} \tilde{W_i}}{2\gamma_i}+\left[\frac{\partial d_i(X)}{\partial X} - \frac{\partial \hat{d_i}(X)}{\partial
 X}\right]^T \nonumber \\
 & & \times \frac{\Theta_i}{2}\left[\frac{\partial d_i(X)}{\partial X} - \frac{\partial \hat{d_i}(X)}{\partial
 X}\right]
 \end{eqnarray}
where, $\beta_i$, $\gamma_i$ $\Theta_i$ are positive definite
quantities.
\begin{eqnarray}\label{eq:LYPN2}
V_i(e_i, \tilde{W_i}) &=& \beta_i \frac{e^2_i}{2} +
\frac{\tilde{{W_i}^T} \tilde{W_i}}{2\gamma_i}+ \tilde{{W_i}^T}
\left[\frac{\partial \Phi_i}{\partial X}\right] \frac{\Theta_i}{2}
{\left[\frac{\partial \Phi_i}{\partial X}\right]}^T\tilde{W_i}\\
\end{eqnarray}
The partial derivatives of the Lyapunov function are
\begin{eqnarray}\label{eq:LYPN1PD}
    \frac{\partial V_i}{\partial e_i} = \beta_i e_i \;\;\; ; \;\;\; \frac{\partial V_i}{\partial \tilde{W_i}} = \frac{\tilde{W_i}}{\gamma_i} + \left[\frac{\partial \Phi_i}{\partial X}\right] \Theta_i {\left[\frac{\partial \Phi_i}{\partial X}\right]}^T \tilde{W_i}
 \end{eqnarray}
Time derivative of Lyapunov function can be written as follows
 \begin{eqnarray}
    \dot{V_i} = \beta_i e_i \dot{e_i} - \frac{\tilde{W}^T_i \dot{\hat{W_i}}}{\gamma_i} - \tilde{W_i} \left[\frac{\partial \Phi_i}{\partial X}\right]  \Theta_i {\left[\frac{\partial \Phi_i}{\partial X}\right]}^T \dot{\hat{W_i}}
 \end{eqnarray}
Substituting the error dynamics from~\ref{eq:ERRORDOTIW} in
above equation
 \begin{eqnarray}
    \dot{V_i} &=& \beta_i e_i\{\tilde{W}^T_i \Phi_i(X) + \epsilon_i - k_{\tau_i} e_i\} - \frac{\tilde{W}^T_i \dot{\hat{W_i}}}{\gamma_i}\nonumber\\
    & & - \tilde{W_i} \left[\frac{\partial \Phi_i}{\partial X}\right]  \Theta_i {\left[\frac{\partial \Phi_i}{\partial X}\right]}^T
    \dot{\hat{W_i}}\nonumber\\
    \dot{V_i} &=& \beta_i e_i\ \tilde{W}^T_i {\Phi_i}(X) + \beta_i e_i \epsilon_i - \beta_i k_{\tau_i} e^2_i - \frac{\tilde{W}^T_i \dot{\hat{W_i}}}{\gamma_i}\nonumber\\
   & &  - \tilde{W_i} \left[\frac{\partial \Phi_i}{\partial X}\right]  \Theta_i {\left[\frac{\partial \Phi_i}{\partial X}\right]}^T \dot{\hat{W_i}}
 \end{eqnarray}
Collecting the coefficients of $\tilde{W}$ and equating the
coefficient of $\tilde{W^T_i}$ to zero following expression is
obtained
 \begin{eqnarray}
    \left(\frac{I_p}{\gamma_i} + \left[\frac{\partial \Phi_i}{\partial X}\right]
    \Theta_i {\left[\frac{\partial \Phi_i}{\partial X}\right]}^T
    \right)\dot{\hat{W_i}} &=&  \beta_i e_i \Phi_i(X) \nonumber
\end{eqnarray}
where $I_p$ is the identity matrix of dimension $p$. Inverting the
coefficient of $\dot{\hat{W_i}}$ lead to the weight update rule in
continuous time.
\begin{eqnarray}\label{eq:WEIGHTUPD}
    \dot{\hat{W_i}} &=& \beta_i e_i {\left(\frac{I_p}{\gamma_i} +
    \left[\frac{\partial \Phi_i}{\partial X}\right]  \Theta_i {\left[\frac{\partial
    \Phi_i}{\partial X}\right]}^T \right)}^{-1} \Phi_i(X)
\end{eqnarray}
The matrix $\bigg[
    \left[\frac{\partial \Phi_i}{\partial X}\right]  \Theta_i {\Big[\frac{\partial
    \Phi_i}{\partial X}\Big]}^T \bigg]$ is singular for $n<p$.
But the matrix is always positive definite, so adding
$\frac{I_p}{\gamma_i}$ will make the matrix nonsingular $\forall
(n,~p)\in \mathcal{N}$. The left over terms from Lyapunov
derivative $\dot{V_i}$ equation are
\begin{eqnarray}
    \dot{V_i} = \beta_i e_i \epsilon_i - k_{\tau_i} \beta_i e^2_i
\end{eqnarray}
For stable system $\dot{V_i} < 0$ \cite{Slotine}, which leads to a condition
\begin{eqnarray}
    |e_i| &>& \frac{|\epsilon_i|}{k_{\tau_i}}
\end{eqnarray}
The above conditions means that if the absolute error in the
$i^{th}$ channel exceeds the value in RHS then the Lyapunov
function becomes negative definite and positive definite
otherwise. Therefore, if the network weights are updated based of
the rule given in ~(\ref{eq:WEIGHTUPD}), then the identification
happens as long as absolute error is greater than certain value.
By increasing $k_{\tau_i}$ error bound can be theoretically
reduced.

\subsection{LQR and Online training}

\begin{figure}
  \centering
  \includegraphics[width=5in]{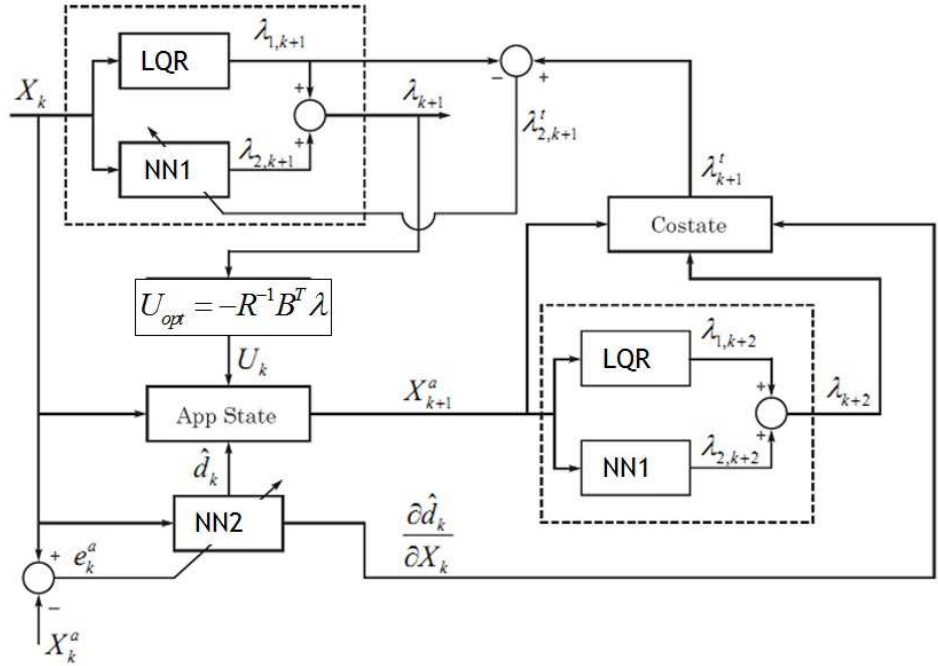}
  \caption{Neural Network Scheme used to train network}
  \label{NN_Training}
\end{figure}

\begin{enumerate}
\item LQR block and $NN_{1}$ are excited with the current state
$\mathbf X_k$ to obtain co-states $\lambda_{1,k+1}$ and $\lambda_{2,k+1}$
respectively. The LQR controller operates twice within LQR block
to calculate $\lambda_{1,k+1}$   \item $NN_{2}$ trained with the
error $e_i$ channel wise \item The Virtual state equation
in~\ref{eq:VRTPLNT} is propagated with $\mathbf X_k$ and $\mathbf U_k$ to obtain
$\mathbf X^a_{k+1}$, where $\mathbf \hat d(X)$ in given by $NN_{2}$ \item LQR and
$NN_{1}$ are again excited with the previously obtained state
$\mathbf X^a_{k+1}$ to obtain $\lambda_{1,k+2}$ and $\lambda_{2,k+2}$
respectively. $\lambda_{k+2}$ is computed using
~\ref{eq:COSTATEBRK} \item The costate equation  is propagated
backward with $X^a_{k+1}$, $\lambda_{k+2}$ using
~\ref{COSTATE_L} to obtain $\lambda^t_{k+1}$ \item
$\lambda^t_{2,k+1}$ is computed as the difference of
$\lambda^t_{k+1}$ and $\lambda_{1,k+1}$ which is the target value
for the training of $NN_{2}$
\end{enumerate}
\section{Results and Discussions}
\label{Sim_Studies} \quad The Extra co-state term in addition to
LQR is evaluated from the back propagation of the co-state
equation assuming the full knowledge of the approximated
disturbance term (from NN2 network which maps the actual
disturbance term). The sum of the costate values from LQR and
back propagation of co-state equation is used in the optimal
control equation to come up with the modified controller to make
the deputy satellite track the desired final orbit. \\
The Hamiltonian for evaluation of the co-state equation is as
follows.
\begin{equation}
{H_{opt}} = \frac{1}{2}\left( {{\mathbf X^T}Q\mathbf X + {\mathbf U^T}Q\mathbf U} \right) +
{\lambda ^T}\left( {A \mathbf X + B\mathbf U + d(\mathbf X)} \right)
\end{equation}
co-state equation is as follows
\begin{equation}
\label{COSTATE_L} \dot \lambda  =  - \frac{{\partial
{H_{opt}}}}{{\mathbf \partial X}}
\end{equation}
\quad For the problem presented in this paper six such co-state
derivatives are to be evaluated i.e. $ \left[
{\begin{array}{*{20}{c}} {{{\dot \lambda }_1}}&{{{\dot \lambda
}_2}}&{{{\dot \lambda }_3}}&{{{\dot \lambda }_4}}&{{{\dot \lambda
}_5}}&{{{\dot \lambda }_6}}
\end{array}} \right]$
For approximating the disturbance term,(Refer section \ref{Dynamic Re-optimization}) a judicious selection of the basis function is done as follows. 
Considering the term $\frac{\mu }{\gamma }$ and expanding using
the definition of $\gamma$ and simplifying the expression can be
written as follows.
\begin{equation}
\begin{array}{l}
\frac{\mu }{\gamma } = \frac{\mu }{{r_c^3}}\left( {{{\left[ {1 - \left( { - \frac{2}{{{r_c}}}x - \frac{{{x^2} + {y^2} + {z^2}}}{{r_c^2}}} \right)} \right]}^{ - \frac{3}{2}}}} \right)\\
{\rm{Defining }}\\
\psi {\rm{ =  }}\left( { - \frac{2}{{{r_c}}}x - \frac{{{x^2} + {y^2} + {z^2}}}{{r_c^2}}} \right)\\
\end{array}
\end{equation}
\quad Power series up to fourth power is considered as the basis
function ${\Phi _i}({\mathbf X_{act}}) = (\psi + {\psi ^2} + {\psi ^3} +
{\psi ^4}){x_1}$ and so on for other disturbance
terms. Trigonometric basis function is considered for J2
perturbation.

\begin{figure}
  \centering
  \includegraphics[width=5in]{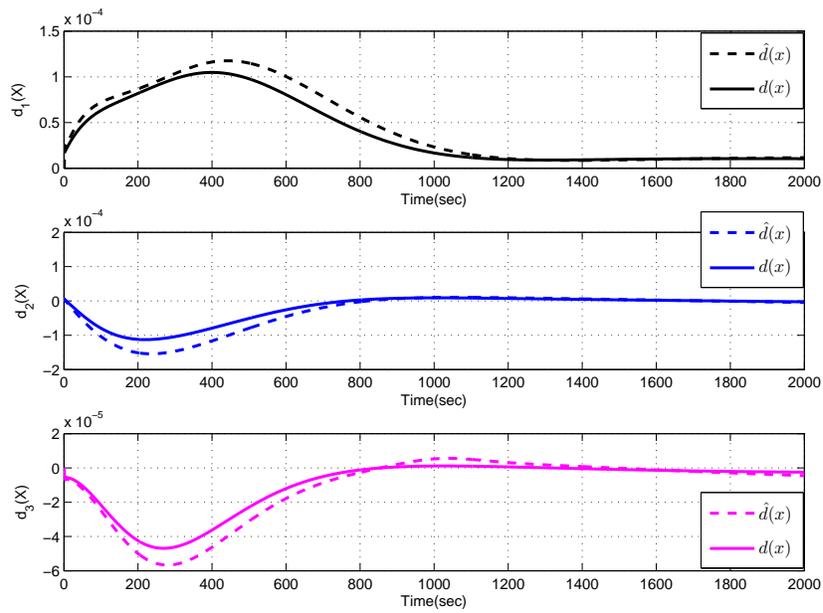}
  \caption{Actual $d(X)$ and Approximated Disturbances $\hat{d}(X)$}
  \label{Disturbance}
\end{figure}

\begin{figure}
  \centering
  \includegraphics[width=4in]{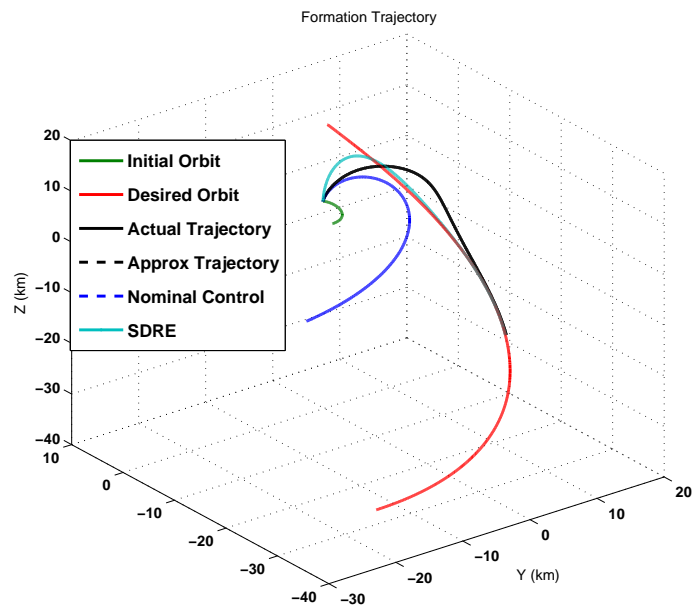}
  \caption{Satellite Orbit transfer trajectory plot: Case:1, Case:2 and Case:3}
  \label{TRAJ}
\end{figure}

\begin{figure}
  \centering
  \includegraphics[width=4.5in]{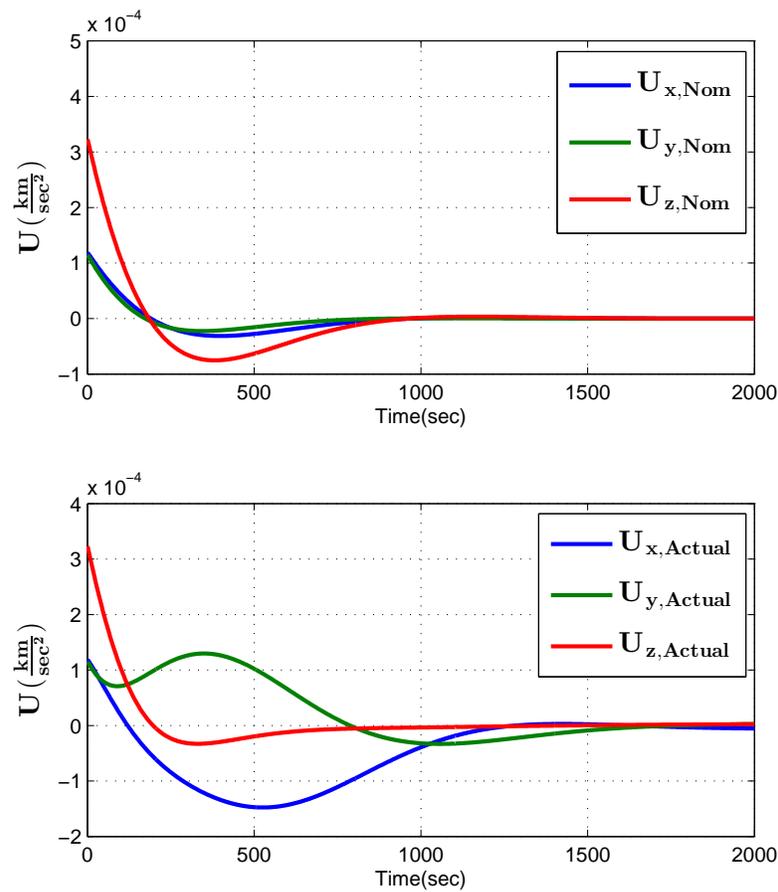}
  \caption{Control $U_x$,$U_y$,$U_z$- Nominal Control(LQR) and Actual Control(LQR+NN)}
  \label{Control}
\end{figure}

The objective of the problem statement is to form the
formation or to reconfigure the formation flying of satellites to
the desired orbit. Mathematically we can put the problem
objectives as follows, main objective is to minimize the terminal
state errors i.e. ${\left[ {\begin{array}{*{20}{c}}
{{x_1}}&{{x_3}}&{{x_5}}
\end{array}} \right]^T} \to {\left[ {\begin{array}{*{20}{c}}
{x_1^*}&{x_3^*}&{x_5^*}
\end{array}} \right]^T}$ at
$t = {t_f}$ . However, since the velocity components should also
match with the desired orbital parameters, one can also impose
${\left[ {\begin{array}{*{20}{c}} {{x_2}}&{{x_4}}&{{x_6}}
\end{array}} \right]^T} \to {\left[ {\begin{array}{*{20}{c}}
{x_2^*}&{x_4^*}&{x_6^*}
\end{array}} \right]^T}$ at $t = {t_f}$ , where ${\left[ {\begin{array}{*{20}{c}}
{x_2^*}&{x_4^*}&{x_6^*}
\end{array}} \right]^T}$ are the corresponding desired orbital
velocity parameters at the position ${\left[
{\begin{array}{*{20}{c}} {x_1^*}&{x_3^*}&{x_5^*}
\end{array}} \right]^T}$.
Initial and final relative orbital satellite of the deputy satellite is given in the Table \ref{Initial_condition_OMLQR}
\begin{table}[h!]
\caption{\small{Deputy Satellite Initial condition for online Optimized LQR solution}}\label{Initial_condition_OMLQR}
\begin{center}\begin{tabular}{|c|c|c|}
    \hline
    \textbf{Orbital} & \textbf{Initial Value}&\textbf{Final Value}\\
    \textbf{Parameters} &                     &\\
    \hline
    $\rho (km)$ & $0.5 km$& $5 km$\\
    \hline
    $\theta (deg) $& $45^0$ & $60^0$\\
    \hline
    $a (km)$& $0$ & $0$\\
    \hline
    $b (km)$& $0$ & $0$\\
    \hline
    $m$ (slope)& $1$ & $1.5$\\
    \hline
    $n$(slope)& $0$ & $0$\\
    \hline
\end{tabular}\end{center}
\end{table}
In this exercise, the chief satellite orbit is considered as $10,000km$ semi-major axis and zero eccentricity circular orbit for LQR control evaluation. The actual values of chief satellite orbit eccentricity and semi-major axis is considered as 0.5
and $11114.51658km$ respectively, which amounts to $50\%$ error in measured eccentricity and $10\%$ error in semi-major axis compared to data accounted for control synthesis using baseline controller(LQR). Figure \ref{Disturbance} illustrates the performance of three neural network used to approximate the nonlinearity of the plant and J2 perturbation disturbance term. Solid line denotes the Actual disturbance and dotted line signifies the neural network approximation of the corresponding disturbance term. Figure \ref{TRAJ} gives the trajectory plot for reconfiguration of the formation.Figure \ref{TRAJ} shows the The initial orbit, commanded orbit and satellite trajectory for given formation flying mission. The Figure \ref{TRAJ} includes the plot for , Actual plant (with Actual disturbance) with Actual Controller and approximate plant and actual Controller and SDRE solution reconfiguration trajectory. Figure \ref{Control} gives the details of the control in all three axis for nominal controller (LQR) and actual controller (LQR + NN).

In the simulation a online optimized LQR controller is compared with at SDRE controller. SDRE controller considered for comparison is assumed to operate on the plant with following information
\begin{itemize}
\item The complete SDC model of plant and $J_2$ perturbation model is considered in system matrices used for computing feedback gain using infinite time SDRE technique.
\item No uncertainty in eccentricity and semi-major axis of chief satellite, thats is actual values of $11114.51658km$ and $0.5$ for semi-major axis and eccentricity respectively is considered in plant and $J_2$ model.
\end{itemize}
Where as the base-line controller LQR considered for online optimization using neural networks
\begin{itemize}
\item Operates on linear plant with circular chief satellite orbit condition $e=0$ and chief satellite orbit radius vector of $10000 km$
\item No external perturbation like $J_2$ gravitational effect.
\end{itemize}
SDRE controller is used to demonstrate the capability of online optimized LQR controller. $LQR + NN$ controller actually in effect behaves as a nonlinear controller and the terminal accuracies achieved over tracking a commanded final orbit are quite close to a controller which operates on the nonlinear model of plant with no uncertainty in plant model and prior information of disturbance in the system.

\begin{figure}
  \centering
  \includegraphics[width=4.5in]{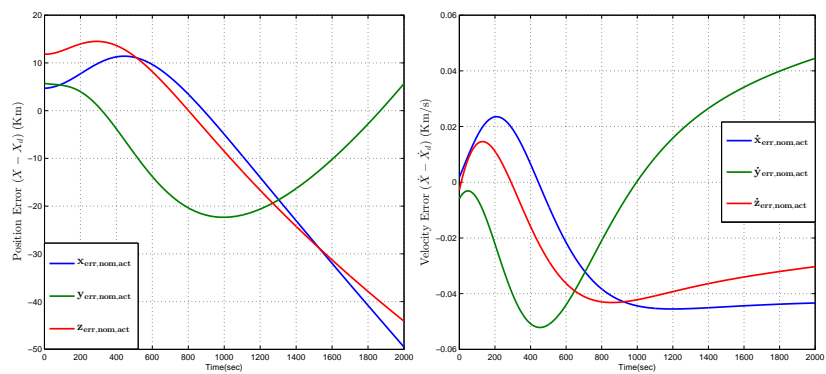}
  \caption{Satellite Position $X$ and Velocity $\dot X$ Error for Actual Plant + Nominal Controller (LQR)}
  \label{NOM_ACT_ERR}
\end{figure}

\begin{figure}
  \centering
  \includegraphics[width=5.5in]{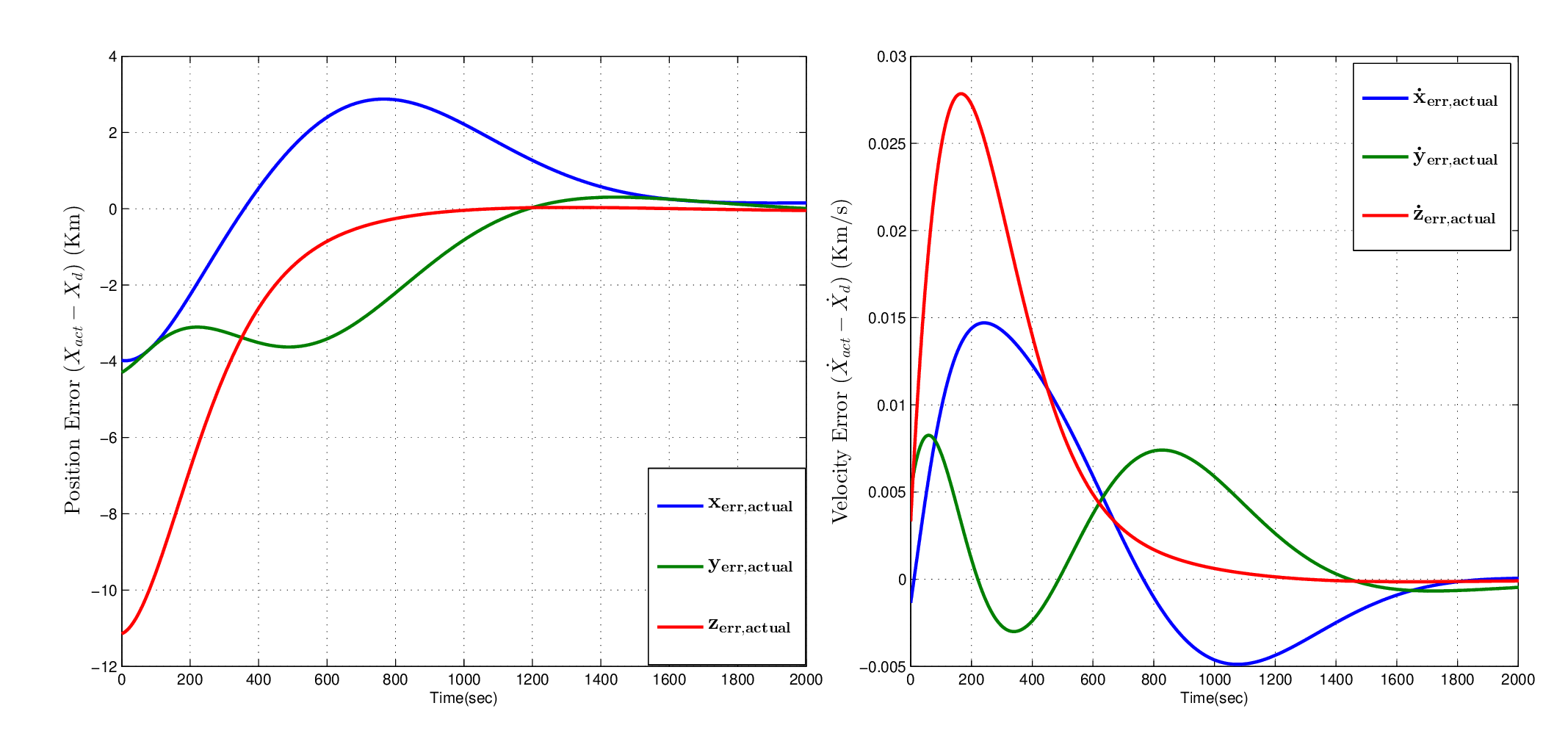}
  \caption{Satellite Position $X$ and Velocity $\dot X$ Error for Actual Plant + Actual Controller}
  \label{ACT_ERR}
\end{figure}

\begin{figure}
  \centering
  \includegraphics[width=5.5in]{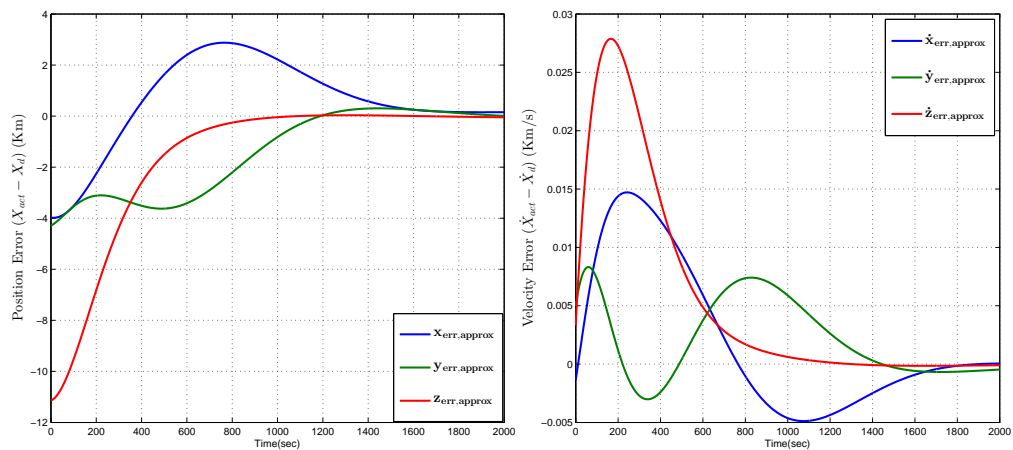}
  \caption{Satellite Position $X$ and Velocity $\dot X$ Error for Approximate Plant + Actual Controller}
  \label{APP_ERR}
\end{figure}

\begin{figure}
  \centering
  \includegraphics[width=5in]{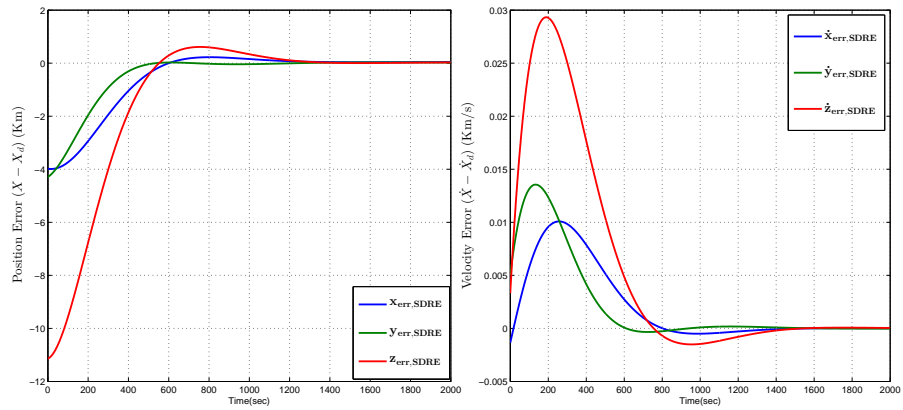}
  \caption{Satellite Position $X$ and Velocity $\dot X$ Error for Actual Plant + SDRE Controller}
  \label{SDRE_ERR}
\end{figure}

\begin{figure}
  \centering
  \includegraphics[width=4.5in]{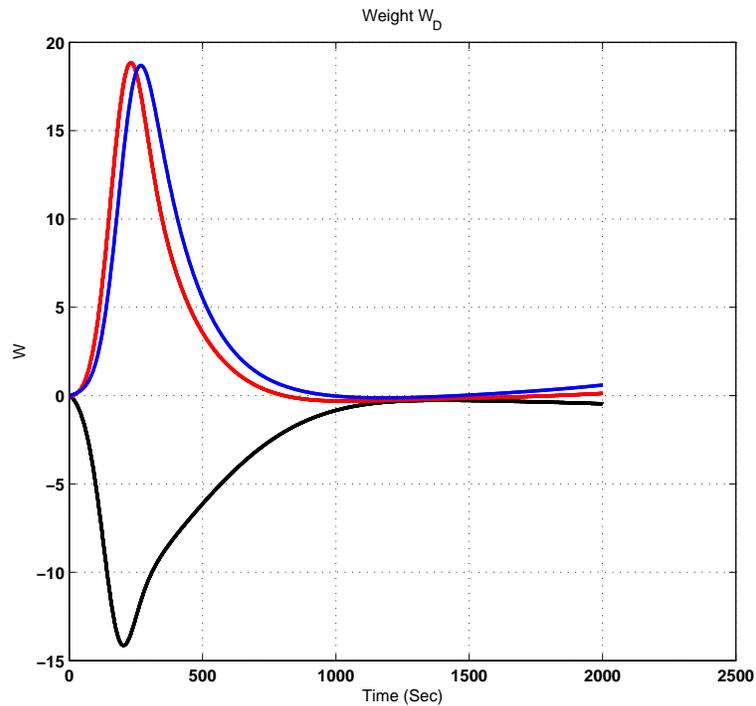}
  \caption{Neural Network Training Weights for disturbance $\hat d(X)$ capturing}
  \label{WEIGHT}
\end{figure}

Figure \ref{NOM_ACT_ERR}, \ref{ACT_ERR}, \ref{APP_ERR} and \ref{SDRE_ERR} gives the plot of position error and velocity error vs time. The Figure \ref{NOM_ACT_ERR} gives the details of state error for nominal controller applied to actual plant, \ref{ACT_ERR} presents the plot of state error of the actual plant model operated with online optimized LQR controller and similarly \ref{APP_ERR} presents the details for online optimized LQR controller implemented with approximate plant model. We can see the terminal state errors are quite close and hence can infer that nonlinear behavior of plant and $J_2$ perturbation are mapped very accurately. Figure \ref{SDRE_ERR} shows the plots of position error and velocity error for SDRE controller + Actual plant. The terminal accuracies of all three cases that is Actual plant + actual controller, approximate plant + actual controller and actual plant + SDRE controller are presented in Table. The neural network training weight for disturbance capturing are shown in the Figure \ref{WEIGHT}

\begin{table} [h!]
\caption{LQR,SDRE and MPSP state error(Eccentric chief satellite orbit solution, Final $\rho = 5\hspace{1mm}km$). (Position errors are in ``$km$`` and velocity in ``$\frac{km}{sec}$``)} 
\centering 
\begin{tabular}{|c|c|c|c|c|} 
\hline
\textbf{Error}  & \textbf{LQR} & \textbf{LQR + NN} & \textbf{LQR + NN}& \textbf{SDRE} \\ [0.5ex] 
\textbf{in States} &  \textbf{Actual Plant} &  \textbf{Actual Plant}    &  \textbf{Approx Plant} &   \textbf{Actual Plant}        \\
\hline 

$x$  & -49.51  & 0.1544 & 0.1543 & 0.04326  \\ 
\hline
$\dot{x}$ & -0.04335 & 5.158e-05& 5.16e-05&  1.56e-05   \\
\hline
$y$ & 5.665 & 0.007096 & 0.006627& 0.02603  \\
\hline
$\dot{y}$ & 0.0445 & -0.00047& -0.000468& -1.95e-05   \\
\hline
$z$& -44.1 & -0.047 & -0.047& 0.02704\\
\hline
$\dot{z}$& -0.03031 & -9.702e-05 & -9.693e-05& 3.97e-05\\ 

\hline 
\end{tabular}
\label{table:1} 
\end{table}


\section{Summary and Conclusions}
This chapter concentrates on synthesis of robust control using LQR as baseline controller. LQR controller is augmented with an extra control to compensate for the un-modeled dynamics, using an online optimized neural network algorithm to map the disturbances and uncertainties. This methodology has been simulated and results have been shown for a spacecraft formation flying problem. The possible application area could be small satellite mission which suffers with limited computation capabilities. Implementing proposed online neural network optimized LQR controller simulates the behavior of a nonlinear controller achieving mission objective with minimum terminal error in case of uncertainty and external disturbance. A comparison with SDRE controller is done to exhibit the capability of the LQR + NN controller to mimic the characteristics and performance of a nonlinear controller.

\cleardoublepage
\chapter{Conclusion}
\label{conclusion}
The main aim of this research is primarily to develop and experiment advanced algorithms for formation flying of small satellites based on various emerging philosophies of efficient solution of nonlinear optimal control problems. In this connection, the objective is to develop highly computationally efficient guidance algorithms based on various emerging techniques on optimal control theory, which can be computed in real-time with limited processing capability. In this thesis optimal control techniques namely LQR, Infinite time and Finite-time SDRE, MPSP, G-MPSP are studied and experimented for formation flying problems. As part of this work, the concept of Robust optimal control 'dynamic re-optimization' is also experimented to make the design potentially robust to un-modeled dynamics and slowly-varying external disturbances.

The biggest challenge for deputy satellites in formation flying mission, rendezvous mission is to remain in the commanded orbit once injected into final orbit. The transition of the reconfiguration trajectory of the deputy satellite should gradually attain the value of the desired orbit with minimum terminal errors. In case of large errors in attaining the position and/or velocity of the desired orbit the satellite veers of the trajectory hence necessitating the need for repeated corrective control action. The controller experimented are of the state feedback in nature. The stated control technique compute the control action proportional to the state error and compute the control to be applied continuously till the desired orbit is attained. The advantage of optimal controllers are that they take into consideration the magnitude of the error to be corrected and minimize the total control effort required for orbit transfer.

In chapter 3 it is shown that for circular chief satellite orbit and for small base-line length formation, the linear Hill's equation are very good approximation of nonlinear CW equations. It is demonstrated that for small orbit reconfiguration, LQR method performs significantly in par with nonlinear controllers and are simpler to implement on board. LQR method experimented in this thesis also forms the guess control for MPSP and G-MPSP techniques used in further chapters.

As the formation mission are considered with large relative separations that is as $\rho$ becomes larger and/or larger chief satellite eccentricity is considered, the LQR control is no more sufficient to address the increasing nonlinear behavior of the plant. A suboptimal control technique namely State Dependent Ricatti Equation(SDRE) solution is experimented in both finite time and infinite time domain. SDRE control is characterized to be a linear looking controller with nonlinear equation of motion written in state dependent coefficient(SDC) form. SDRE controller are not unique since the solution depends on the designer to choose the way plant can be rewritten in SDC form. Two SDC models are discussed in the thesis, performance of the both controller using different SDC model of plant is studied and compared to arrive at the SDC form of plant which retains as much possible nonlinearity of the plant, yet render the plant dynamics in linear looking form. The finite time SDRE solution is used as comparative method for MPSP and G-MPSP results in further chapters.

Two suboptimal guidance logics are presented in this thesis for formation flying of small satellites using the recently developed
MPSP and G-MPSP techniques. The final conditions have been put as hard conditions, because of which the solution turns out to be highly
accurate in ensuring the desired orbit for the deputy satellite is met. Comparison with the finite time SDRE solution reveals that MPSP/G-MPSP
guidance achieves the objective with tighter tolerance and with lesser amount of control usage. It was also found that the
proposed MPSP/G-MPSP guidance is computationally efficient and hence can possibly be used onboard the deputy satellites.

Final part of this thesis presents a novel ''online optimized LQR controller'' a robust controller for satellite formation flying mission in presence of uncertainties. The controller presented in this chapter uses LQR as baseline controller with linear system model. An optimization technique using online trained neural network is implemented to approximate the disturbances and uncertainties to synthesize an extra control to compensate for the un-modeled dynamics. This methodology has been simulated and results have been shown for a spacecraft formation flying problem. The possible application area could be small satellite mission which suffers with limited computation capabilities. Implementing proposed online neural network optimized LQR controller simulates the behavior of a nonlinear controller achieving mission objective with minimum terminal error in
case of uncertainty as well.

Finally it can be inferred that optimal control techniques experimented in purview of this work, prove themselves to be good platform for satellite formation flying missions involving distributed and multiple agents (deputy satellites). Nevertheless this thesis did not exhaust the numerous nonlinear control techniques applicable to the problem of formation flying. These nonlinear control techniques should be explored and evaluated in comparison with optimal control techniques before anything to be concluded about the superiority of the two.

\cleardoublepage
%

\bibtitle{References}
\input{../Bibliography/thesis.bbl}

@article{Harshal,
  author = {H. Oza and R. Padhi},
  title = {Impact Angle Constrained Suboptimal MPSP Guidance of Air-to-Ground Missiles},
  Journal = {AIAA Journal of Guidance, Control, and Dynamics},
  volume = 35,
  number = 1,
  pages = {153-164},
  month = { },
  year = 2012
}

@conference{PDas,
  author = {P. G. Das and R. Padhi},
  title = {Nonlinear Model Predictive Spread Acceleration Guidance with Impact Angle Constraint for Stationary
 Targets},
  booktitle = {Proceedings of $17^{th}$ World Congress of the International Federation of Automatic Control},
  month = { },
  year = 2008,
  address = {Seoul, South Korea}
}

@book{bryson_book,
  author =  {A. E. Bryson and  Y. C. Ho},
  title  =  {Applied Optimal Control},
  publisher  =  {Hemisphere Publishing Corporation},
  year  =  1975,
}

@book{Elbert,
  author =  {T. F. Elbert},
  title  =  {Optimal Theory: Estimation and Control System},
  publisher  =  {Von Nostord Reinhard Company},
  year  =  1984
}

@book{Slotine,
  author =  {Jean-Jacques Slotine and Weiping Li},
  title  =  {Applied Nonlinear Control},
  publisher  =  {Prentice Hall},
  year  =  1991
}

@book{rossiter,
  author =  {J. A. Rossiter},
  title  =  {Model based predictive control: A Practical Approach},
  publisher  =  {CRC},
  address = {New York},
  year  =  2003
}

@book{Werbos,
  author =  {P. J. Werbos},
  title  =  {Approximate Dynamic Programming for Real-Time Control and Neural Modeling},
  booktitle = {Handbook of Intelligent Control},
  publisher  =  {White D. A., and Sofge D.A(Eds.)},
  address = {New York: Van Nostrand Reinhold},
  pages = {65-89},
  year  =  1992
}

@book{Curtis,
  author =  {H. D. Curtis},
  title  =  {Orbital Mechanics for Engineering Students},
  publisher  =  {ELSEVIER)},
  year  =  2010
}

@article{Clohessy,
  author = {W. H. Clohessy and R. S. Wiltshire},
  title = {Terminal guidance for Satellite Rendezvous},
  Journal = {Journal of the Aerospace Sciences},
  volume = 27,
  number = 5,
  pages = {653-658,674},
  month = { },
  year = 1960
}

@article{Hill,
  author = {G. W. Hill},
  title = {Researches in Lunar theory},
  Journal = {American journal of Mathematics},
  volume = 1,
  number = 1,
  pages = {5-26,29-147,245-260},
  month = { },
  year = 1878
}

@article{Park,
  author = {H.-E. Park and S.-Y. Park and K.-H. Choi},
  title = {Satellite Formation
Reconfiguration \& Station Keeping using SDRE technique},
  Journal = {Aerospace Science And Technology},
  volume = 15,
  Issue = 16,
  pages = {440-452},
  month = { },
  year = 2011
}

@article{Ahn,
  author = {H. S. Ahn and K. L. Moore and Y. Q. Chen},
  title = {Trajectory keeping in satellite formation flying via robust periodic control},
  Journal = {International Journal Of Robust And Non-Linear Control},
  volume = 20,
  Issue = 14,
  pages = {1655-1666},
  month = { },
  year = 2010
}

@conference{Vadali,
  author = {S. R. Vadali and H. Schaub and K. T. Alfriend},
  title = {Initial condition and fuel optimal control for formation flying satellites},
  booktitle = {AIAA Guidance, Navigation and Control Conference},
  month = {August},
  year = 1999,
  address = {Portland, USA}
}

@article{Irvin,
  author = {D. J. Irvin and D. R. Jacques},
  title = {A study of linear versus nonlinear control techniques
for the reconfiguration of satellite formations},
  Journal = {Advances in the Astronautical Sciences},
  pages = {589-608},
  month = { },
  year = 2002
}

@misc{Lorenz,
  author = {M-M Burlacu and P. Lorenz},
  title = {{A survey of small satellites domain:challenges, applications and communications key issues}},
  howpublished = "\url{http://icast.icst.org/2010/09/survey-small-satellites-domain-challenges-applications-and-communications-key-issues}",
  year = {},
  note = "[Online; Accessed on June 22, 2012]"
}

@article{Alf,
  author = {D. W. Gim and K. T. Alfriend},
  title = {State transition matrix of relative motion for the perturbed
noncircular reference orbit},
  Journal = {Journal of Guidance, Control, and Dynamics},
  volume = 26,
  number = 6
  pages = {956-971},
  month = { },
  year = 2003
}

@article{de2000adaptive,
  title={Adaptive nonlinear control of multiple spacecraft formation flying},
  author={De Queiroz, Marcio S and Kapila, Vikram and Yan, Qiguo},
  journal={Journal of Guidance, Control, and Dynamics},
  volume={23},
  number={3},
  pages={385--390},
  year={2000}
}

@article{schaub2001impulsive,
  title={Impulsive feedback control to establish specific mean orbit elements of spacecraft formations},
  author={Schaub, Hanspeter and Alfriend, Kyle T},
  journal={Journal of Guidance, Control, and Dynamics},
  volume={24},
  number={4},
  pages={739--745},
  year={2001}
}

@article{mishne2004formation,
  title={Formation control of satellites subject to drag variations and J2 perturbations},
  author={Mishne, David},
  journal={Journal of guidance, control, and dynamics},
  volume={27},
  number={4},
  pages={685--692},
  year={2004}
}

@article{prussing1986optimal,
  title={Optimal multiple-impulse time-fixed rendezvous between circular orbits},
  author={Prussing, John E and Chiu, J-H},
  journal={Journal of Guidance, Control, and Dynamics},
  volume={9},
  number={1},
  pages={17--22},
  year={1986}
}
\end{document}